\documentclass[10pt]{article}
\usepackage{graphicx}
\usepackage[left=2.5cm, top=3.5cm, right=2.5cm]{geometry}
\usepackage{latexsym}
\usepackage{algorithmic}
\usepackage{algorithm}
\newtheorem{lemma}{Lemma}
\newtheorem{openquestion}{Open question}
\newtheorem{conjecture}{Conjecture}
\def\qed{$\Box$}

\newtheorem{theorem}{Theorem}

\title{Tromino Tilings of Domino Deficient Rectangles}
\author{Mridul Aanjaneya\footnote{Department of Computer Science and 
Engineering, IIT Kharagpur, 721302, India [Email: mridul@cse.iitkgp.ernet.in]}} 
\date{}

\begin{document}
\maketitle
\begin{abstract}
We consider tromino tilings of $m\times n$ domino-deficient rectangles, 
where $3|(mn-2)$ and $m,n\geq0$, and 
characterize all cases of domino removal   
that admit such tilings, thereby settling the open problem  
posed by J. M. Ash and S. Golomb in \cite {marshall}. 
Based on this characterization, we design a procedure
for constructing such a tiling  
if one exists. We also consider the 
problem of counting such tilings and derive the exact formula for the number 
of tilings for $2\times(3t+1)$ rectangles, the exact generating function for 
$4\times(3t+2)$ rectangles, where $t\geq0$, and an upper bound on the number of tromino 
tilings for $m\times n$ domino-deficient rectangles. 
We also consider general 2-deficiency in $n\times4$ rectangles, where $n\geq8$, and characterize all 
pairs of squares which do not permit a tromino tiling.  
\end{abstract}

\section{Introduction}
\label{intro}

Tiling the plane is an interesting field of recreational
mathematics. In a 1953 talk at the Harvard Math Club, Solomon Golomb defined a
class of geometric figures called {\it polyominoes}, namely, connected figures formed
of congruent squares placed so each square shares one side with at least one other square.
{\it Dominoes}, which use two squares, and {\it Tetrominoes} (the {\it Tetris} pieces), which use
four squares, are well known to game players. Golomb first published a paper about
polyominoes in {\it The American Mathematical Monthly} \cite{gol}. Polyominoes were 
later popularized by Martin Gardner in his {\it Scientific American} columns called ``Mathematical
Games" (see, for example, \cite{mar1}, \cite{mar2}). 
A region is {\it tiled} with a given tile if it is completely covered by its copies 
without any overlap. 
Several results about tiling regular shapes with polyominoes
are mentioned by Stanley and Ardila \cite{ardila}, and Do \cite{math}.
Many of the initial questions asked about polyominoes concern the 
number of {\it n-ominoes} (those formed from $n$ squares), and what shapes can be tiled 
using just one of the polyominoes, possibly leaving one or two squares uncovered. 

In this paper we consider tilings of rectangles using 3-ominoes or {\it trominoes} of which 
there are two basic shapes, namely a $1\times3$ rectangle and an L-shaped figure (more commonly known 
as the {\it right tromino}). 
We restrict ourselves to tilings only with the 
right tromino. From now on, we 
will simply say {\it tromino} to mean the right tromino. 
In the past few years, tromino tilings of rectangles have also been studied quite extensively. 
Chu and Johnsonbaugh \cite{chu}, first characterized all $m\times n$ rectangles that 
permit a tromino tiling. 

\begin{theorem}{\bf (Chu-Johnsonbaugh Theorem \cite{chu})}
An $m\times n$ rectangle
can always be tiled by trominoes if $3|mn$, $2\leq m\leq n$,  
except for $3\times(2k+1)$ rectangles where $k\geq 1$. 
\end{theorem}
 
A rectangle 
from which one square has been removed is called a {\it deficient rectangle}. Golomb \cite{gol} 
proved that deficient squares whose side length is a power of two can be tiled 
by trominoes. 
Chu and Johnsonbaugh extended Golomb's work to the general cases of deficient squares \cite{john}. 
Ash and Golomb \cite{marshall} considered the problem of tiling a deficient rectangle. 
They characterized all positions of
a square whose removal still permits a tromino tiling. Their result is as follows:

\begin{theorem}{\bf (Deficient Rectangle Theorem \cite{marshall})}
An $m\times n$ deficient rectangle, $2\leq m\leq n$, $3|(mn-1)$, has a tiling, regardless of the position 
of the missing square, if and only if (a) neither side has length 2 unless both of them do, and 
(b) $m\neq 5$. 
\end{theorem}

A slightly weaker version of this theorem was proved by Chu and Johnsonbaugh in \cite{chu}. 
Now consider the problem of tiling a rectangle from which two squares have been removed. 
This problem was posed as an open problem in \cite{marshall}.  
A rectangle from which two squares are missing is called a 
{\it 2-deficient rectangle}. 
In this paper, we consider tromino tilings of a special class of 2-deficient rectangles, 
namely, the {\it domino-deficient rectangles}. 
These are rectangles from which a domino 
has been removed. We characterize all cases of domino removal in such rectangles which 
do not permit a tromino tiling. 
It turns out that there are only 16 cases of domino 
removal which prevent a tiling for any $m\times n$ rectangle, 
where $m,n\geq7$ and $3|(mn-2)$. These positions are   
$\{(1,2),(2,2)\}$, $\{(2,1),(2,2)\}$, $\{(2,3),(2,4)\}$, $\{(3,2),(4,2)\}$ 
and their symmetric counterparts (viewed w.r.t. the other three corners). 
Based on this characterization, we design a procedure 
for constructing a tiling of a given domino-deficient rectangle 
if one exists. We also consider the 
problem of counting such tilings and derive the exact formula for the number 
of tilings for $2\times(3t+1)$ rectangles and the exact generating function for 
$4\times(3t+2)$ rectangles, where $t\geq0$. We derive an upper bound on the number of tromino 
tilings of arbitrary $m\times n$ domino-deficient rectangles. 
We also consider general 2-deficiency in $n\times4$ rectangles, where $n\geq8$, and characterize all 
pairs of squares which do not permit a tromino tiling. 

\subsection{Definitions and notation}

\begin{figure}[htbp]
\centerline{
{\scalebox{.25}{\includegraphics{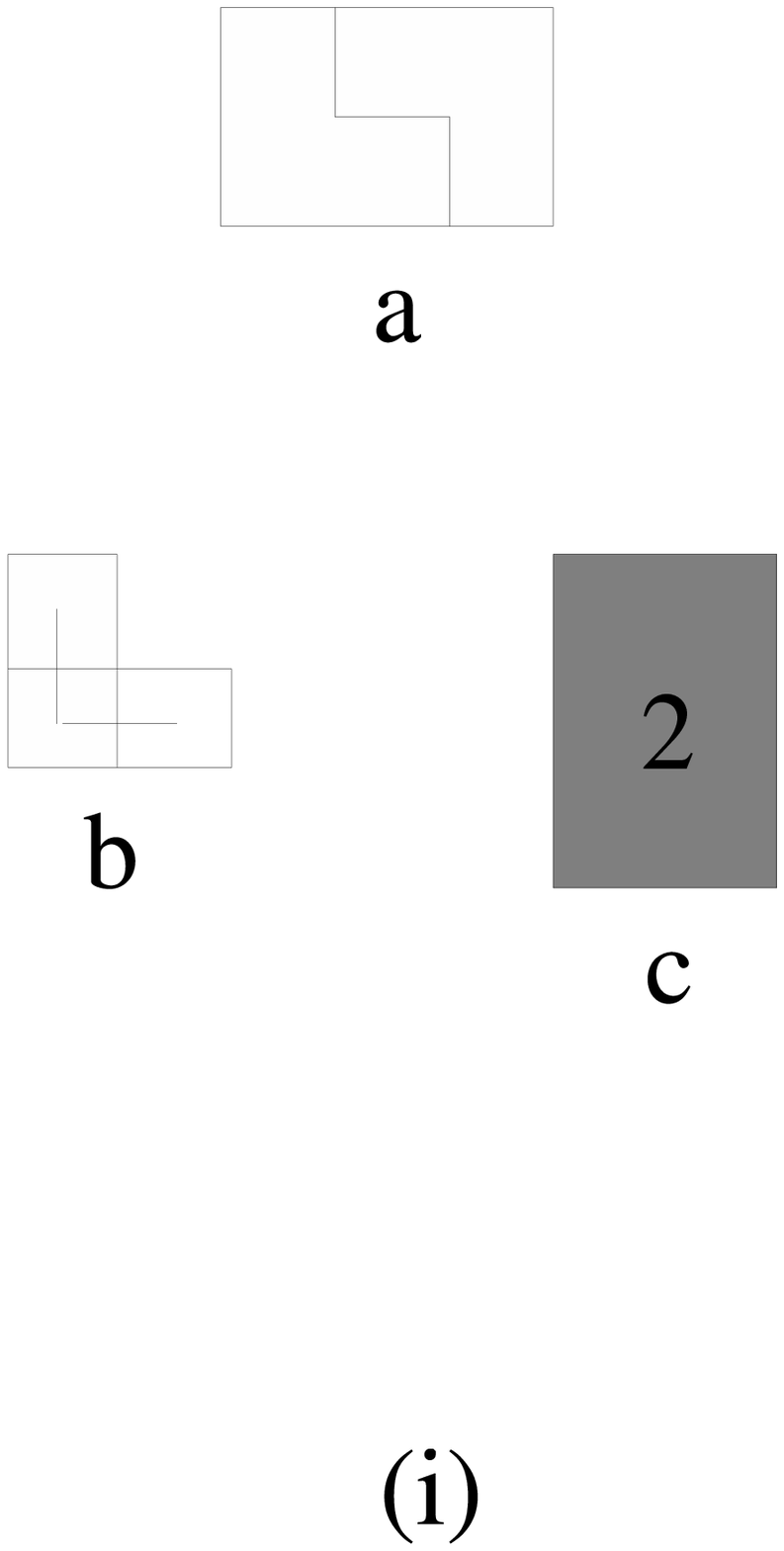}}}
{\scalebox{.2}{\includegraphics{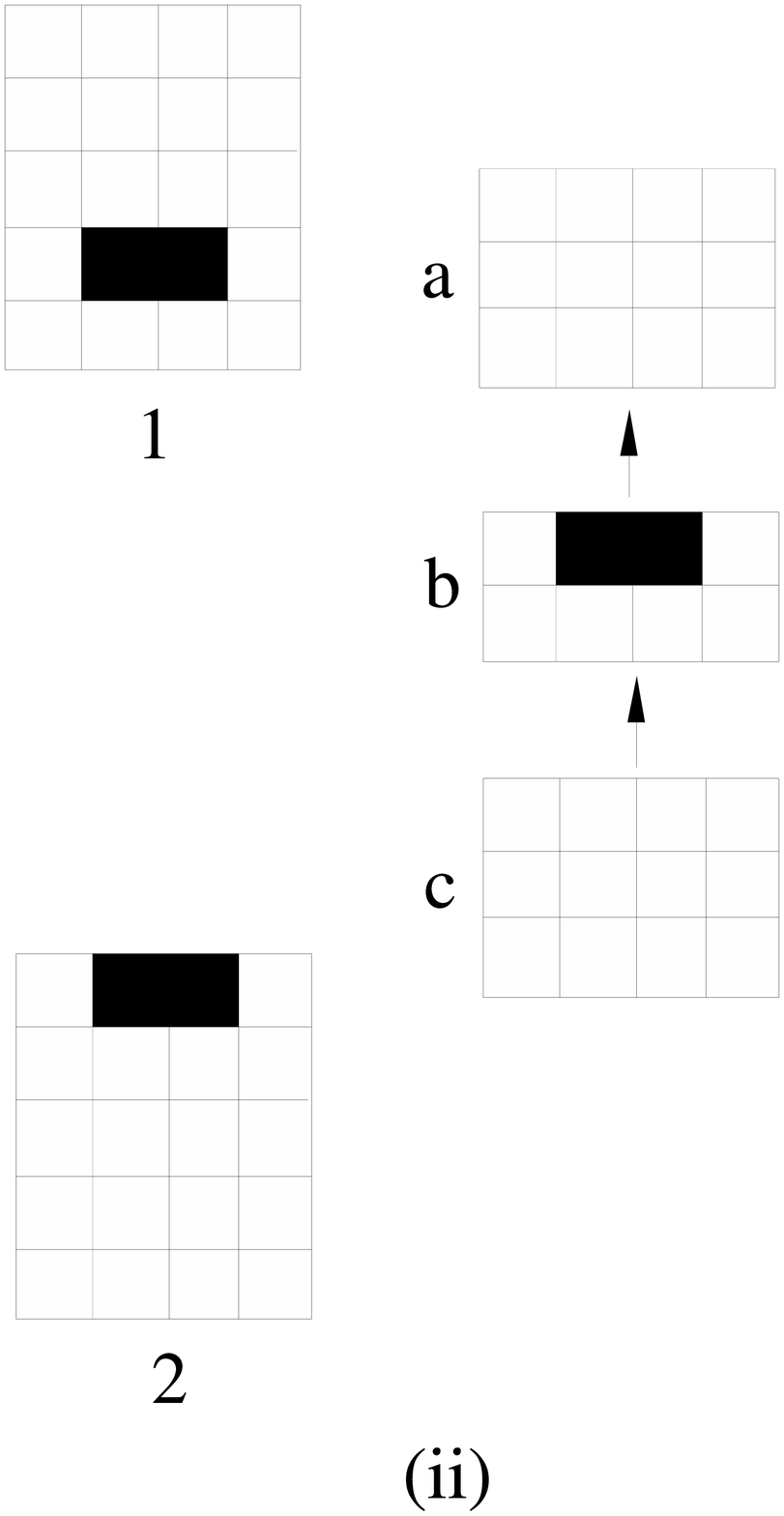}}}
{\scalebox{.2}{\includegraphics{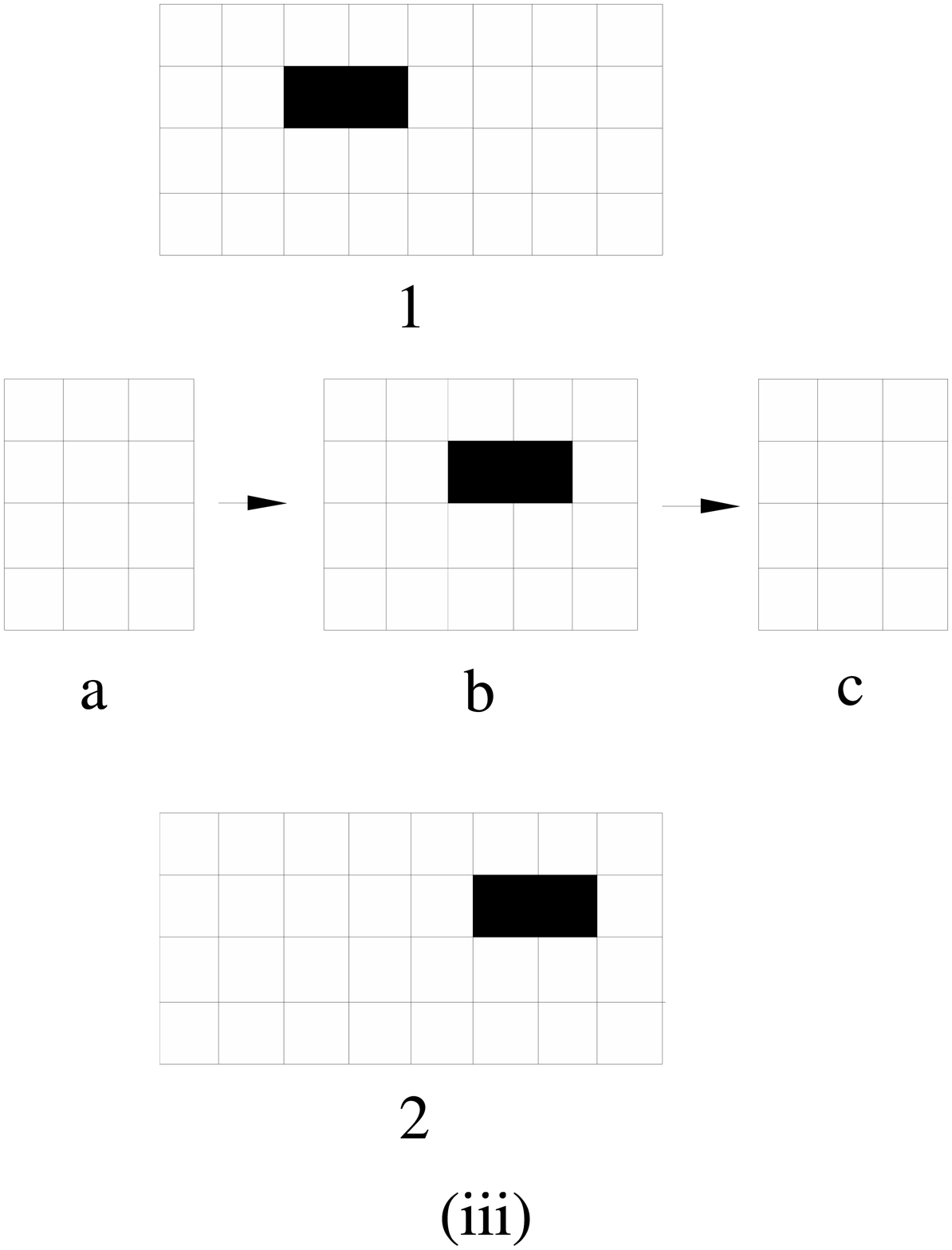}}}
}
\caption{(i) Notations for a tromino tiling of $R(i,j)$. (ii) A $(3,4)$-{\it vquad shift}.
(iii) A $(4,3)$-{\it hquad shift}.}
\end{figure}

Firstly, the reader should note that a $2\times3$ rectangle can be tiled with trominoes
(as shown in Figure 1(i)(a)). We denote a rectangle with $i$ rows and $j$ columns by $R(i,j)$.
We will indicate decompositions into non-overlapping subrectangles by means of an additive 
notation. For example, a $3i\times 2j$ rectangle can be decomposed into $ij$ 
subrectangles of dimension $3\times 2$  
and we write this fact as $R(3i,2j) = \sum_{a=1}^{i}\sum_{b=1}^{j}R(3,2) = ijR(3,2)$. It follows 
from this and Figure 1(i)(a) that any $2i\times 3j$ or $3i\times 2j$ rectangle can be tiled with trominoes. 
For such rectangles, a very ``trivial" tiling rule 
would be to place $2\times3$ or $3\times2$ rectangles lengthwise. 
From now on, any rectangle decomposed into a combination of $3i\times 2j$ subrectangles, $2i\times 3j$ 
subrectangles and trominoes will be considered as successfully tiled by trominoes. The square lying in 
row $i$ and column $j$ is denoted as $(i,j)$. To make the notation simple, trominoes are depicted, 
in the rest of the paper, as a composition of two lines forming an L-shape across an actual tromino 
(as shown in Figure 1(i)(b)). Domino-deficient rectangles with $i$ rows and $j$ columns are denoted in 
general by $R(i,j)^{--}$. All $R(3,2)$ and $R(2,3)$ rectangles 
are shown as gray-shaded rectangles labeled by the number $2$, since 
they can be tiled in two ways 
(as shown in Figure 1(i)(c)). 
Any pair of missing squares that does not allow a tiling of the resultant structure is referred to as a 
{\it bad pair}. We will always specify bad pairs with respect to the {\it top left corner} of the given $m\times n$ 
rectangle. The reader should note that the bad pairs for 
$R(m,n)^{--}$ are the same as those for $R(n,m)^{--}$ (the only 
difference is that their corresponding coordinates 
change w.r.t. the 
top left corner). For a given $m\times n$ rectangle, we define a 
$(m,k)$-{\it hquad shift} to be a process of detaching the leftmost (rightmost) $k$ columns and attaching 
an $m\times k$ rectangle from the right (left). For rectangles with no deficiency, such an operation has 
no meaning. However, for domino-deficienct rectangles, such an operation changes the position of the 
missing domino w.r.t. the top left corner of the rectangle, although size of the given rectangle remains the same. 
Figure 1(iii) shows a $(4,3)$-hquad shift for $R(4,8)^{--}$. 
Similarly, for 
an $m\times n$ rectangle, we define a $(k,n)$-{\it vquad shift} to be a process of detaching the uppermost 
(lowermost) $k\times n$ rectangle and attaching a $k\times n$ rectangle from the bottom (top). 
Figure 1(ii) shows a $(3,4)$-vquad shift for $R(5,4)^{--}$. 
We will use a vquad (hquad) shift to change an untileable configuration of a rectangle into a tileable one. 
The reader should note that we do not specify explicitly the direction associated with such shifts, however, 
the direction will be clear from the context, since we will only apply such shifts 
to rectangles where shifting is possible in exactly one direction. 

\subsection{Organization of the Paper}

We present some elementary results for rectangles in Section 2, thereby giving an 
informal proof of the Chu-Johnsonbaugh Theorem. 
In Section 2.1, we present a special case of domino-deficiency, 
where a domino is removed from a corner of the given rectangle. 
We characterize all rectangles that admit such tilings. 
We then move on to consider general domino-deficient rectangles, where 
the position of the missing domino is no more restricted to a corner. 
For an $m\times n$ domino-deficient rectangle to be tileable by trominoes, 
the resultant area $(mn-2)$ must be divisible by 3, i.e., $mn\equiv2$(mod 3). 
Taking into consideration a rotation by one right angle we assume without 
loss of generality, that $m\equiv1$(mod 3) and $n\equiv2$(mod 3). 
So, $m$ can assume the values $1, 4, 7,...(1+3t)$, where $t\geq0$, and $n$ 
can assume the values $2, 5, 8,...(2+3t)$, where $t\geq0$. Since no tromino 
can fit in one row or column, $m\geq4$. We present the entire analysis 
of characterizing bad cases of domino removal in $m\times n$ rectangles 
in four exhaustive subcases as follows. 
In Section 3, we consider the simplest of  
these four cases, where we consider tiling a $R(2,3t+4)^{--}$ rectangle, where $t\geq0$. 
We also address the problem of counting the number of tilings for such rectangles 
in Section 3.1, and derive the exact closed-form formula. 
In 
Section 4, we consider tiling $R(4,3t+8)^{--}$ rectangles. 
We also consider counting 
tilings for such rectangles in Section 4.1, 
and derive the exact generating function. 
In Section 5, we characterize domino-deficiency in $R(5,3t+4)^{--}$ 
rectangles and then proceed to 
consider tilings in arbitrary $m\times n$ domino-deficient rectangles in Section 6. 
We prove that there are only 16 bad cases of domino removal in any $m\times n$ 
domino-deficient rectangle, where $m,n\geq7$; we also design a 
procedure for constructing such a tiling if one exists. 
In Section 7, we derive an 
upper bound on the number of tromino tilings of arbitrary $m\times n$ domino-deficient rectangles. 
Finally, in Section 8, we present an approach to study general $2$-deficiency in rectangles and 
characterize all bad pairs for $n\times4$ rectangles, where $n\geq8$.

\section{Elementary results for rectangles}

First consider some rectangles that cannot be tiled. 
We show by contradiction that a $3\times3$ square $Q$ cannot be tiled. 
There are three possible ways of covering the square $(3,1)$, 
as shown in Figure 2(i)(a)-(c). 
Orientation 2(i)(c) is immediately ruled out, 
since square $(1,1)$ cannot be tiled. In the cases 2(i)(a) and (b), a feasible tiling must 
tile the  leftmost $3\times2$ subrectangle of $Q$, so that it is also 
a tiling of the third column $R(3,1)$ of $Q$, a contradiction. Similarly, 
one can show by contradiction that a $3\times5$ rectangle $R$ cannot be tiled. 
The above argument shows that a feasible
tiling must tile the first two columns of $R$, and hence, also the rightmost three columns 
of $R$, a contradiction since we have just shown a $3\times3$ square to be untileable. 
Iterating this procedure, one can show that no $R(3,odd)$ can be tiled. 
It turns out that there are no 
other untileable rectangles with area divisible by 3. 
Consider the following three decompositions: 

\begin{eqnarray}
R(3t,2k)     & = & tk\cdot R(3,2), t,k\geq1 \\
R(6t,2k+3)   & = & R(6t,2k) + R(6t,3), t,k\geq1 \\
R(9+6t,2k+5) & = & R(9+6t,2k) + R(9,5) + R(6t,2) + R(6t,3), t,k\geq0 
\end{eqnarray}

The reader should verify that any $m\times n$ rectangle can be 
written in the form $R(3k,even)$, $R(6k,odd)$ or $R(9+6k,n)$, $n\geq5$ and $k\geq0$, 
where $2\leq m\leq n$, such that 
if one of $m$ and $n$ is 3 then the other is not odd. 
Equations (1)-(3) show the corresponding tiling rules in each of these cases, the tiling 
of $R(5,9)$ (and hence $R(9,5)$) is shown in Figure 2(i)(d). 
Note that apart from $R(9,5)$, each of the subrectangles obtained in equations (1)-(3) has dimensions 
$3i\times2j$ or $2i\times3j$, where $i,j\geq1$, and so is tileable. 
So we have an 
informal proof of the Chu-Johnsonbaugh Theorem. 

\subsection{Dog-Eariness in domino-deficient rectangles}
\label{codes}

We now consider tromino tilings of a special class of domino-deficient rectangles, namely, 
the {\it domino-deficient dog-eared rectangles}, where a domino is removed from a 
corner of the given rectangle. 
In our present discussion, we assume that the domino was removed from the 
{\it top right corner}. We will denote an $m\times n$ domino-deficient dog-eared rectangle 
by $R(m,n)^{--}$. If this rectangle is rotated by $\pi$ radians, a similar figure with 
missing lower left-hand corner is created. If it is reflected about a central vertical 
(resp., horizontal) axis, a similar figure with missing lower right-hand (resp., upper left-hand) 
corner is created. The problem of tiling the original figure is clearly equivalent to tiling any 
of the other six cases (note that the missing domino can be either vertical or horizontal). 
We have the following result: 

\begin{theorem}{\bf [Domino-Deficient Dog-Eared Rectangle Theorem]} \\
$R(m,n)^{--}$ can always be tiled by trominoes if a domino is removed from a corner, 
provided $mn\equiv2$(mod $3$) and $m,n\geq4$. 
\end{theorem}
\begin{proof}
Without loss of generality, assume $m=3j'+1$ and $n=3k'+2$, where $j',k'\geq1$.
Set $j'=j+1, k'=k+1$, and so $m=3j+4$ and $n=3k+5$, where $j,k\geq0$. 
In each of the cases shown we will break the given $m\times n$ rectangle into subrectangles, 
which either satisfy the conditions of the Chu-Johnsonbaugh Theorem \cite{chu}, 
and so are tileable by trominoes, 
or are one among the cases shown in Figures 2(ii)(a)-(d), in which case we apply the corresponding tiling 
shown in Figure 2(ii). We first consider the cases when $m=4,7$. 
When $m=4$, we have only one case 
$R(4,3k+5)^{--}$ with $k\geq0$. When $m=7$, since 7 has different parity, in order to satisfy the 
criterions of the Chu-Johnsonbaugh Theorem, we divide this case 
into two subcases, accordingly as $k$ is even or odd. So, 
we have to tile either $R(7,6l+5)^{--}$, where $k=2l$ and $l\geq0$, 
or $R(7,6l+8)^{--}$, where $k=2l+1$ and $l\geq0$. 
There correspond these three decompositions:  

\begin{figure}[htbp]
\centerline{
\scalebox{.2}{\includegraphics{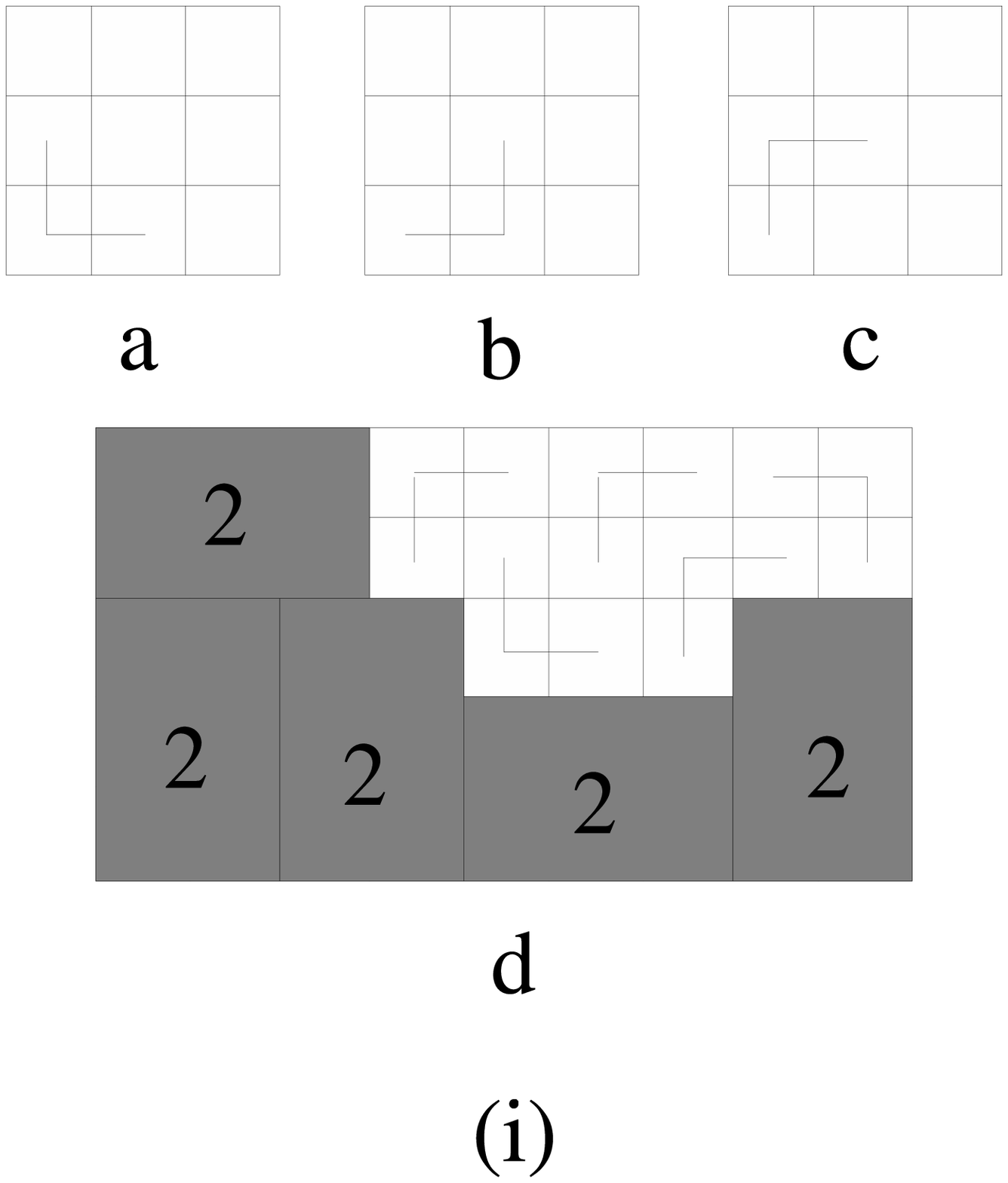}}
\scalebox{.2}{\includegraphics{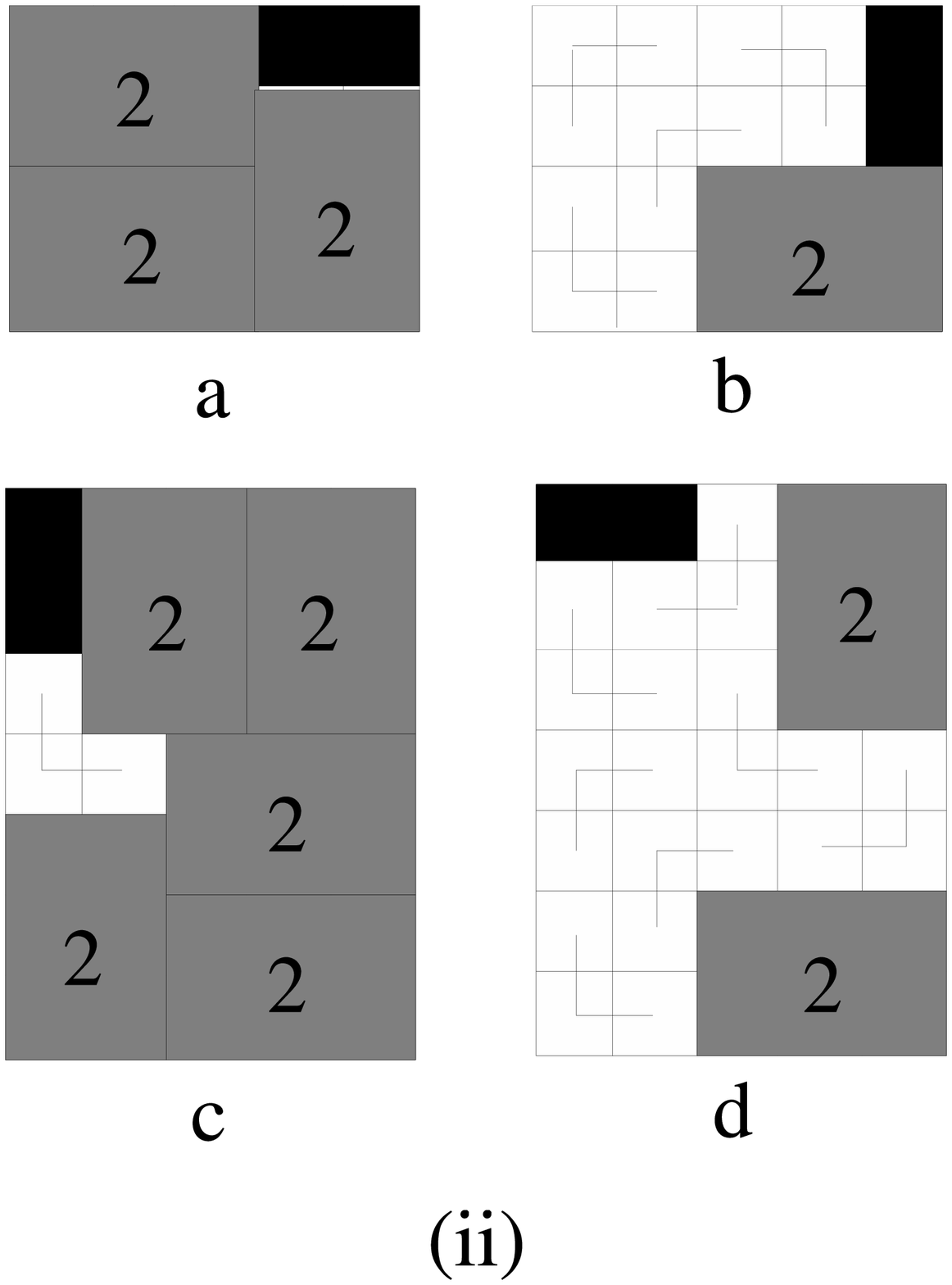}}
}
\caption{(i) (a)-(c) An impossibility proof. (d) Tiling of $R(9,5)$.  
(ii) (a) and (b) Tilings of $R(4,5)^{--}$. (c) and (d) Tilings of $R(7,5)^{--}$
(Symmetric cases are also possible). }
\end{figure}

\begin{eqnarray}
R(4,3k+5)^{--} & = & R(4,3k) + R(4,5)^{--} \\ 
R(7,6l+5)^{--} & = & R(7,6l) + R(7,5)^{--} \\
R(7,6l+8)^{--} & = & R(7,6l) + R(7,8)^{--}  \nonumber \\
               & = & R(7,6l) + R(3,8) + R(4,3) + R(4,5)^{--}
\end{eqnarray}

For the algebraically inclined reader, these decompositions need no further explanation. However, the 
geometrically inclined reader should draw pictures to visualize them. (All the similar decompositions 
appearing below have straightforward geometrical interpretations.) Here, in the first two cases, a 
large rectangle was stripped from the left side of the figure. In the third case, a large rectangle 
was first stripped from the left side of the figure to obtain $R(7,8)^{--}$, from which the rectangle 
$R(3,8)$ was removed from the bottom, to get $R(4,8)^{--}$. $R(4,3)$ was then stripped from its left 
side to finally get $R(4,5)^{--}$. All the full rectangles are tileable since they 
satisfy the conditions of the Chu-Johnsonbaugh Theorem, $R(4,5)^{--}$ and 
$R(7,5)^{--}$ are tiled as in Figure 2(ii).  
We now consider the case when $m=3j+4, n=3k+5$, where $j\geq2, k\geq0$. 
Since $m$ can have different parity, in order to satisfy the 
conditions of the Chu-Johnsonbaugh Theorem, we 
divide the above case into two subcases accordingly as $k$ is even or odd.
We have the following decompositions: 

\begin{eqnarray}
R(3j+4,6l+5)^{--} & = & R(3j+4,6l) + R(3j+4,5)^{--} \nonumber \\
                  & = & R(3j+4,6l) + R(3j,5) + R(4,5)^{--}  \\
R(3j+4,6l+8)^{--} & = & R(3j+4,6l) + R(3j+4,8)^{--} \nonumber \\
                  & = & R(3j+4,6l) + R(3j,8) + R(4,3) + R(4,5)^{--}
\end{eqnarray}

In these cases too, we strip the $m\times n$ rectangle  
as above, obtaining $R(4,5)^{--}$ (tileable as in Figure 2(ii)(a)-(b)) and full rectangles; 
$R(3j+4,6l)$, $R(3j,5)$, $R(3j,8)$ and $R(4,3)$, that satisfy the criterions of the 
Chu-Johnsonbaugh Theorem, and so are tileable by trominoes. 
The reader should note that $R(3j,5)$ in equation (7) does not violate the conditions of 
the Chu-Johnsonbaugh Theorem. Since $j\geq2$, $R(3j,5)$ can never denote $R(3,5)$, which 
is untileable. A similar reasoning holds for the case when $m=3j'+2$ and $n=3k'+1$, where $j',k'\geq1$, and 
so our result follows. \hfill\qed 
\end{proof}

We now consider general domino-deficient rectangles, 
where the position of the missing domino is no more restricted to a corner.
We will denote an $m\times n$ domino-deficient rectangle by $R(m,n)^{--}$. 
For an $m\times n$ domino-deficient rectangle to be tileable by trominoes, the resultant 
area ($mn-2$) must be divisble by 3, i.e., $mn$ $\equiv$ $2$(mod $3$). 
Taking into consideration a rotation by one right angle 
we assume without loss of generality, that 
$m\equiv 1$(mod 3), and $n\equiv2$(mod 3). 
So, $m$ can assume the values 1, 4, 7,...$(1+3t)$, where $t\geq0$, 
and $n$ can assume the values 2, 5, 8,...$(2+3t)$, where $t\geq0$. 
Since no tromino can fit in one row or column, 
$m\geq4$. 
We present the entire analysis of characterizing bad 
cases of domino removal in $m\times n$ rectangles in four exhaustive subcases as follows.  
In Section 3, we consider tilings of $R(2,3j+4)^{--}$ rectangles, in Section 4, we consider 
tilings of $R(4,3j+8)^{--}$ rectangles, in Section 5, we 
consider tilings of $R(5,3j+4)^{--}$ rectangles, and finally in Section 6, 
we consider tilings of $R(3j+7,3k+8)^{--}$ rectangles, 
where $j,k\geq0$. 

\section{Tromino tilings of $R(2,3t+4)^{--}$ rectangles}
\label{counts}

We start with the simplest of the four cases of domino-deficiency enumerated in the previous section, 
namely, when one dimension of the given $m\times n$ rectangle (say $m$) is 2. 
First consider tromino tilings of $2\times3t$ rectangles 
instead of domino-deficient $2\times(3t+1)$ rectangles, where $t\geq1$. 
Consider the orientations of the tromino covering $(1,1)$ 
(as shown in Figure 3(a)-(b)). 
If it covers $(2,1)$ (see Figure 3(a)) then a $2\times3$ rectangle 
($R(2,3)$) is completed by the tromino covering $(2,2)$, 
and if it covers $(2,2)$ (see Figure 3(b)) then $R(2,3)$ is completed 
by the tromino covering $(2,1)$. So, any tiling of 
$R(2,3t)$ consists of a tiling of $R(2,3)$ on the leftmost side followed 
by a tiling of $R(2,3(t-1))$ (see Figure 3(c)). 
Let $T(2,3t)$ denote the number of tromino tilings of $R(2,3t)$. 
Taking $R(2,3)$ as the basis case and performing induction on $t$, 
using the above argument in the inductive step, we 
conclude that $T(2,3t)=2^t$.
We state this result in the following lemma: 

\begin{figure}[htbp]
\centerline{\scalebox{.2}{\includegraphics{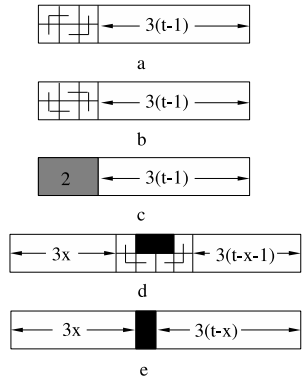}}}
\caption{Tilings of $R(2,3t)$ and $R(2,3t+1)^{--}$.}
\end{figure}

\begin{lemma}
The number of tromino tilings of a $2\times3t$ rectangle, $T(2,3t)=2^t$. 
\end{lemma}

We now consider tromino tilings of $R(2,3t+1)^{--}$ rectangles, 
where $t\geq1$. Consider the situation when 
the removed domino is vertical (see Figure 3(e)). 
This domino divides the given $2\times(3t+1)$ rectangle into two 
subrectangles. These two smaller subrectangles must be completely tileable by trominoes, since 
each tromino occupies an area of 3, so 
the area of both these rectangles must be divisible by 3. 
So, we conclude that a vertical domino can be removed only from the columns $x=3k+1$, 
where $k\geq0$. Now consider the case when a horizontal domino is removed from $R(2,3t+1)$. 
Without loss of generality, assume that it is present in the $1st$ row and occupies $(1,r)$ 
and $(1,r+1)$. The reader can easily see that the only way of covering $(2,r)$ and $(2,r+1)$ 
is as shown in Figure 3(d). Again, we get two smaller rectangles which must be tileable by trominoes. 
We conclude that there areas must also be divisible by 3. So, the only 
possible values of $r=3x+2$, where $x\geq0$. We summarize these results in the following theorem:  
 
\begin{theorem}{\bf [Domino-Deficient Direc Theorem]} 
In case of a vertical domino removal from $R(2,3t+4)$, the remaining area permits a 
tromino tiling if and only if the missing domino occupies the position $\{(1,3k+1),(2,3k+1)\}$, 
where $k\geq0$. In case of a horizontal domino removal from $R(2,3t+4)$, the remaining 
area permits a tiling if and only if the missing domino occupies either the position 
$\{(1,3x+2),(1,3x+3)\}$ or the position $\{(2,3x+2),(2,3x+3)\}$, where $x\geq0$. 
\end{theorem}

\subsection{Counting tilings of $R(2,3t+4)^{--}$ rectangles}

Apart from proving existence of tilings for $R(2,3t+1)^{--}$ rectangles, $t\geq1$, the Domino-Deficient 
Direc Theorem is also constructive in nature. This fact can also be used for counting the 
number of tromino tilings of $R(2,3t+1)$ with $2t$ trominoes and one domino. Let $T(2,3t+1)^{--}$ 
denote this number. First consider the case when the removed domino is vertical 
and let $T_V(2,3t+1)^{--}$ be the number of tilings in this case. From the 
Domino-Deficient Direc Theorem, the domino can only occupy the columns $x=3k+1$, where $k\geq0$. 
So the two smaller rectangles on either side are $R(2,3k)$ and $R(2,3(t-k))$. 
Using this fact and Lemma 1, we immediately arrive at the recurrence:

\begin{eqnarray}
T_V(2,3t+1)^{--} & = & \sum_{k=0}^{t}T(2,3k)\times T(2,3(t-k)) \nonumber \\ 
                 & = & \sum_{k=0}^{t}2^k\times 2^{t-k} \nonumber \\
                 & = & (t+1)\cdot2^t
\end{eqnarray}

Now consider the case when the removed domino is horizontal and let $T_H(2,3t+1)^{--}$ 
denote the number of tilings in this case. The Domino-Deficient Direc Theorem 
states that the removed domino can only occupy the pair of columns $(3x+2,3x+3)$, where $x\geq0$. 
So the smaller rectangles on either side are $R(2,3x)$ and $R(2,3(t-x-1))$. 
The removed domino can be either in the $1st$ or the $2nd$ row, introducing 
an additional factor of 2. 
From the above conditions and Lemma 1, we get the following recurrence: 

\begin{eqnarray}
T_H(2,3t+1)^{--} & = & 2\sum_{x=0}^{t-1}T(2,3x)\times T(2,3(t-x-1)) \nonumber \\ 
                 & = & 2\sum_{x=0}^{t-1}2^x\times 2^{t-x-1} \nonumber \\
                 & = & t\cdot2^t
\end{eqnarray}

The removed domino can be either horizontal or vertical. So, from (9) and (10), we conclude that 
$T(2,3t+1)^{--}=T_H(2,3t+1)^{--}+T_V(2,3t+1)^{--}=(t+1).2^t+t.2^t=(2t+1).2^t$. 
We summarize this result in the following theorem: 

\begin{theorem}
The number of tromino tilings of $R(2,3t+1)^{--}$, where $t\geq1$, is 
\begin{eqnarray}
T(2,3t+1)^{--} & = & (2t+1)\cdot2^t
\end{eqnarray} 
\end{theorem}

The reader may think that this result is 
surprising and contrary to his expectations. 
Just by removing a domino from the $2\times n$ rectangle, 
the complexity of the number of tromino tilings changes from $\mathcal{O}(2^t)$ to $\mathcal{O}(t.2^t)$, 
although our intuition says that adding deficiency should be pretty restrictive on the 
orientations of some trominoes, and so the number of tilings should decrease! 
The catch lies in observing that the size of the 
rectangle being considered also increases (from $R(2,3t)$ 
to $R(2,3t+1)$), so the number of trominoes remains the same, i.e., $2t$. 
Moreover, we also take into account all permissible positions of 
the missing domino. 

\section{Tromino tilings of $R(4,3t+8)^{--}$ rectangles}

Consider the second case when one dimension of the given $m\times n$ rectangle (say $m$) is 4. 
The bad pairs for the case $n=5$ will be enumerated separately in another section, so for now we will 
assume that $n\geq8$. See Figure 4. The bad pairs for $R(4,8)^{--}$ have been 
indicated by dark squares. It 
turns out that these are the only pairs that do not permit a tromino tiling. As the reader can see, the 
pairs of squares $\{(1,2),(2,2)\}$, $\{(2,1),(2,2)\}$, $\{(3,1),(3,2)\}$, $\{(3,2),(4,2)\}$, 
$\{(2,7),(2,8)\}$, $\{(1,7),(2,7)\}$, $\{(3,7),(3,8)\}$, $\{(3,7),(4,7)\}$ make the cornermost squares 
$(1,1), (1,8), (4,1), (4,8)$ inaccessible, and so, do not permit a tromino tiling. Now consider the four 
pairs $\{(2,3),(2,4)\}$, $\{(2,5),(2,6)\}$, $\{(3,3),(3,4)\}$, $\{(3,5),(3,6)\}$. 
In the pair $\{(2,3),(2,4)\}$ the tromino covering $(1,3)$ 
makes the square $(1,1)$ inaccessible. Similarly, in the case of the 
pairs $\{(2,5),(2,6)\}$, $\{(3,3),(3,4)\}$, $\{(3,5),(3,6)\}$ the trominoes covering 
$(1,6)$, $(4,3)$, and $(4,6)$ make the squares $(1,8)$, $(4,1)$ and $(4,8)$ inaccessible. So, we conclude 
that these pairs are also bad (refer Figure 5(ii)). 
Finally, consider the pairs $\{(2,3),(3,3)\}$ and $\{(2,6),(3,6)\}$. 
Since the two pairs are symmetric (reflections of each other in the vertical axis), 
we need only consider the badness of $\{(2,3),(3,3)\}$. If the tromino covering $(1,3)$ covers $(1,2)$, 
then the square $(1,1)$ becomes inaccessible. So this tromino must cover $(1,4)$ (see Figure 5(i)).
Similarly, the tromino 
covering $(4,3)$ must cover $(4,4)$. But now we have isolated a $4\times2$ rectangle which does not 
satisfy the conditions of the Chu-Johnsonbaugh Theorem, and so is untileable. We conclude that the pairs 
$\{(2,3),(3,3)\}$ and $\{(2,6),(3,6)\}$ are also bad. We are now ready to prove the following theorem:	

\begin{figure}[htbp]
\centerline{
{\scalebox{.2}{\includegraphics{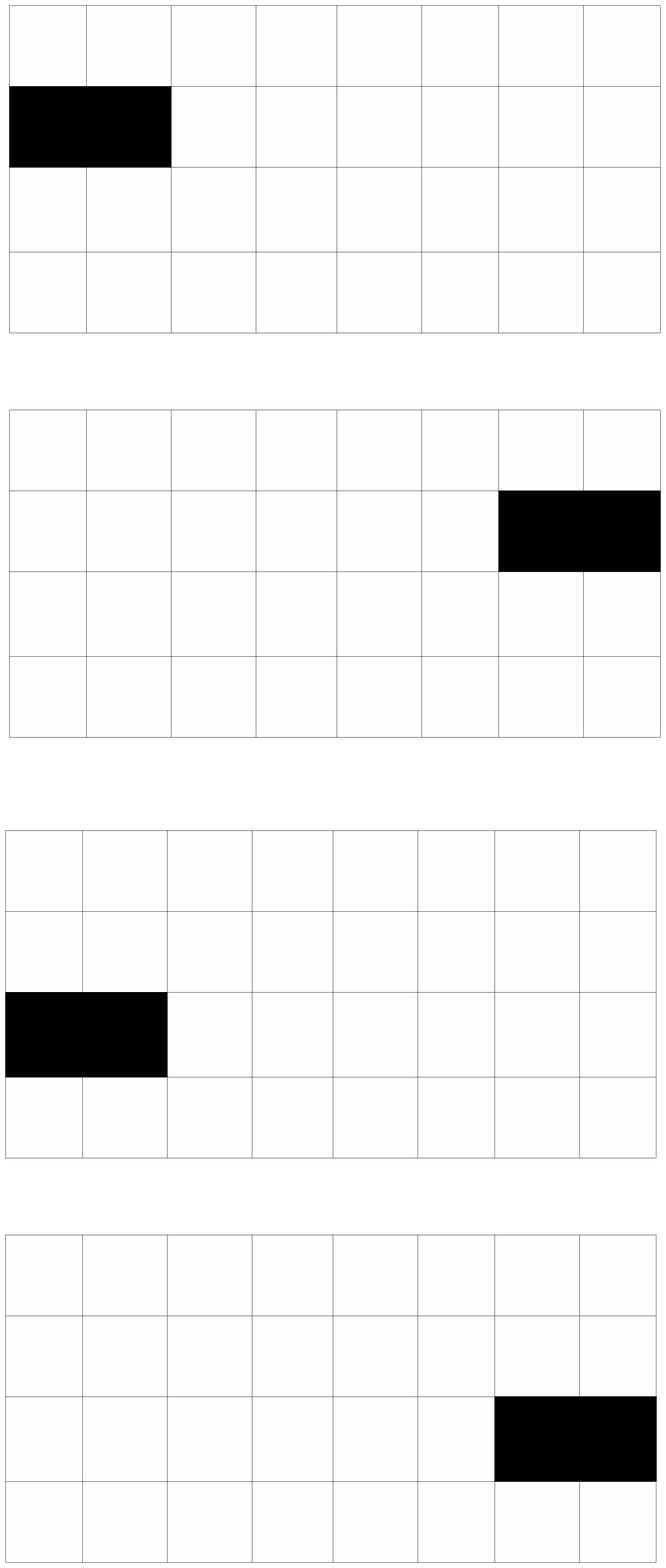}}}
{\scalebox{.2}{\includegraphics{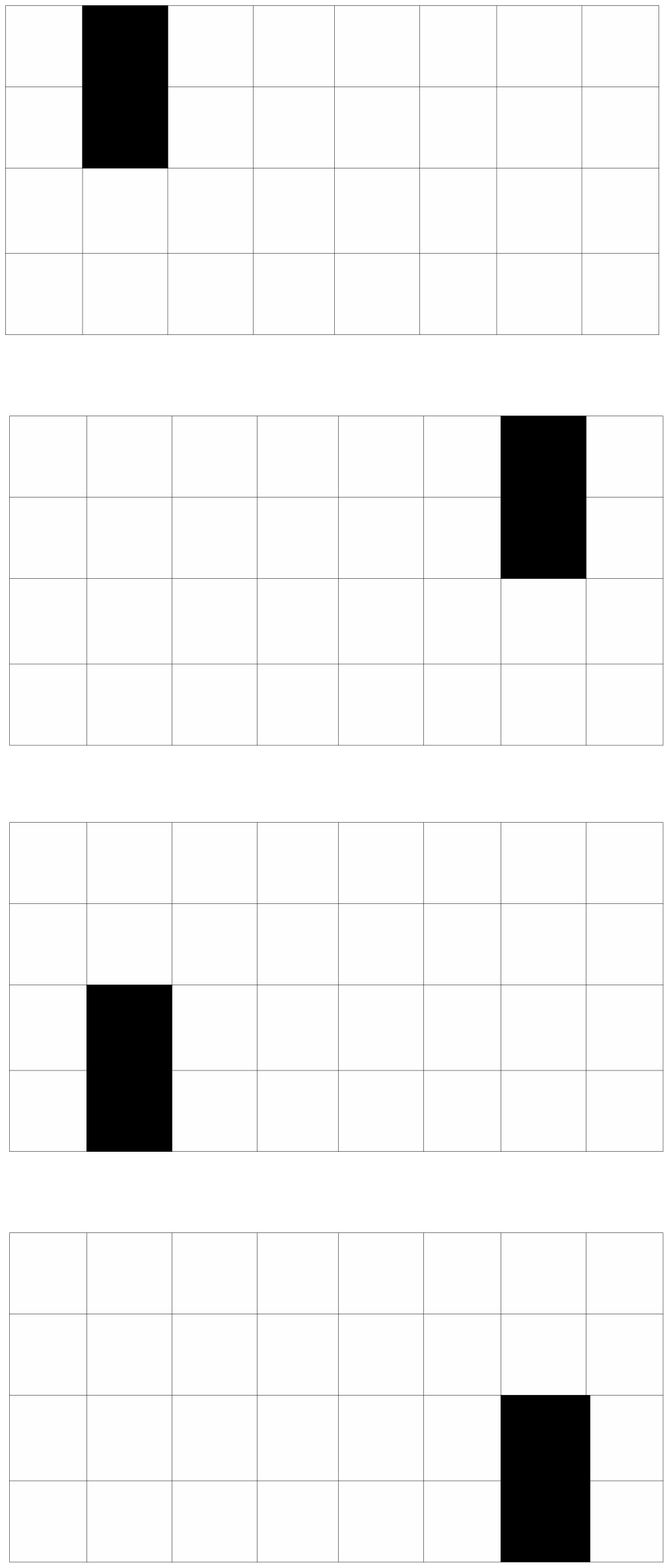}}}
{\scalebox{.2}{\includegraphics{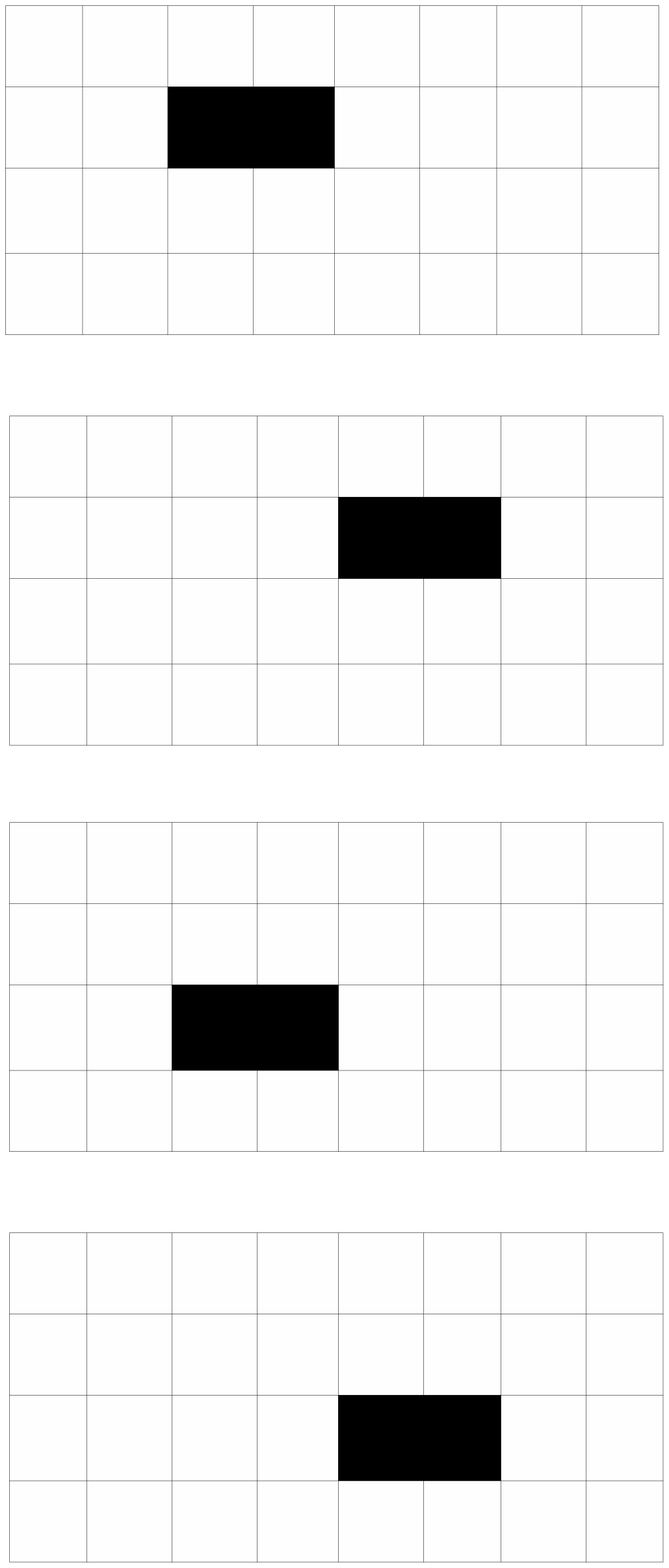}}}
{\scalebox{.2}{\includegraphics{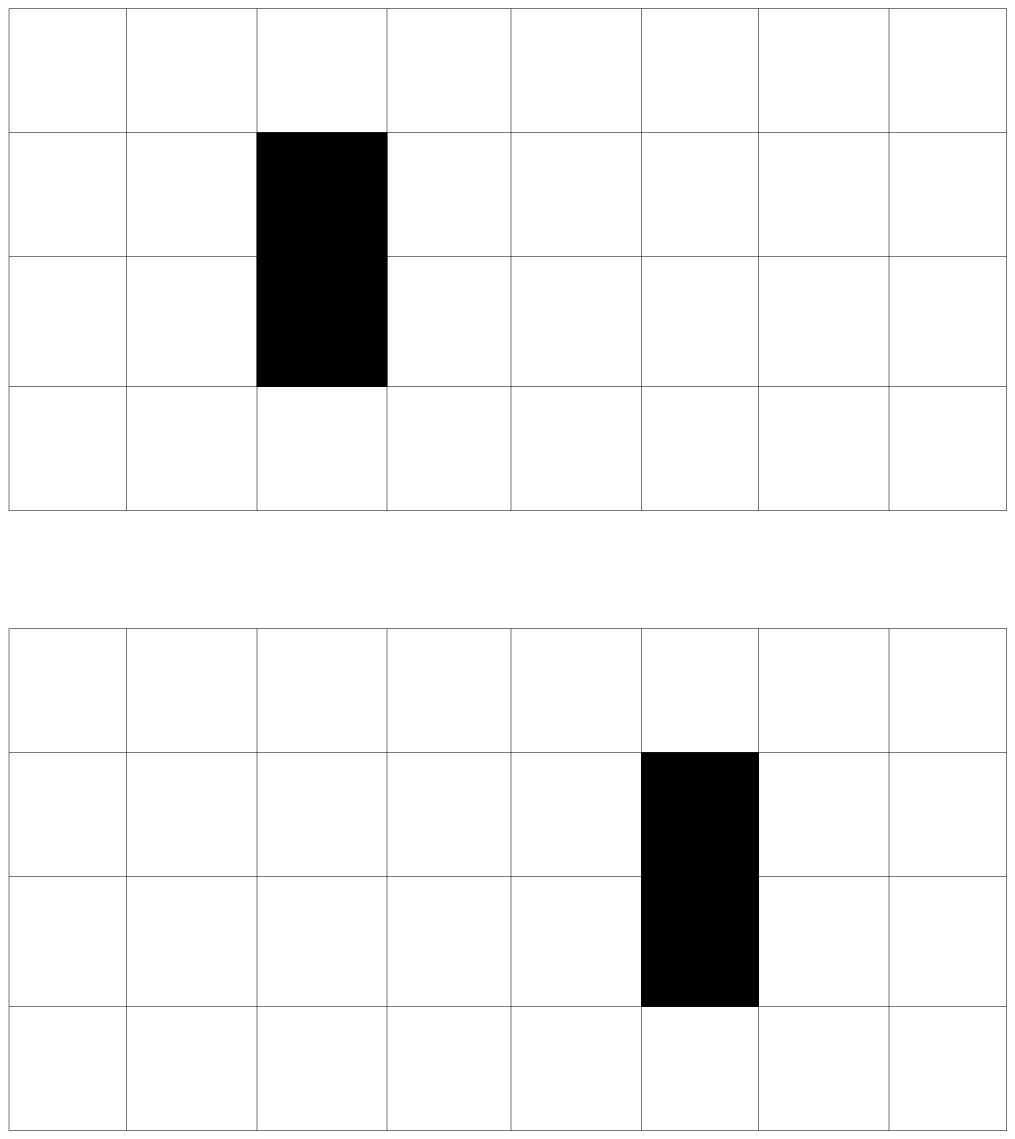}}}
}
\caption{Bad squares for $R(4,8)^{--}$.}
\end{figure}  

\begin{figure}[htbp]
\centerline{
{\scalebox{.2}{\includegraphics{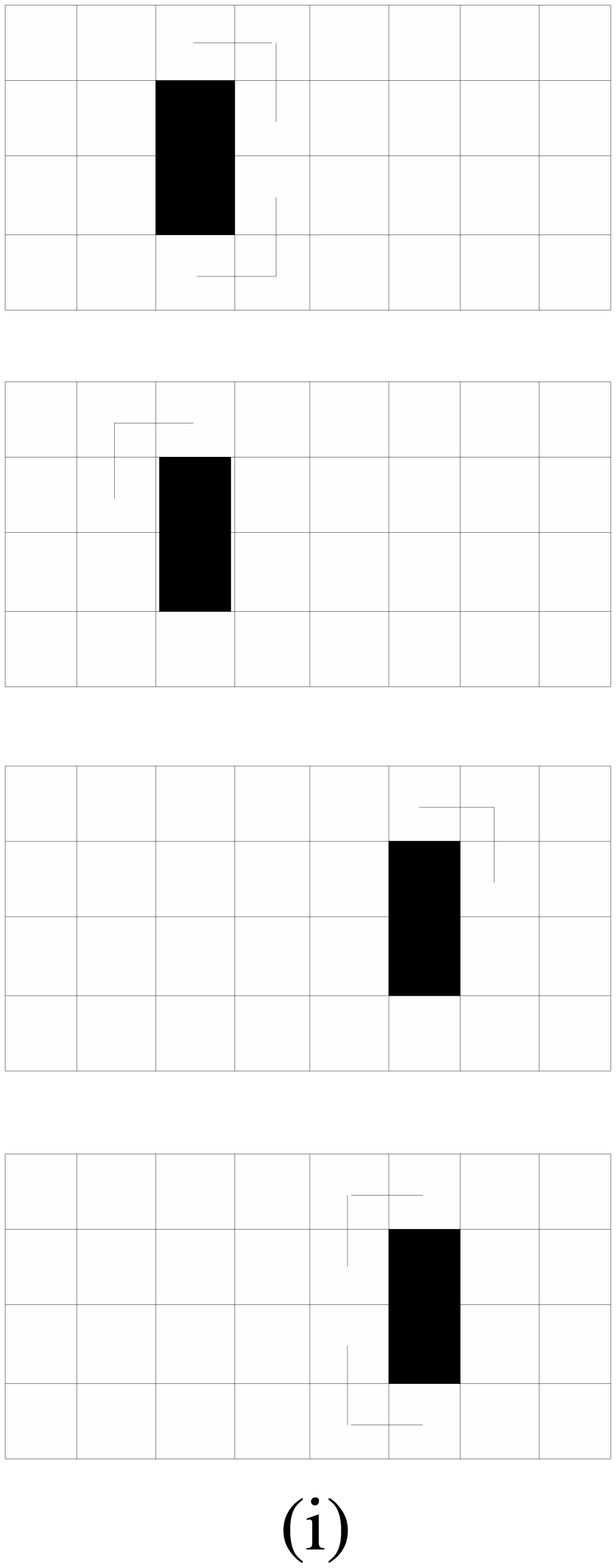}}}
{\scalebox{.2}{\includegraphics{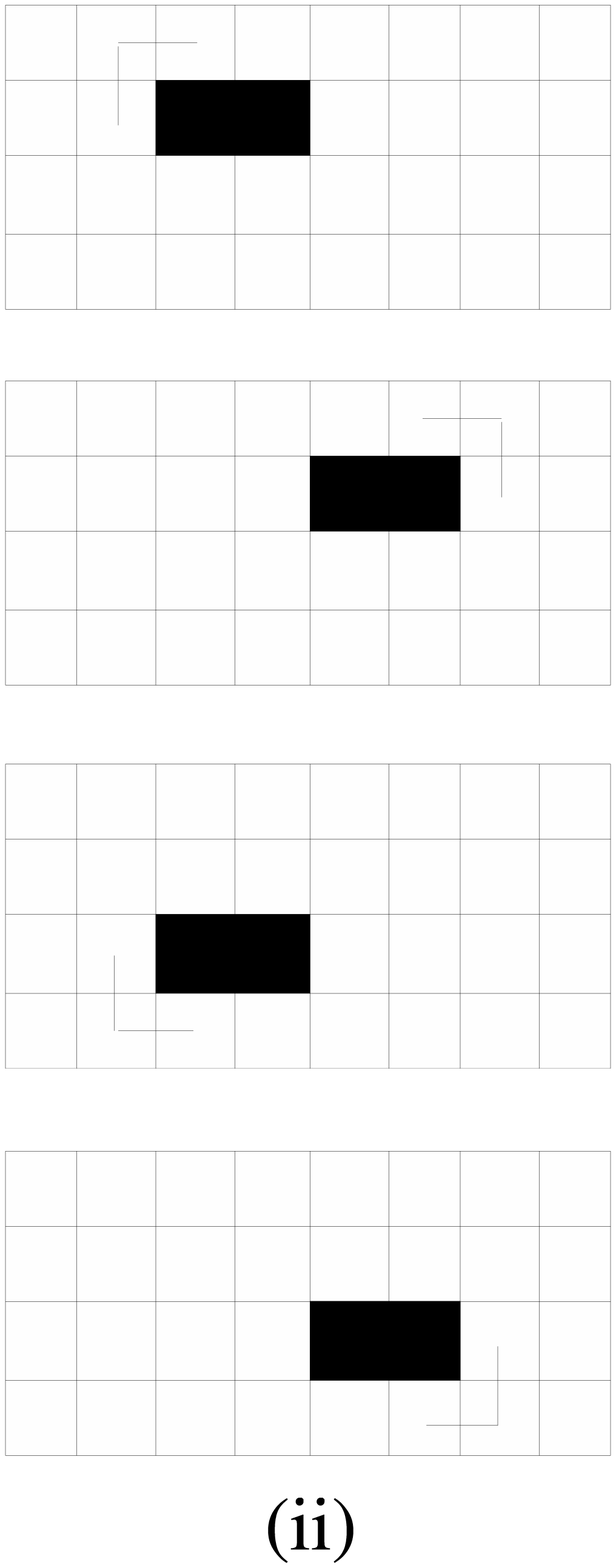}}}
{\scalebox{.2}{\includegraphics{pic67}}}
{\scalebox{.2}{\includegraphics{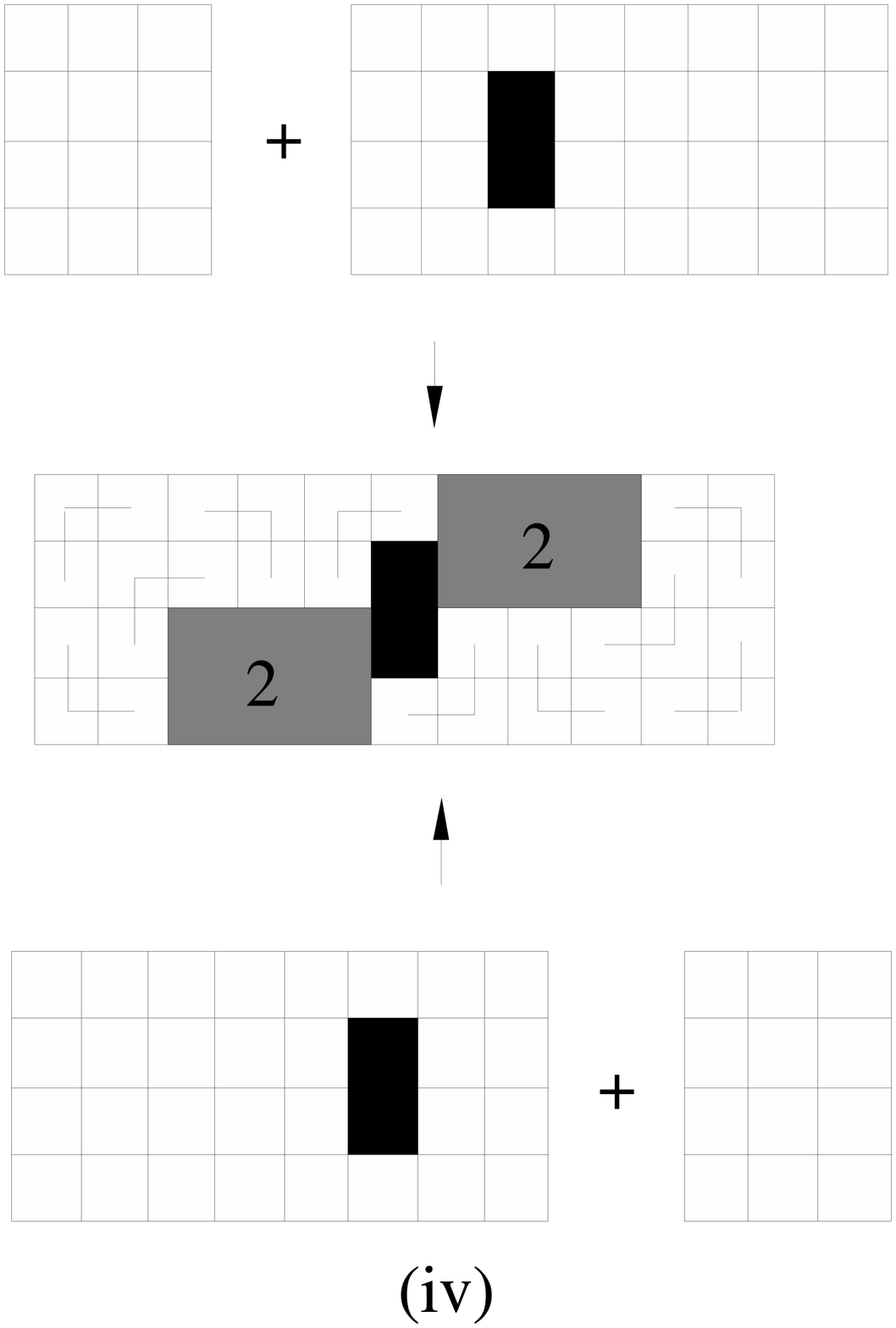}}}
}
\caption{(i) and (ii) Untileability for some bad pairs for $R(4,8)^{--}$. 
(iii) A $(4,3)$-{\it hquad shift}. (iv) Alternate measure when a $(4,3)$-{\it hquad shift} does not help.}
\end{figure}	

\begin{theorem}{\bf [Domino-Deficient Quadrec Theorem]} 
The only bad pairs for $R(4,3t+8)^{--}$ are $\{(2,1),(2,2)\}$, $\{(1,2),(2,2)\}$, 
$\{(2,3t+7),(2,3t+8)\}$, $\{(1,3t+7),(2,3t+7)\}$, $\{(3,1),(3,2)\}$, $\{(3,2),(4,2)\}$, 
$\{(3,3t+7),(3,3t+8)\}$, $\{(3,3t+7),(4,3t+7)\}$, $\{(2,3),(3,3)\}$, $\{(2,3t+6),(3,3t+6)\}$, 
$\{(2,3),(2,4)\}$, $\{(2,3t+5),(2,3t+6)\}$, $\{(3,3),(3,4)\}$ and $\{(3,3t+5),(3,3t+6)\}$. 
\end{theorem}
\begin{proof}
The proof for the badness of the cases of domino-removal enumerated 
in the theorem statement is similar to that for 
$R(4,8)^{--}$, and so is left for the reader as an exercise. We will prove the existence of a tromino 
tiling in all other cases. 
Consider tilings of $R(4,3t+8)^{--}$, where $t\geq1$. This big domino-deficient rectangle can be 
viewed as $t$ subrectangles of dimension $4\times3$ and one 
domino-deficient $R(4,8)^{--}$ rectangle joined together. Following our 
additive notation, we write this as:

\begin{eqnarray}
R(4,3t+8)^{--} & = & t\cdot R(4,3) + R(4,8)^{--}
\end{eqnarray}

If the missing domino in the rectangle $R(4,8)^{--}$ does not form a bad pair, then we can tile it. A 
$4\times3$ rectangles satisfies the conditions of the 
Chu-Johnsonbaugh Theorem, and so is tileable. So, in this 
case, we achieve a tiling of $R(4,3t+8)^{--}$. 
Consider the situation when the missing domino in $R(4,8)^{--}$ (as 
above) forms a bad pair. In this case, we perform a $(4,3)$-{\it hquad shift}.  
The reader should note that we are only considering pairs of squares 
which are not enumerated in the theorem statement, so such a shift is always possible. 
If such a shift removes the ``badness" of $R(4,8)^{--}$, then we are done. 
The only case in which a $(4,3)$-hquad shift will not 
work, is when $R(4,8)^{--}$ has $\{(2,3),(3,3)\}$ or $\{(2,6),(3,6)\}$ as its bad pair. These pairs 
are symmetric with respect to a shift by three columns, and so just get interchanged. 
In this case, we join a removed $4\times3$ rectangle from the left (or right) as shown in Figure 5(iv). The new 
rectangle $R(4,11)^{--}$ is tileable (as is evident from the figure). 
Each of the $(t-1)$ subrectangles of dimension $4\times3$ is also tileable, 
so we get a tiling of $R(4,3t+8)^{--}$.  \hfill\qed   
\end{proof}

\subsection{Counting tilings of $R(4,3t+8)^{--}$ rectangles}

We now proceed to enumerate all tilings of $R(4,3t+8)$ rectangles, where $t\geq0$ with 
one domino and $4t+10$ trominoes. Let $T(4,3t+8)^{--}$ denote this number. Let the number 
of such tilings be $T_V(4,3t+8)^{--}$ ($T_H(4,3t+8)^{--}$) when the missing domino is 
vertical (horizontal). Consider the three interfaces as shown in Figure 6(i).
These three interfaces are called {\it straight}, {\it deep jog}, and 
{\it shallow jog}.
For these three interfaces, we define $N(t)$, $N_1(t)$ and $N_2(t)$ respectively as the number 
of tilings when there are $n=3t$ columns to the left of the dotted line. Obviously 
$N_1(t)$ and $N_2(t)$ remain the same if we count tilings of their vertical reflections 
instead. To express these, we write them as generating functions 

\begin{eqnarray*}
G(z) & = & \sum_tN(t)z^t
\end{eqnarray*}

and similarly for $G_1(z)$ and $G_2(z)$. Finally, $N(0)=1$, since there is exactly one 
way to tile a $4\times0$ rectangle. 
Now consider the case when the missing domino is 
vertical. We set our coordinate system so that the domino occupies the $4th$ column. 
Assume the domino occupies the position $\{(1,4),(2,4)\}$. Several cases arise depending 
on whether a single tromino covers the squares $(3,4)$ and $(4,4)$, or two separate 
trominoes cover them. Consider the former case first. If a tromino covers $(3,4)$, $(4,4)$, 
$(3,3)$, then a $2\times3$ rectangle is completed by the tromino covering $(4,3)$ 
(see Figure 6(ii)(d)). 
If this tromino covers $(4,3)$, then the tromino covering $(3,3)$ has three permissible 
orientations. If it covers $(3,2)$ and $(4,2)$ then a $2\times3$ rectangle is again 
completed. If it covers $(3,2)$ and $(2,2)$ (resp. $(3,2)$ and $(2,3)$), then a $3\times2$ rectangle 
is completed by the tromino covering $(2,3)$ (resp. $(1,3)$)
(see Figure 6(ii)(c)). Now consider the case when two different trominoes 
cover $(3,4)$ and $(4,4)$. If the tromino covering $(3,4)$ covers $(3,5)$ and $(4,5)$, then 
a $2\times3$ rectangle is completed by the tromino covering $(4,4)$ (see Figure 6(ii)(b)). 
If a tromino covers 
$(4,4)$, $(4,5)$ and $(3,5)$, then there are two possible orientations for the tromino 
covering $(3,4)$. If it covers $(3,3)$ and $(4,3)$, then a $2\times3$ rectangle is again 
completed. If it covers $(3,3)$ and $(2,3)$, then we get the case shown in Figure 6(ii)(a). 
(The reader should note that these arguments will also hold when the vertical domino is 
present in the last two rows.)  
Assume now that the domino occupies the position $\{(2,4),(3,4)\}$. The reader can easily see 
that only two cases are possible in this situation (see Figures 6(ii)(e) and (f)). Based on the 
above case analysis, we get equation (13). Suppose $G_V(z) = \sum_{t=1}T_V(4,3t+2)^{--}z^t$. 
From (13), it can be seen that $G_V(z)$ is basically a sum of convolutions of 
$G(z)$, $G_1(z)$ and $G_2(z)$. We express this fact in equation (14).  

\begin{figure}[htbp]
\centerline{
{\scalebox{.2}{\includegraphics{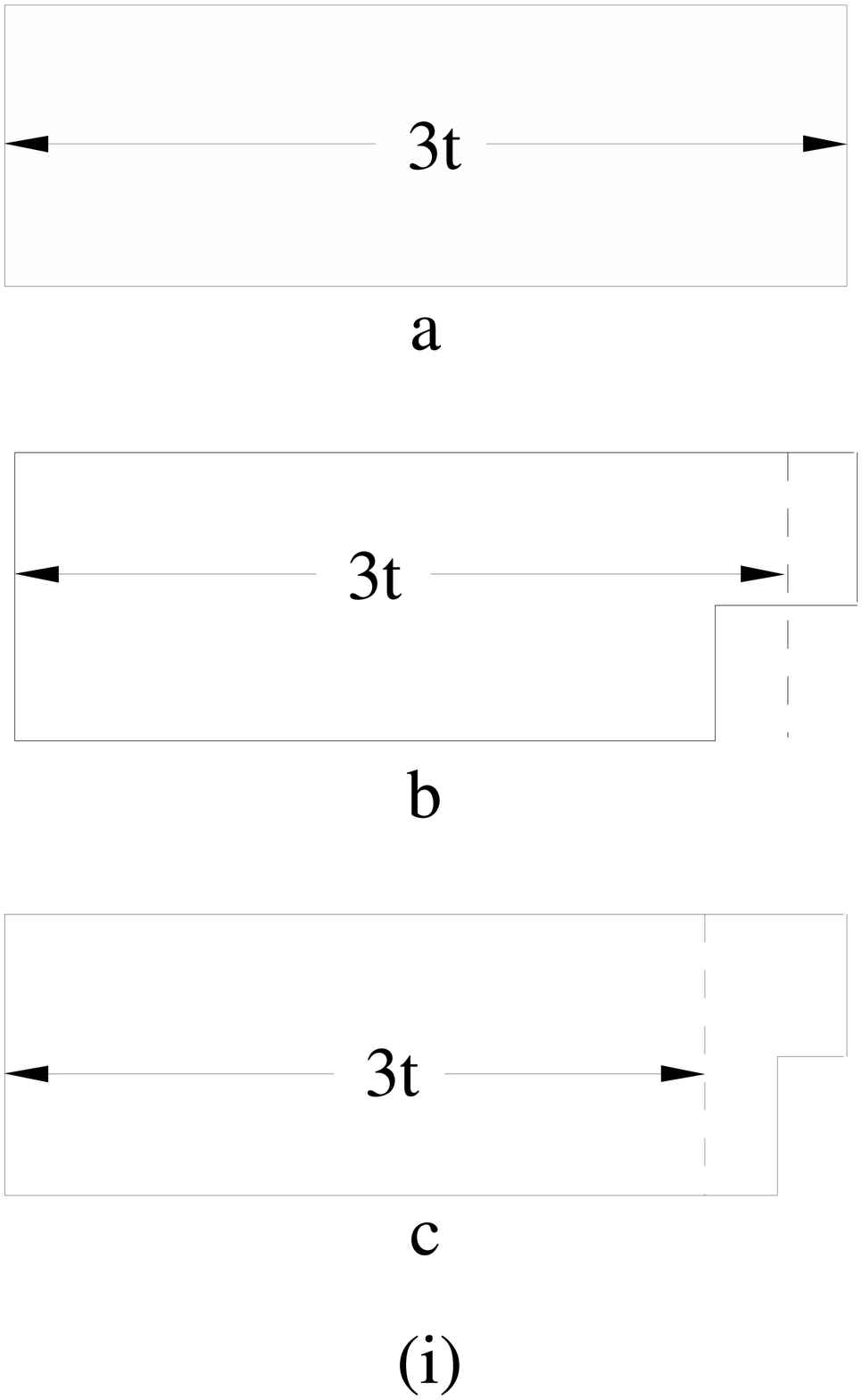}}}
{\scalebox{.2}{\includegraphics{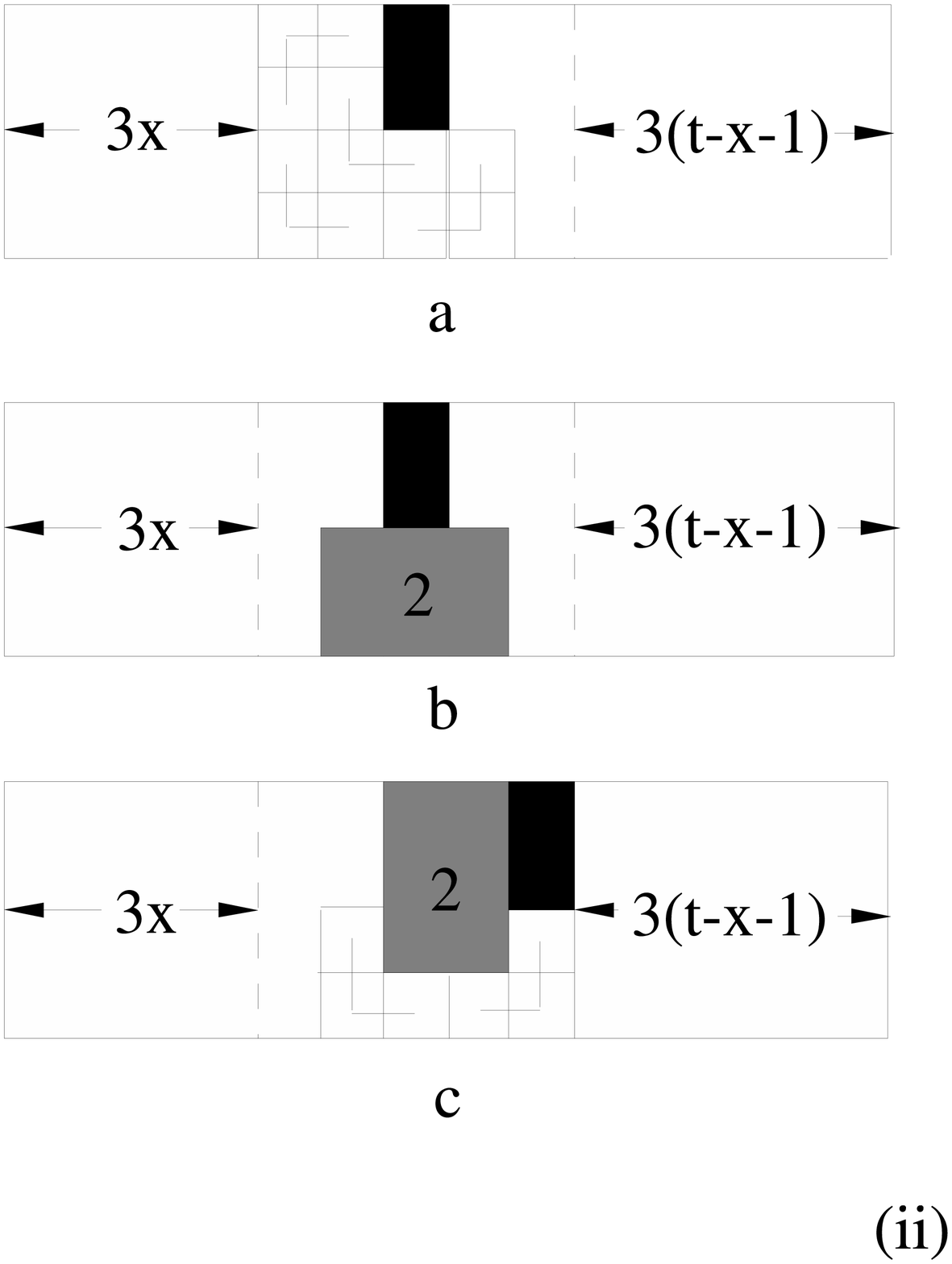}}}
{\scalebox{.2}{\includegraphics{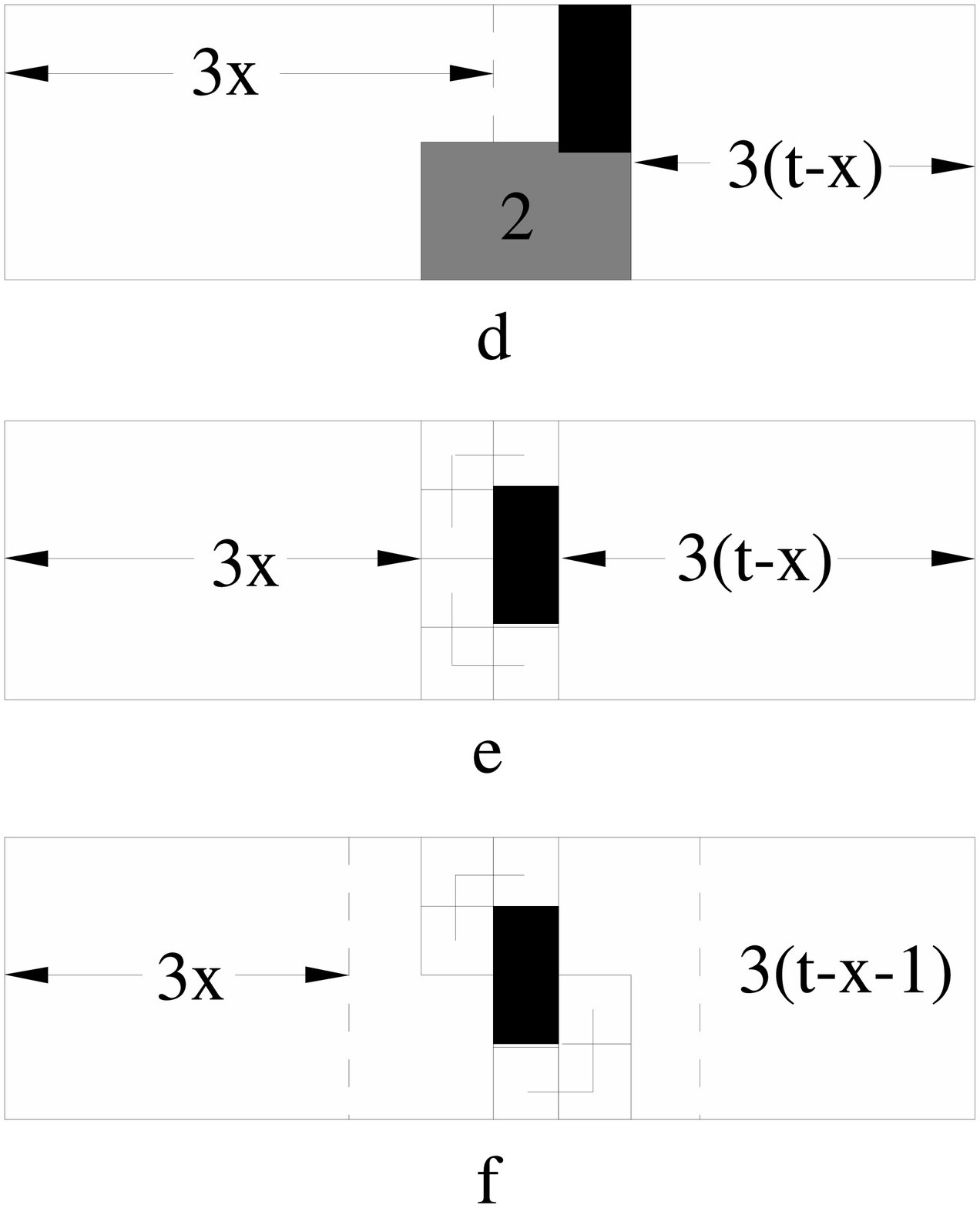}}}
}
\caption{(i) Three kinds of interfaces on rectangles of width 4, (a) straight, 
(b) deep jog, (c) shallow jog. (ii) Various cases when the domino is vertical.}
\end{figure}

\begin{eqnarray}
T_V(4,3t+2)^{--} & = & 8\times\{\sum_{x=0}^{t-1}N_2(x)\cdot N(t-x-1) + \sum_{x=1}^tN_1(x)\cdot N(t-x)\} \nonumber \\ 
                 & + & 4\times\{\sum_{x=0}^{t-1}N(x)\cdot N_2(t-x-1) + \sum_{x=1}^{t-2}N_2(x)\cdot N_2(t-x-1)\} \nonumber \\
                 & + & 2\times\{\sum_{x=0}^tN(x)\cdot N(t-x) + \sum_{x=1}^{t-2}N_2(x)\cdot N_2(t-x-1)\} 
\end{eqnarray}
\begin{eqnarray}
G_V(z) & = & 12zG_2(z)G(z) + 8G_1(z)G(z) + 6z{G_2}^2(z) + 2G^2(z)
\end{eqnarray}
  
\begin{figure}[htbp]
\centerline{
{\scalebox{.15}{\includegraphics{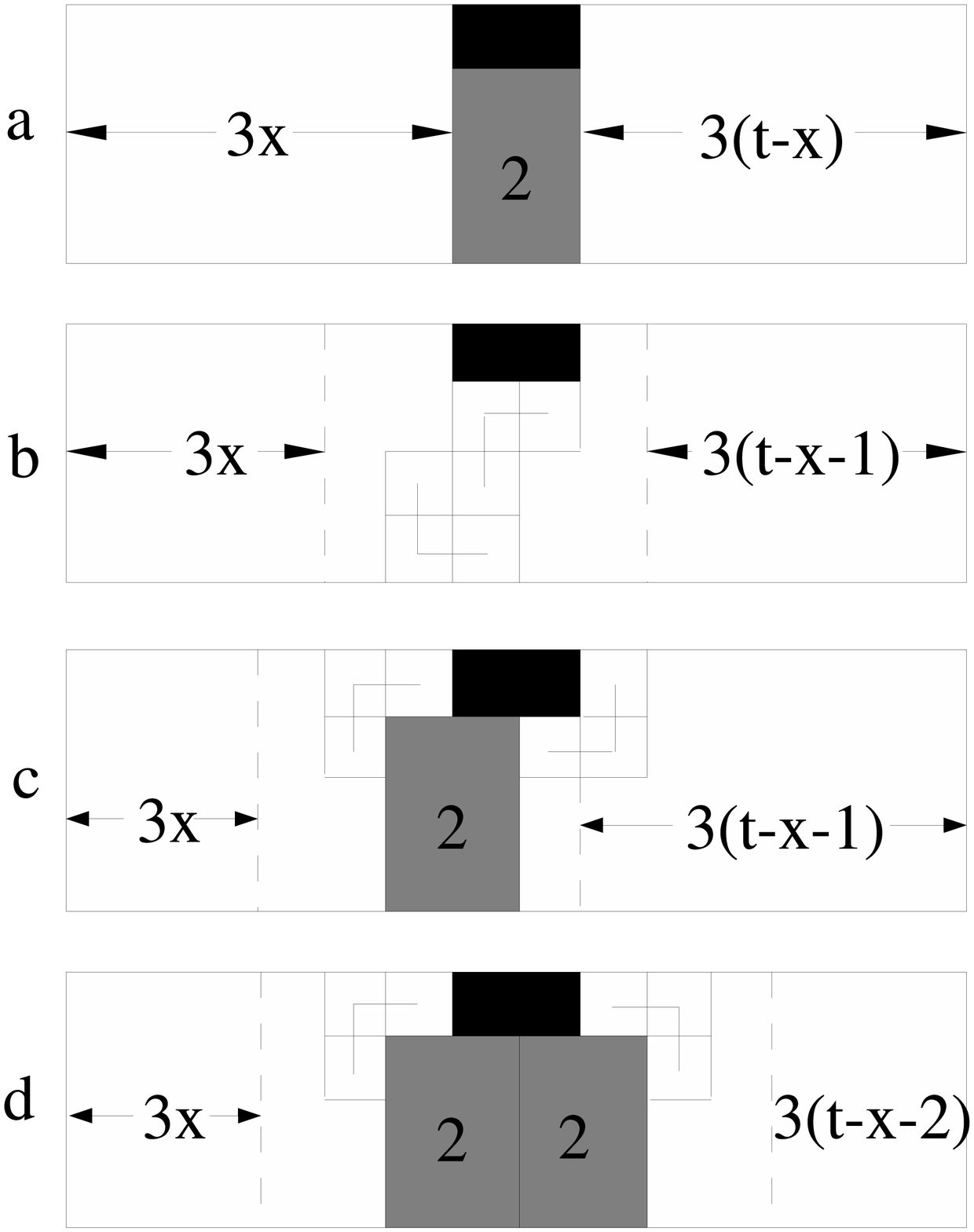}}}
{\scalebox{.15}{\includegraphics{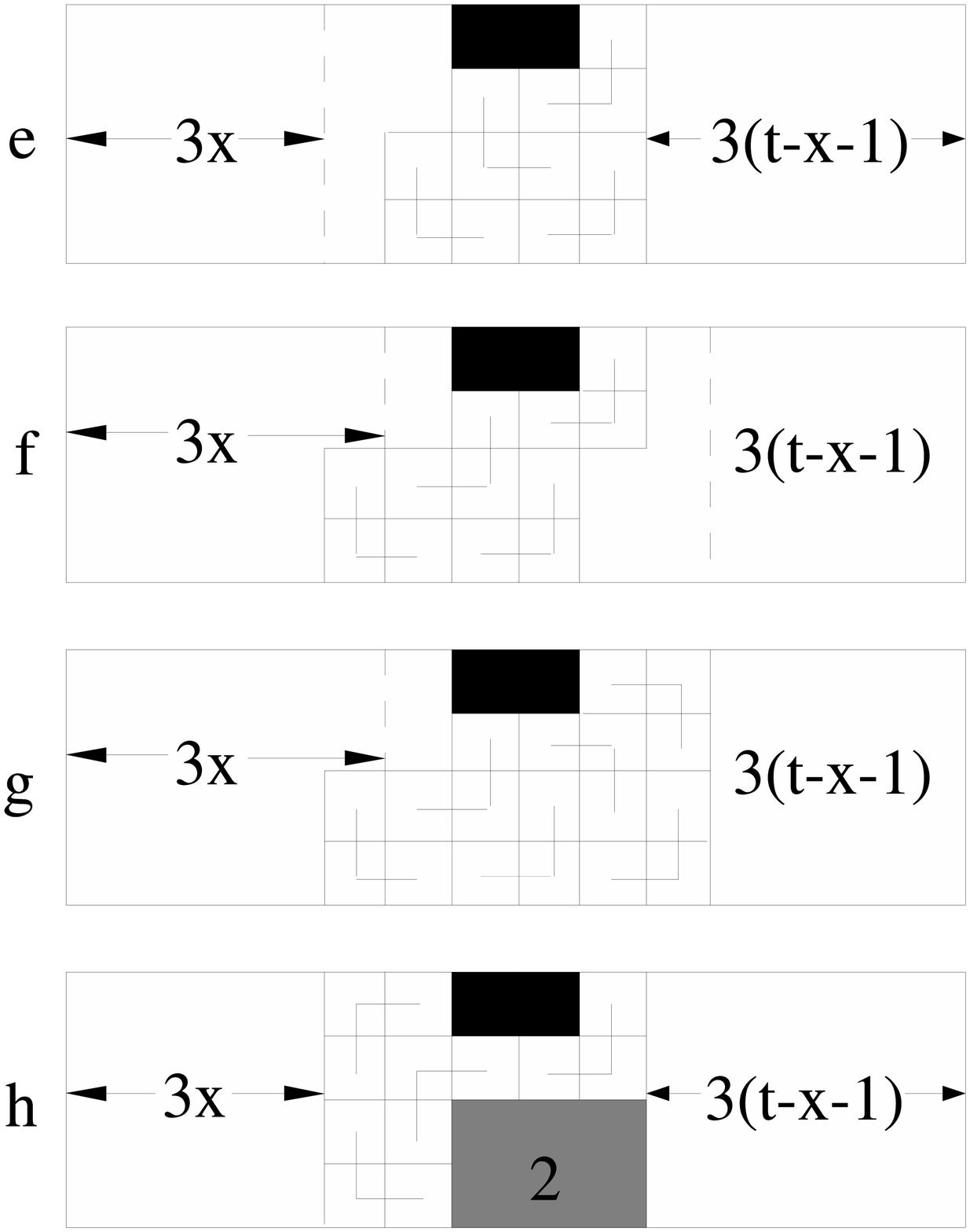}}}
{\scalebox{.15}{\includegraphics{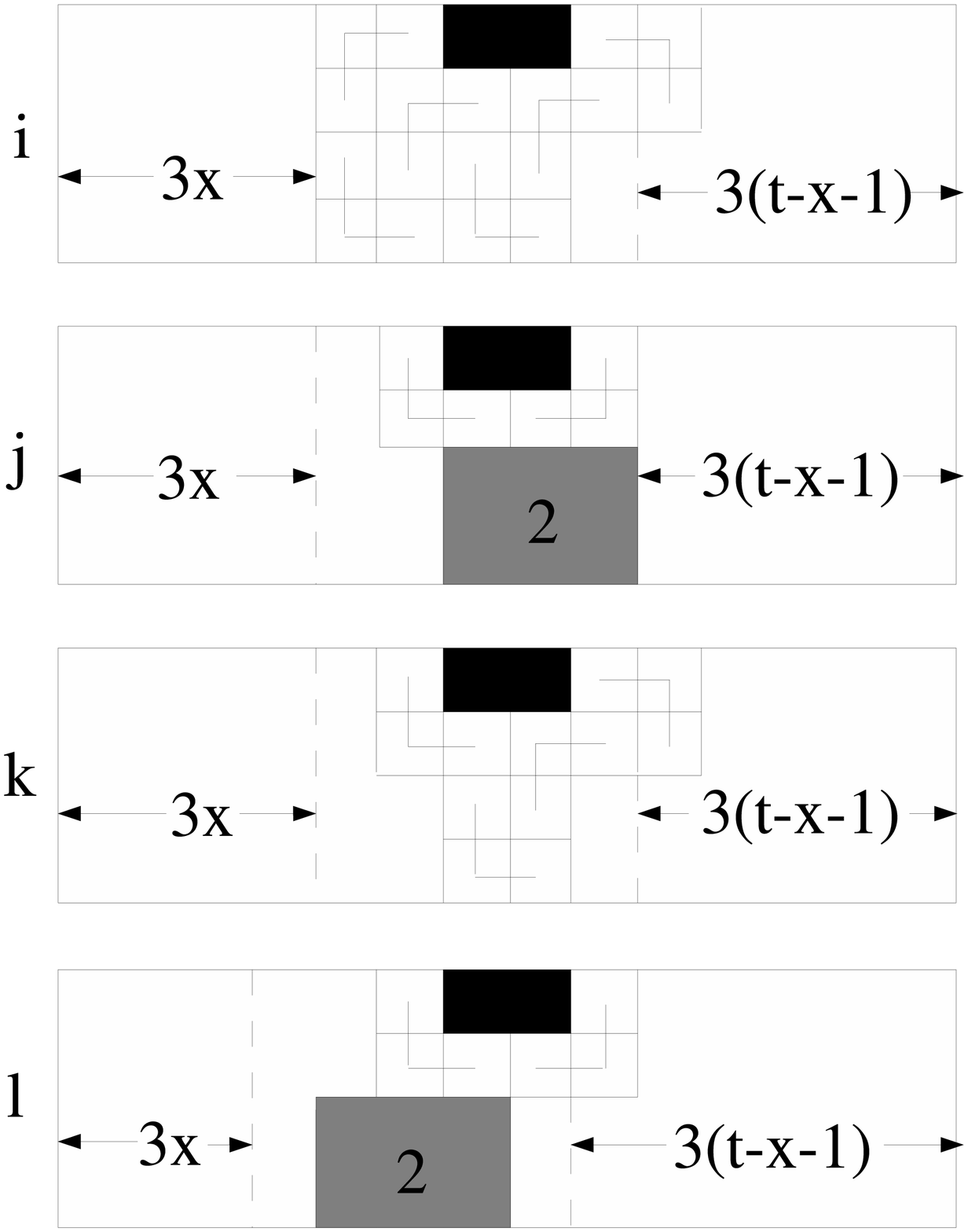}}}
{\scalebox{.15}{\includegraphics{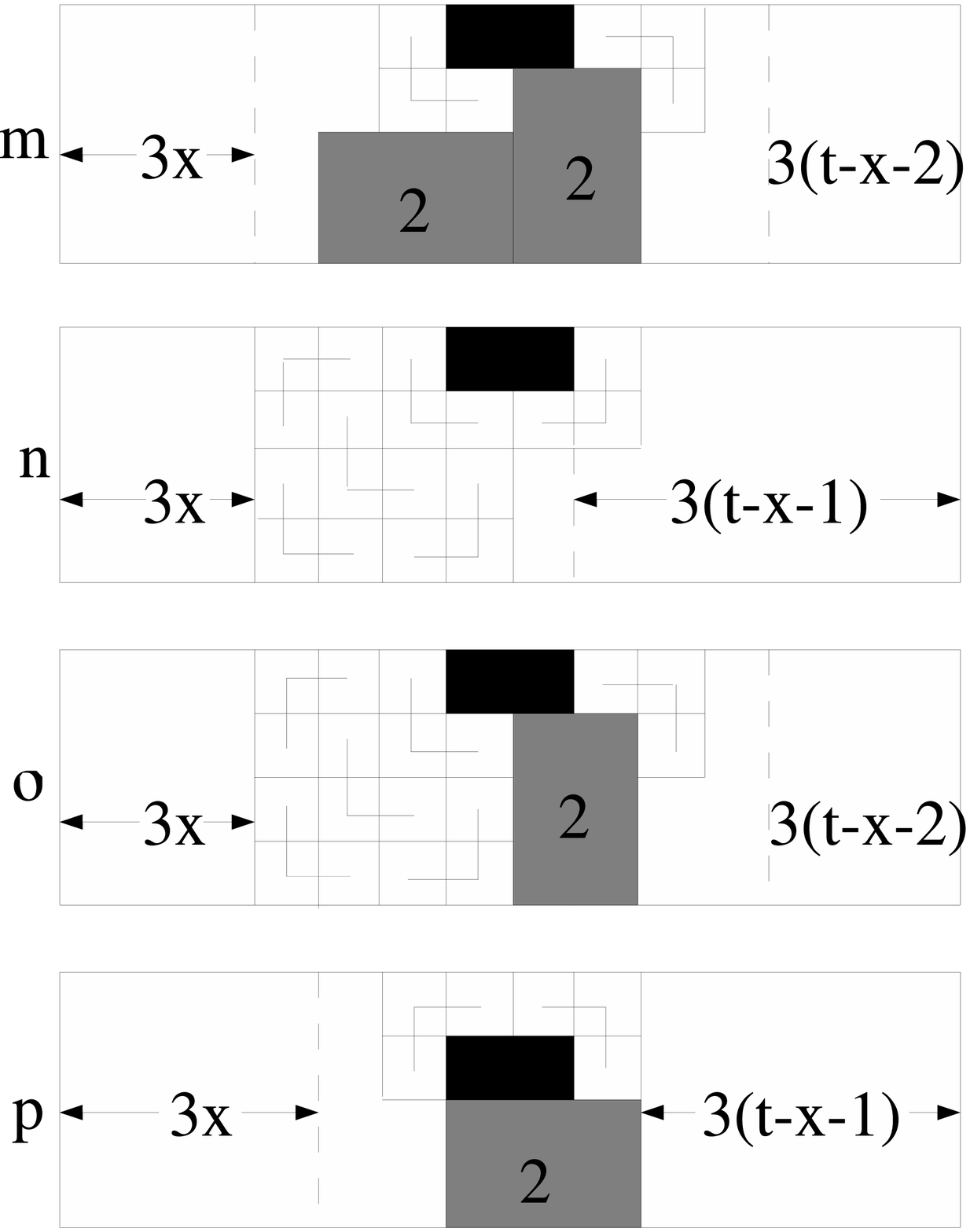}}}
{\scalebox{.15}{\includegraphics{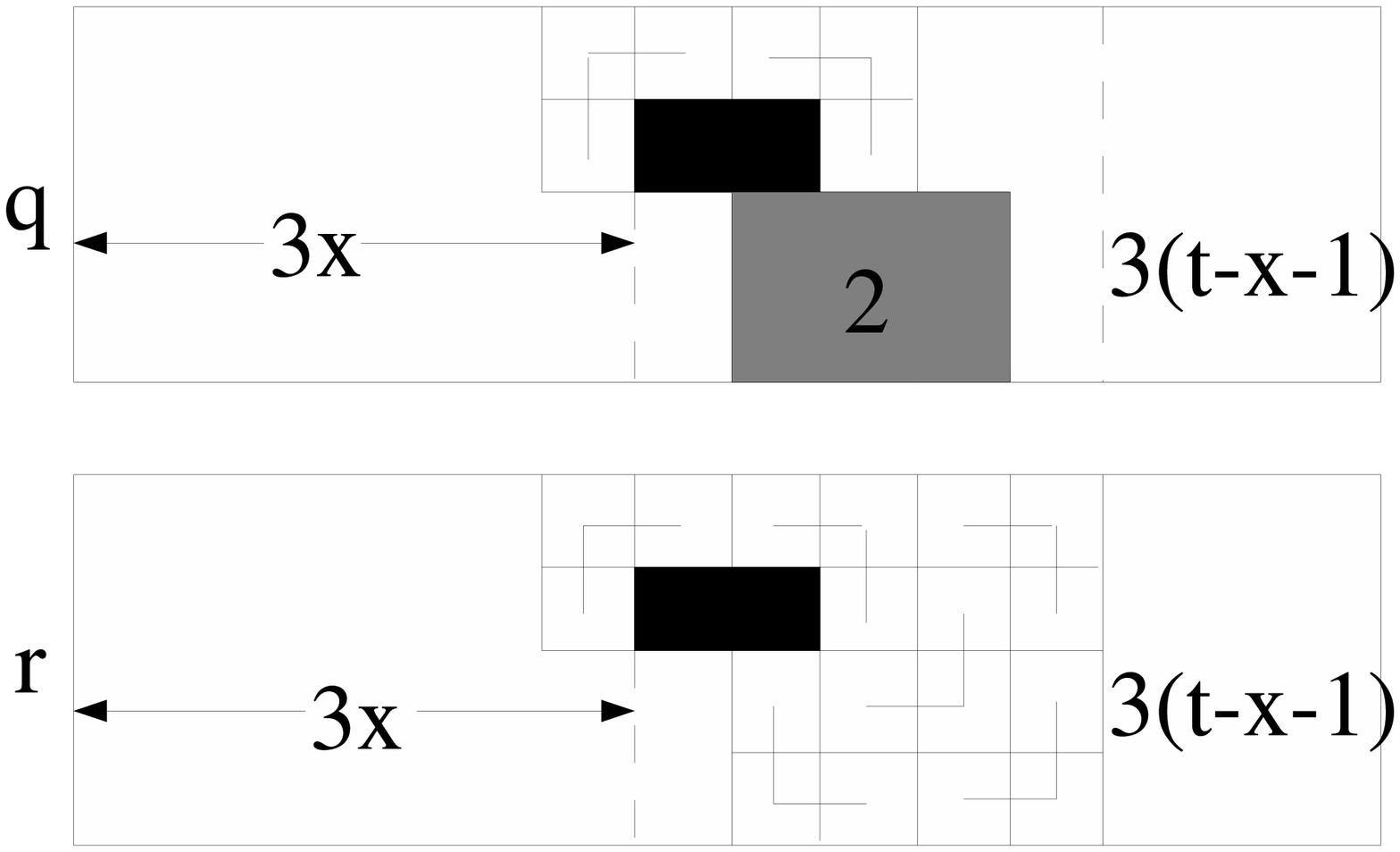}}}
}
\caption{Various cases when the domino is horizontal.}
\end{figure}  

We now consider the case when the missing domino is horizontal. Setting our coordinate 
system such that this domino is present in columns $4-5$. Assume first that it is present 
in the first row, i.e., it occupies the position $\{(1,4),(1,5)\}$. Several cases arise 
depending upon the orientation of the tromino covering $(2,4)$, which we call the 
{\it crucial tromino} for our present discussion. We shall refer to reflections 
in the Y-axis as {\it vertical reflections} and those in the X-axis as {\it horizontal 
reflections}. If the crucial tromino covers $(2,5)$ and $(3,4)$, 
then there are two possible orientations of the tromino covering $(4,4)$, as shown in Figure 
7(a) and (b). If the crucial tromino covers $(3,4)$ and $(3,5)$, then we get the case shown 
in Figure 7(e). The case when the crucial tromino covers $(2,5)$ and $(3,5)$ is symmetric to the 
case when it covers $(2,5)$ and $(3,4)$. Now assume that the crucial tromino covers $(2,3)$ and 
$(3,4)$. If the tromino covering $(4,4)$ covers $(4,3)$ and $(3,3)$, then there are two cases 
possible, as shown in Figure 7(c) and (d). If the tromino covering $(4,4)$ covers $(4,5)$ and 
$(3,5)$, then two cases are possible, which are the vertical reflections of Figure 7(i) and (k). 
If the crucial tromino covers $(3,4)$ and $(3,3)$, then we get the cases shown in 
Figure 7(f) and (g). Three cases arise when the crucial tromino covers $(2,3)$ and $(3,3)$. These 
are shown in Figures 7(h), 7(i), and the vertical reflection of Figure 7(g). Finally, consider 
the case when the crucial tromino covers $(2,3)$ and $(1,3)$. If the tromino covering $(3,4)$ 
covers $(2,5)$ and $(3,5)$, then we get the vertical reflection of Figure 7(e). If the tromino 
covering $(3,4)$ covers $(3,5)$ and $(4,5)$, then there are two possible cases, which are the 
vertical reflections of Figures 7(h) and (j), and if it covers $(4,4)$ and $(3,5)$, then we get 
the case in Figure 7(j). If the tromino covering $(3,4)$ covers $(4,4)$ and $(4,5)$, then three 
cases arise, which are those in Figures 7(j), (k) and the vertical reflection of Figure 7(f). 
Now assume that the tromino covering $(3,4)$ covers $(4,4)$ and $(3,3)$, then the two cases 
shown in Figures 7(l) and (m) arise depending on how $(2,5)$ is covered. If the tromino covering 
$(3,4)$ covers $(4,4)$ and $(4,3)$, then we get the cases shown in Figure 7(l), (n) and (o), and 
if it covers $(3,3)$ and $(4,3)$, then we get the vertical reflection of Figure 7(j). So, Figures 
7(a) -(o) (and their horizontal reflections), show all the various cases which arise when the 
missing domino is present in the first or fourth row. We now consider the case when the 
missing domino is present in the second row. Again setting our coordinate system such that this 
domino occupies the position $\{(2,4),(2,5)\}$. The reader can easily see that the tromino covering 
$(1,4)$ ($(1,5)$) must cover $(1,3)$ and $(2,3)$ ($(1,6)$ and $(2,6)$). Depending on how $(3,5)$ is 
covered, three cases essentially arise (or their vertical reflections), as shown in Figure 7(p)-(r).  
Thus, Figure 7 (and its reflections) shows all the possible cases which can arise when 
the missing domino is horizontal. Based on the above case analysis, and assuming 
$G_H(z) = \sum_{t=1}T_H(4,3t+2)^{--}z^t$, we get the following equation: 

\begin{eqnarray}
G_H(z) & = & 4(1+2z)G^2(z) + 4z(1+6z)G_2^2(z) + 32zG_1(z)G_2(z) + 28zG(z)G_2(z) \nonumber \\
       & + & 16zG(z)G_1(z)
\end{eqnarray} 

The reader should note that the missing domino can be either horizontal or vertical, so we have 
$T(4,3t+2)^{--} = T_H(4,3t+2)^{--} + T_V(4,3t+2)^{--}$. Assuming $F(z) = \sum_{t=1}T(4,3t+2)^{--}z^t$, 
$F(z) = G_V(z) + G_H(z)$. So from equations (11) and (12), we have,

\begin{eqnarray}
F(z) & = & 6G^2(z) + 10zG_2^2(z) + 32zG_1(z)G_2(z) + 24z^2G_2^2(z) \nonumber \\ 
     & + & 40zG(z)G_2(z) + 16zG(z)G_1(z) + 8zG^2(z) + 8G(z)G_1(z)
\end{eqnarray}

In \cite{moore} Moore had derived the generating functions $G(z)$, $G_1(z)$ and $G_2(z)$ as:

\begin{eqnarray}
G(z)   & = & \frac{1-6z}{1-10z+22z^{2}+4z^{3}} \\
G_1(z) & = & \frac{z(1-2z)}{1-10z+22z^{2}+4z^{3}} \\
G_2(z) & = & \frac{2z}{1-10z+22z^{2}+4z^{3}}
\end{eqnarray} 

Putting these values in (16), we get, 

\begin{eqnarray}
F(z) & = & \frac{6-56z+152z^{2}-120z^{3}+160z^{4}}{(1-10z+22z^{2}+4z^{3})^{2}}
\end{eqnarray}

The asymptotic growth of $N(t)$ 
(recall that $N(t)$ is the number of tilings for the {\it straight} interface, and 
the value of $G(z)=\sum_{t}N(t)z^t$ is as in (17)) 
is the reciprocal of the radius of convergence of $G's$ Taylor 
series. Thus $N(t)\propto\lambda^t$ where $\lambda$ is the largest positive root of 

\begin{eqnarray*}
\lambda^{3} - 10\lambda^{2} + 22\lambda + 4 = 0
\end{eqnarray*} 

Numerically, we have 

\begin{eqnarray*}
\lambda & = & 6.54560770847481152029.......
\end{eqnarray*}

Now $F(z)$ is basically a convolution of $G(z)$, $G_1(z)$ and $G_2(z)$. 
It is easy to see that both $N_1(t),N_2(t)\propto\lambda^t$ also.  
So, we have,

\begin{eqnarray*}
T(4,3t+2)^{--} & \propto & \sum_{r=0}^{t}\lambda^{r}\times\lambda^{t-r}  \\
               & \propto & (t+1)\lambda^{t} 
\end{eqnarray*}

The reader should once again note that, similar to our observation in the previous section, 
in this case also, by introducing a small deficiency of 2 in the $4\times n$ rectangle, 
the complexity of the number of tromino tilings changes from $\mathcal{O}(\lambda^t)$ to 
$\mathcal{O}(t\cdot\lambda^t)$. This leads us to proposing the following conjecture: 

\begin{conjecture}
The complexity of the number of tromino tilings of an $m\times n$ rectangle, 
changes from $N_T(m,n)$ to 
$\mathcal{O}((m+n)\cdot N_T(m,n))$ in tiling an extended domino-deficient 
rectangle with minimal dimensions $m\times n'$, where $n'>n$.
\end{conjecture}

An interesting question would be to analyze the complexity of the 
number of tromino tilings $N_T(m,n)$ of an $m\times n$ rectangle on introducing a deficiency of 
$k$ dominoes. How does $N_T(m,n)$ vary when the 
number of trominoes remains the same, but $k$ increases? What happens if the size of the 
rectangle remains the same? We state the following problem for the interested reader: 

\begin{openquestion}
How does the number of tromino tilings $N_T(m,n)$ of an $m\times n$ rectangle, 
where $m,n,k\geq1$ and $3|(mn-2k)$, vary with $k$, the number of dominoes 
which are removed, when (a) The number of trominoes 
remains the same, and (b) The size of the rectangle remains the same?
\end{openquestion}

\section{Tromino Tilings of $R(5,3t+4)^{--}$ Rectangles}

Consider now the case when one dimension of the given rectangle is $5$. 
It turns out that the bad pairs for this case are different from rest of the cases
considered so far. First consider the simplest rectangle 
of this case, a $4\times5$ rectangle. Figure 8(1) shows all the bad pairs for $R(4,5)^{--}$. 
The bad pairs $\{(2,2),(2,3)\}$, $\{(2,3),(2,4)\}$, $\{(3,2),(3,3)\}$ and $\{(3,3),(3,4)\}$ 
are symmetric, and the reader can see that the tromino covering $(1,3)$ (in the first two 
cases) and $(4,3)$ (in the last two cases) make the squares $(1,5)$ (resp., $(1,1)$, $(4,5)$, 
and $(4,1)$) inaccessible. See Figure 8(1)(a)-(d). 
So we conclude that these pairs are bad. Now consider the bad 
pairs $\{(1,3),(2,3)\}$ and $\{(3,3),(4,3)\}$. Since these pairs are symmetric, we will only 
prove the badness of the first pair. There are two possible orientations for the tromino 
covering $(1,1)$ and $(1,2)$. If it covers $(2,2)$, then a $3\times2$ rectangle is completed 
by the tromino covering $(2,1)$. It can be easily seen that the square $(4,1)$ becomes 
inaccessible in this case. So in order to permit a tiling, the tromino 
covering $(1,1)$ and $(1,2)$ must not complete 
a $3\times2$ rectangle. So the only possible case is that shown in Figure 8(1)(f). The reader 
can see that $(4,5)$ becomes inaccessible in this case. So we conclude that these two pairs are 
also bad. Finally, consider the pair $\{(2,3),(3,3)\}$. In this case at least two corner squares 
are made inaccessible by the trominoes covering $(1,3)$ and $(4,3)$. 
So this pair is also bad.   
We now present a very interesting observation regarding bad pairs for a $5\times7$ rectangle. 
The reader should note that the bad pairs for $R(m,n)^{--}$ are same as those for $R(n,m)^{--}$, 
so we will be alternately considering either of the above two cases.

\begin{figure}[htbp]
\centerline{
{\scalebox{.19}{\includegraphics{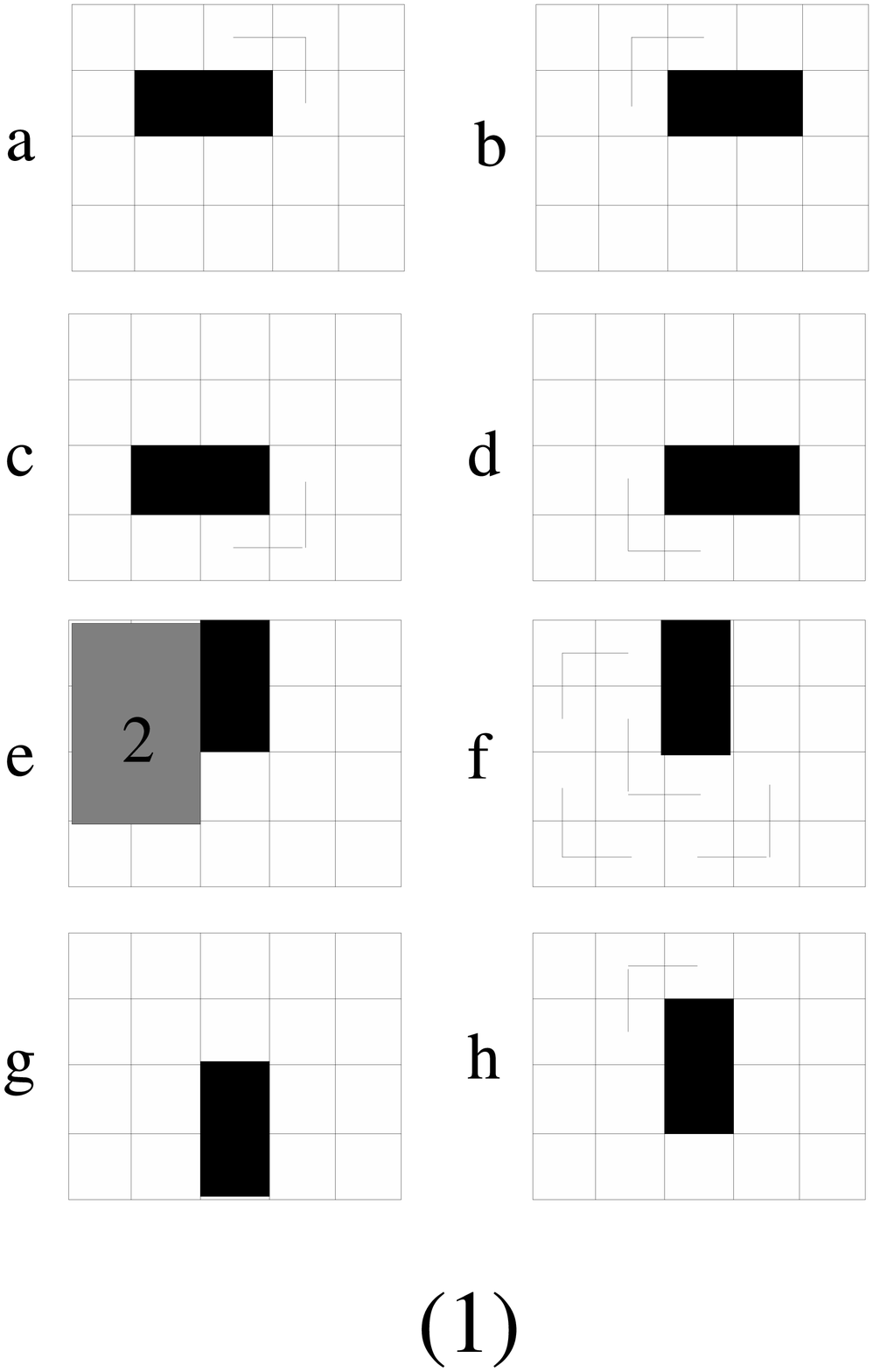}}}
{\scalebox{.19}{\includegraphics{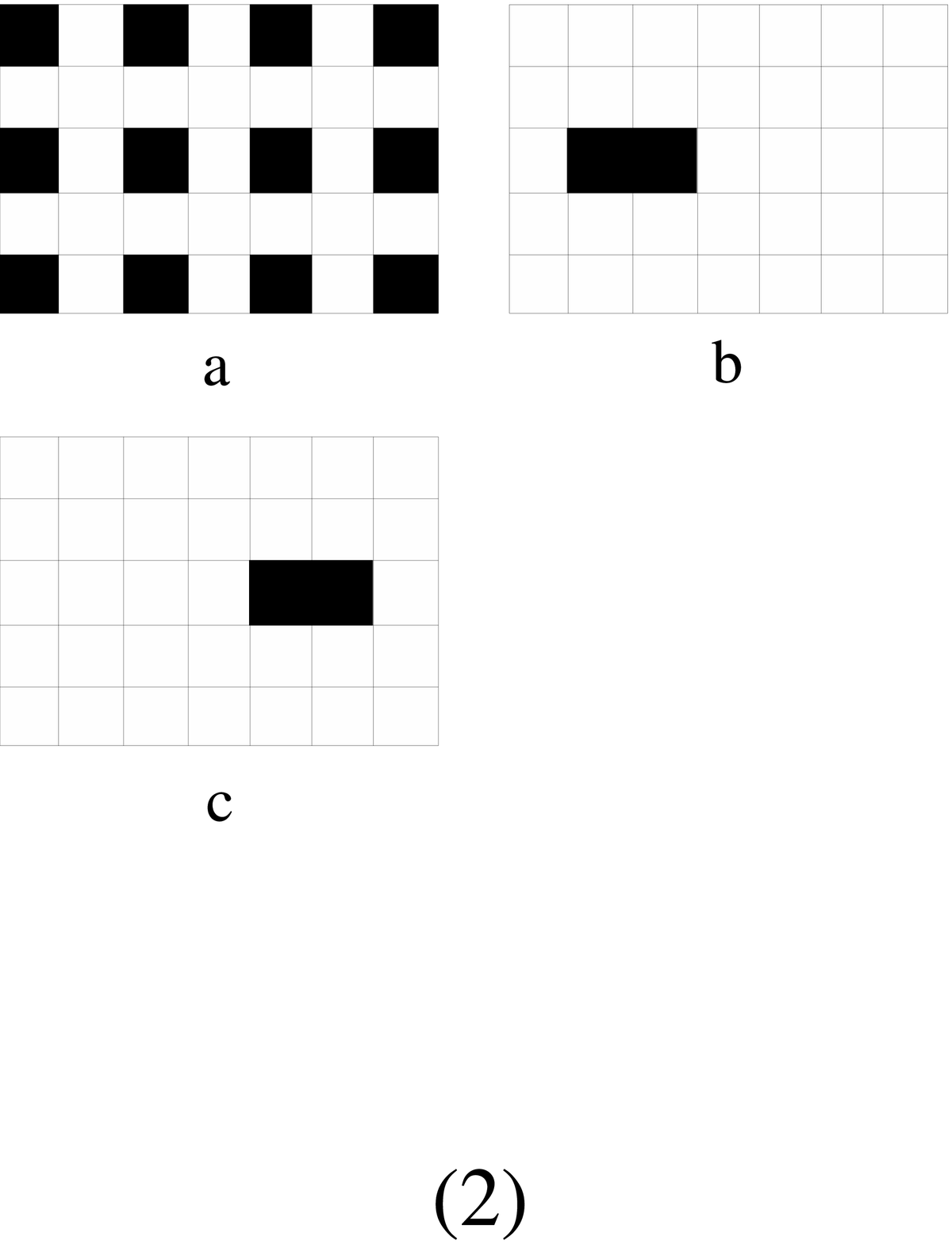}}}
{\scalebox{.19}{\includegraphics{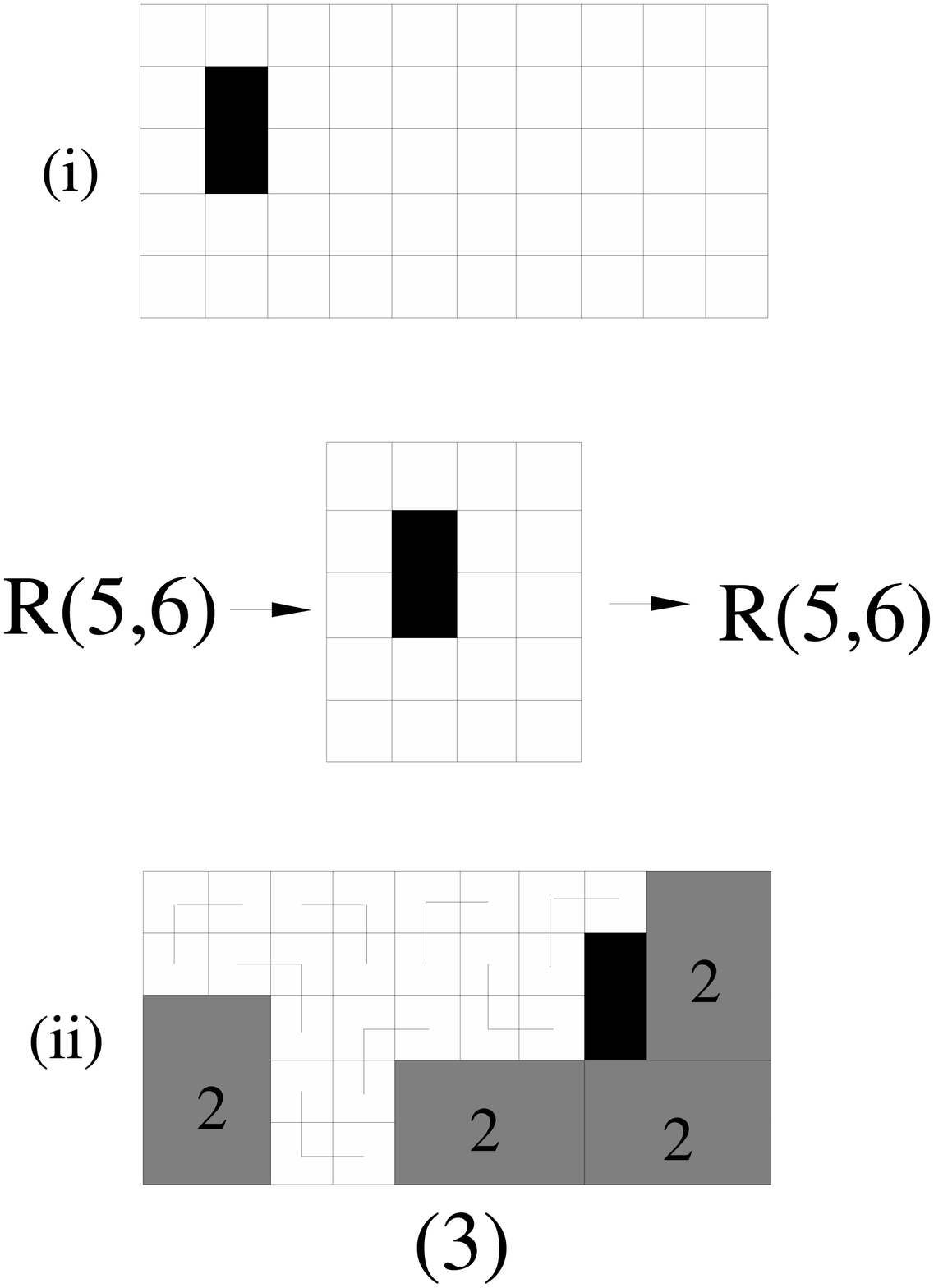}}}
{\scalebox{.19}{\includegraphics{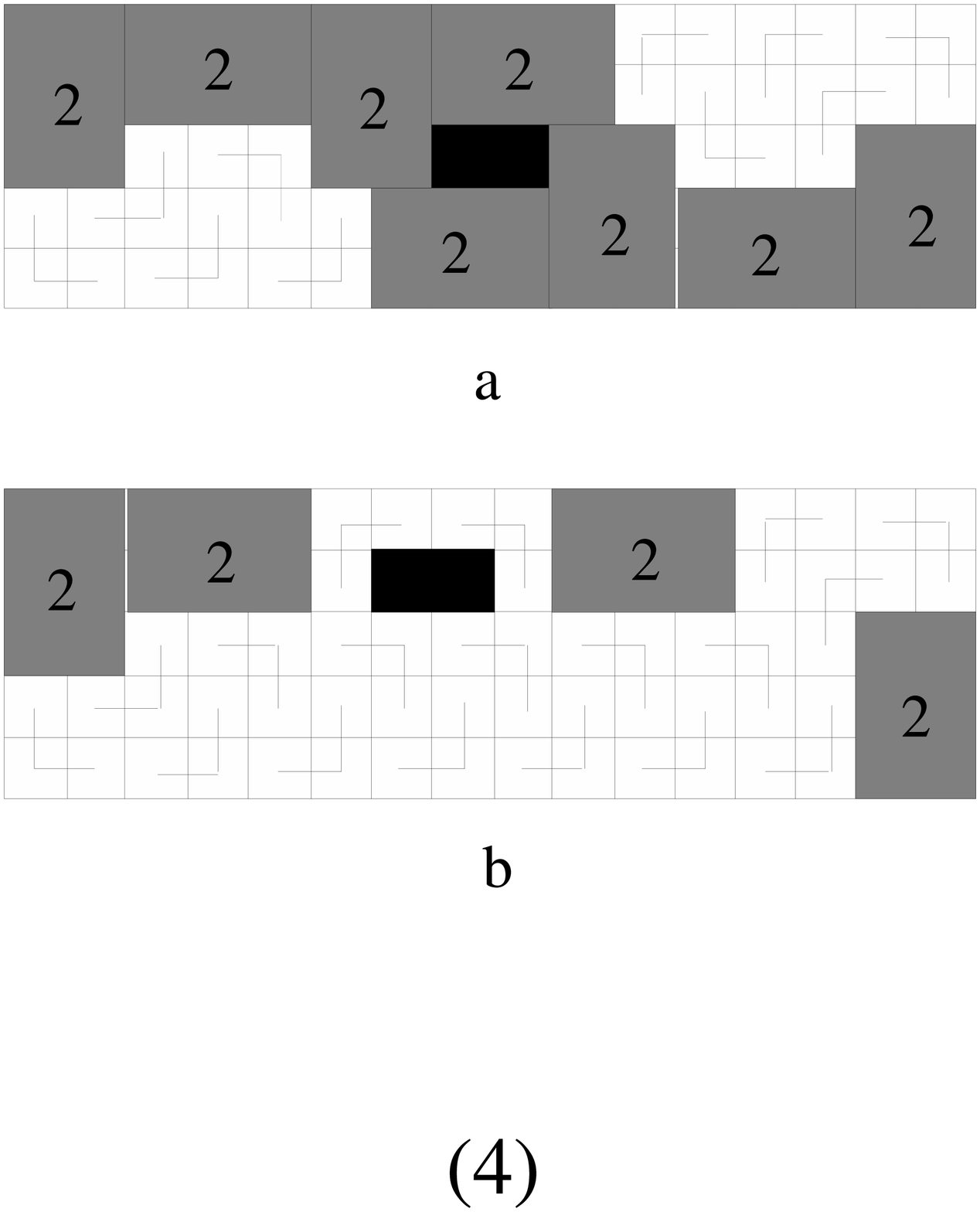}}}
}
\caption{(1) Bad pairs for $R(4,5)^{--}$. (2) Bad pairs for $R(5,7)^{--}$. 
(3) A $(5,6)$-{\it hquad shift}. (4) Alternate approach when a $(5,6)$-{\it hquad shift} fails.}
\end{figure}

\begin{lemma}{\bf [Deficient $5\times7$ Lemma]}
If both the $x$ or $y$ coordinates of the position of the domino removed from $R(5,7)$ are even, 
then the resulting shape is not tileable.
\end{lemma}
\begin{proof}
We form a kind of checkerboard by marking each of the $12$ squares as shown in Figure 8(2)(a). 
If both the $x$ or $y$ coordinates of the removed domino are even, then any tiling of 
$R(5,7)^{--}$ must contain one tromino for each of the $12$ marked squares, so that the 
tiling must have area at least $12\cdot3=36$, which is absurd since the area of $R(5,7)^{--}$ is 
$33$. Thus, all pairs which satisfy the above criterion are bad. \hfill\qed
\end{proof}

Apart from the pairs proved bad by the {\it Deficient $5\times7$ Lemma}, we have two more bad 
pairs for $R(5,7)^{--}$, namely $\{(3,2),(3,3)\}$ and $\{(3,5),(3,6)\}$ 
(as shown in Figure 8(2)(b)-(c)). Due to symmetry, 
we need consider the badness of the first pair only. Consider the tromino covering $(3,1)$ 
in the former case. The reader can easily see that either $(1,1)$ or $(5,1)$ becomes 
inaccessible. We now move on to larger rectangles. For $R(5,10)^{--}$, 
the only bad pairs are $\{(2,1),(2,2)\}$, $\{(2,9),(2,10)\}$, $\{(4,1),(4,2)\}$, $\{(4,9),(4,10)\}$, 
$\{(1,2),(2,2)\}$, $\{(1,9),(2,9)\}$, $\{(4,2),(5,2)\}$, $\{(4,9),(5,9)\}$, $\{(2,3),(2,4)\}$, 
$\{(2,7),(2,8)\}$, $\{(4,3),(4,4)\}$, $\{(4,7),(4,8)\}$, $\{(2,2),(3,2)\}$, $\{(3,2),(4,2)\}$, 
$\{(2,9),(3,9)\}$, $\{(3,9),(4,9)\}$, $\{(3,2),(3,3)\}$ and $\{(3,8),(3,9)\}$. We leave it as an 
exercise for the reader to prove the badness of the pairs mentioned above. It turns out that for 
$R(5,13)^{--}$, the only bad pairs are the corresponding analogues for those of $R(5,10)^{--}$ (as mentioned 
above). We now move on to prove something stronger, that the bad pairs for any $R(5,3t+10)^{--}$, 
where $t\geq0$, are the corresponding analogues for the pairs mentioned above. 

\begin{theorem}{\bf [Domino-Deficient Pentrec Theorem]}
The only bad pairs for $R(5,3t+10)^{--}$, where $t\geq0$, are 
$\{(2,1),(2,2)\}$, $\{(2,3t+9),(2,3t+10)\}$, $\{(4,1),(4,2)\}$, $\{(4,3t+9),(4,3t+10)\}$, 
$\{(1,2),(2,2)\}$, $\{(1,3t+9),(2,3t+9)\}$, $\{(4,2),(5,2)\}$, $\{(4,3t+9),(5,3t+9)\}$, $\{(2,3),(2,4)\}$, 
$\{(2,3t+7),(2,3t+8)\}$, $\{(4,3),(4,4)\}$, $\{(4,3t+7),(4,3t+8)\}$, $\{(2,2),(3,2)\}$, $\{(3,2),(4,2)\}$, 
$\{(2,3t+9),(3,3t+9)\}$, $\{(3,3t+9),(4,3t+9)\}$, $\{(3,2),(3,3)\}$ and $\{(3,3t+8),(3,3t+9)\}$.
\end{theorem}
\begin{proof}
The proof for the badness of the above cases of domino-removal is similar to that for $R(5,10)^{--}$ and 
$R(5,13)^{--}$, and so is left for the reader as an exercise. We will only prove the 
existence of a tiling in all other cases. Since $R(5,3)$ does not admit 
a tromino tiling, in order to satisfy the 
criterions of the Chu-Johnsonbaugh Theorem, 
we divide the case of tiling $R(5,3t+10)^{--}$ into two subcases, 
accordingly as $t$ is even or odd. First consider the case when $t$ is even, i.e., $t=2l$. 
Any tiling of $R(5,6l+10)^{--}$ can be viewed as a tiling of $l$ subrectangles of dimension 
$5\times6$ and 
one domino-deficient $R(5,10)^{--}$ rectangle. This can be written as:

\begin{eqnarray}
R(5,6l+10)^{--} & = & l\cdot R(5,6) + R(5,10)^{--}
\end{eqnarray} 

A $5\times6$ rectangle satisfies the conditions of the Chu-Johnsonbaugh Theorem and so is tileable. 
If the missing domino in $R(5,10)^{--}$ does not form a bad pair, then we can tile it, 
thereby achieving a tiling of $R(5,6l+10)^{--}$. So consider the situation when this missing domino 
forms a bad pair. In this case, we perform a $(5,6)$-hquad shift (see Figure 8(3)). If by using this 
technique, the given $R(5,10)^{--}$ rectangle becomes tileable, then we are done. However, if 
such a shift 
fails to remove the ``badness" of $R(5,10)^{--}$,then the bad pairs must be one of $\{(2,1),(2,2)\}$
(resp., $\{(2,9),(2,10)\}$, $\{(4,1),(4,2)\}$, $\{(4,9),(4,10)\}$), $\{(2,3),(2,4)\}$ 
(resp., $\{(2,7),(2,8)\}$, $\{(4,3),(4,4)\}$, $\{(4,7),(4,8)\}$), or $\{(3,2),(3,3)\}$ 
(resp., $\{(3,8),(3,9)\}$). The bad pair $\{(2,1),(2,2)\}$ changes to $\{(2,7),(2,8)\}$ 
on a $(5,6)$-hquad shift, while the pair $\{(3,2),(3,3)\}$ changes to $\{(3,8),(3,9)\}$ (the other bad pairs 
change analogously). In this case, we join a removed $5\times6$ rectangle from the left (or right) as shown in 
Figure 8(4) (symmetric cases are also possible). Note that such 
a join is always possible since the missing domino does not occupy the positions 
enumerated in the theorem statement. So, we get a tileable $R(5,16)^{--}$ and $(l-1)$ 
subrectangles of dimension $5\times6$, thereby achieving a tiling of $R(5,6l+10)^{--}$. 

Now consider the case when $t$ is odd. Assuming $t=2l+1$, where $l\geq0$, we have to tile $R(5,6l+13)^{--}$. 
Similar to the above case, we can view a tiling of $R(5,6l+13)^{--}$ as a tiling of $l$ 
subrectangles of dimension $5\times6$ and 
one domino-deficient $R(5,13)^{--}$ rectangle. Following our additive notation,

\begin{eqnarray}
R(5,6l+13)^{--} & = & l\cdot R(5,6) + R(5,13)^{--}
\end{eqnarray}

If this $R(5,13)^{--}$ rectangle is tileable, then we are done. Otherwise, the 
reader should note that none of the bad pairs (enumerated in the theorem statement) changes to another bad 
pair on applying a $(5,6)$-hquad shift. So, we can always change an untileable configuration 
of $R(5,13)^{--}$ (as above) into a tileable one. 
\end{proof}\hfill\qed

\section{Tilings of Arbitrary Domino-Deficient Rectangles}

We now move on to proving our major result. We characterize bad pairs for arbitrary $m\times n$ 
domino-deficient rectangles, 
where $m,n\geq7$ and $3|(mn-2)$. 
We first note that the only bad pairs for $R(7,8)^{--}$ are $\{(2,1),(2,2)\}$, $\{(6,1),(6,2)\}$, 
$\{(2,7),(2,8)\}$, $\{(6,7),(6,8)\}$, $\{(1,2),(2,2)\}$, $\{(6,2),(7,2)\}$, $\{(1,7),(2,7)\}$, 
$\{(6,7),(7,7)\}$, $\{(2,3),(2,4)\}$, $\{(2,5),(2,6)\}$, $\{(6,3),(6,4)\}$, $\{(6,5),(6,6)\}$, 
$\{(3,2),(4,2)\}$, $\{(4,2),(5,2)\}$, $\{(3,7),(4,7)\}$, and $\{(4,7),(5,7)\}$. It turns out that 
the corresponding analogues for $R(7,11)^{--}$ and $R(10,8)^{--}$ are the only bad pairs  
for these two rectangles also. Based on 
this observation, we have the following lemma:

\begin{figure}[htbp]
\centerline{
{\scalebox{.2}{\includegraphics{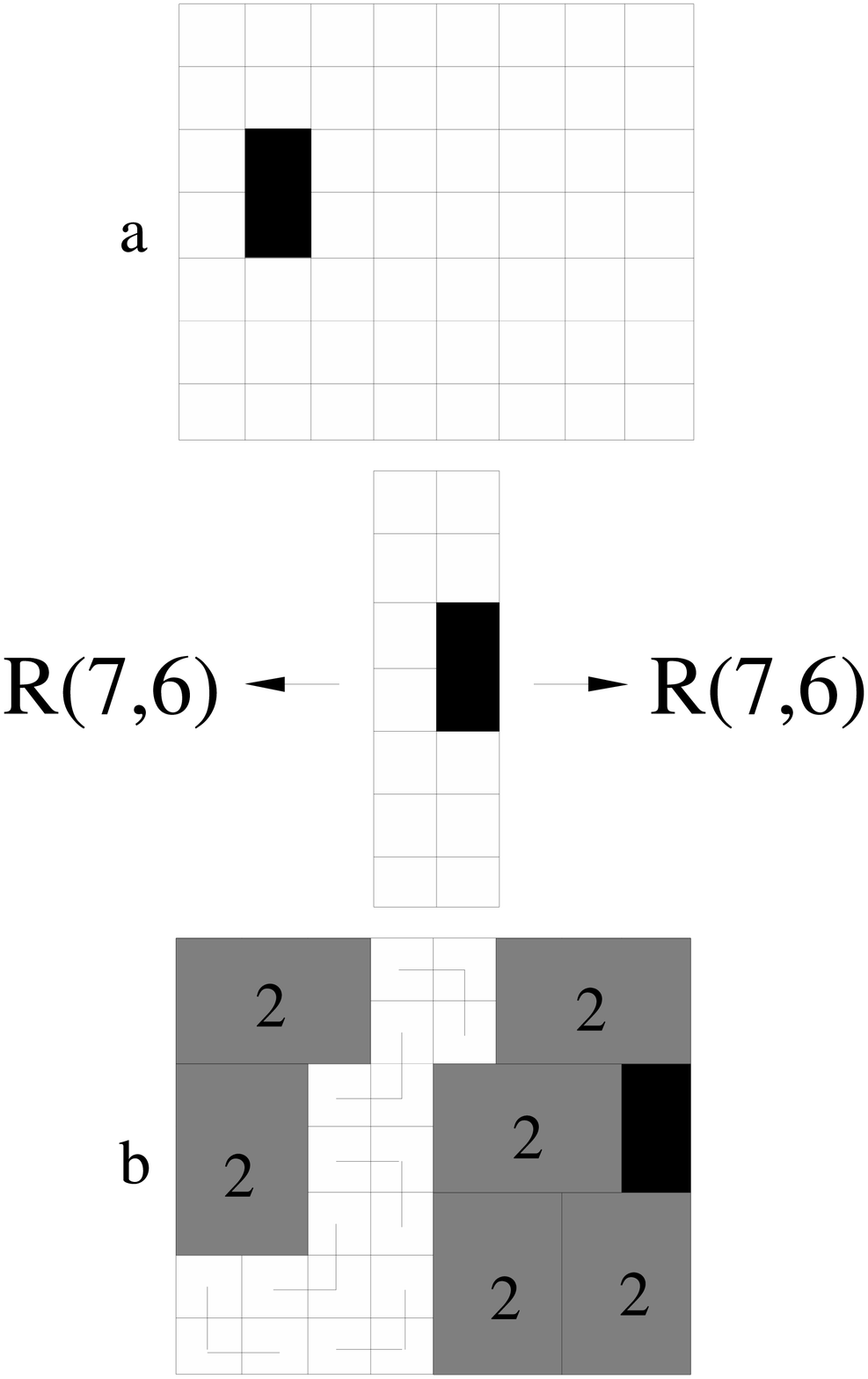}}}
{\scalebox{.2}{\includegraphics{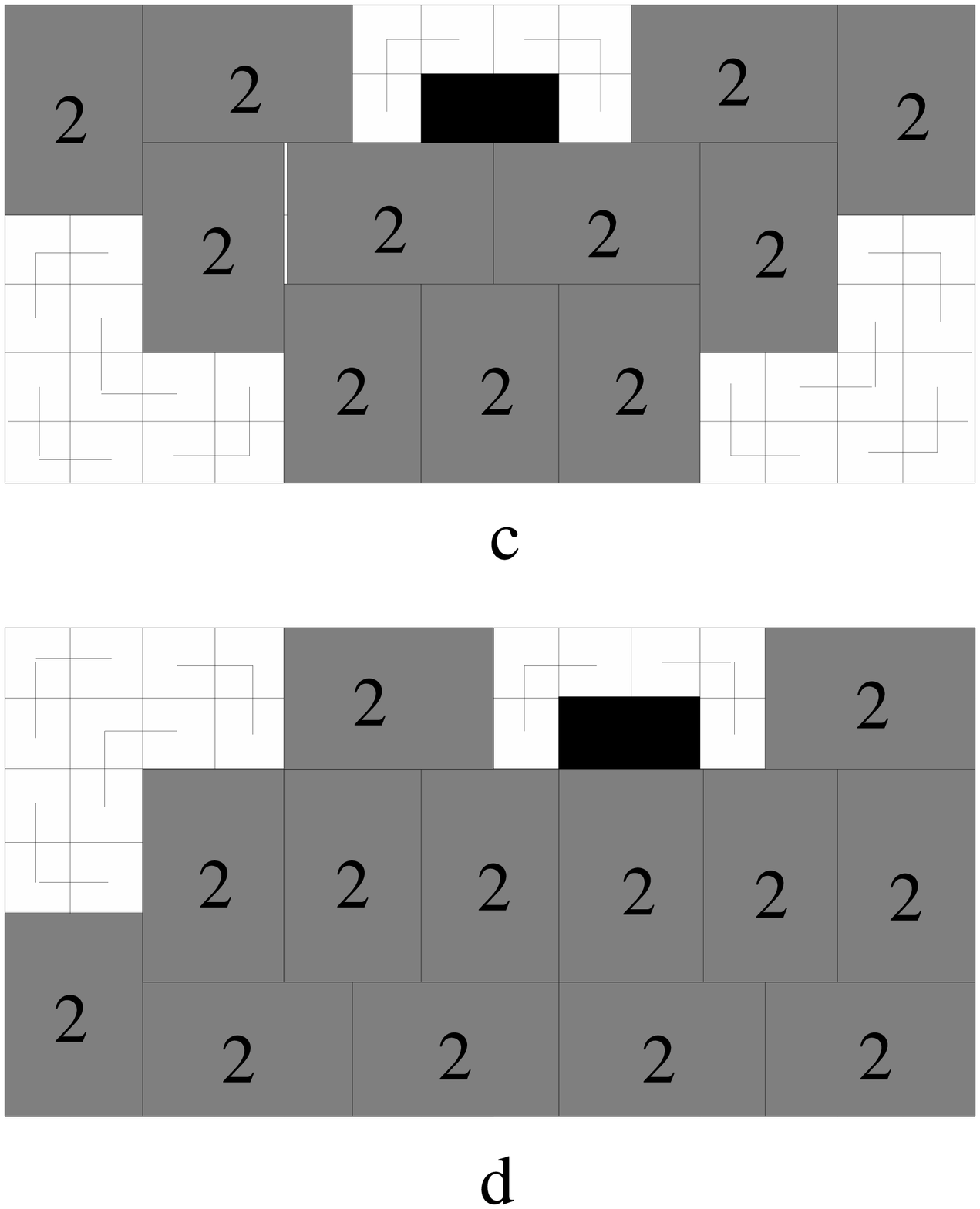}}}
}
\caption{(a) and (b) A $(10,3)$-{\it hquad shift}. (c) and (d) Tilings of $R(7,14)^{--}$ obtained by 
joining $R(7,6)$ to $R(7,8)^{--}$ having the missing domino at $\{(2,1),(2,2)\}$ or $\{(2,3),(2,4)\}$.}
\end{figure}

\begin{lemma}
The only bad pairs for $R(7,3t+8)^{--}$, where $t\geq0$, are 
$\{(2,1),(2,2)\}$, $\{(6,1),(6,2)\}$, 
$\{(2,3t+7),(2,3t+8)\}$, $\{(6,3t+7),(6,3t+8)\}$, 
$\{(1,2),(2,2)\}$, $\{(6,2),(7,2)\}$, $\{(1,3t+7),(2,3t+7)\}$, 
$\{(6,3t+7),(7,3t+7)\}$, $\{(2,3),(2,4)\}$, $\{(2,3t+5),(2,3t+6)\}$, 
$\{(6,3),(6,4)\}$, $\{(6,3t+5),(6,3t+6)\}$, 
$\{(3,2),(4,2)\}$, $\{(4,2),(5,2)\}$, $\{(3,3t+7),(4,3t+7)\}$, and $\{(4,3t+7),(5,3t+7)\}$. 
Furthermore, the corresponding analogues are the only bad pairs for $R(10,3t+8)^{--}$. 
\end{lemma}
\begin{proof}
The proof for the badness of the above cases of domino-removal is similar 
to that given in previous sections and 
so is left for the reader as an exercise.
Since $R(7,3)$ does not admit a tromino tiling, 
in order to satisfy the criterions of the Chu-Johnsonbaugh Theorem, 
we divide the case of tiling $R(7,3t+8)^{--}$ rectangles into two subcases, 
accordingly as $t$ is even or odd.  
We first consider tiling $R(7,6l+11)^{--}$ and $R(10,3t+8)^{--}$ rectangles, where $t,l\geq0$. 
Following our additive decomposition notation, we have,

\begin{eqnarray}
R(7,6l+11)^{--} & = & l\cdot R(7,6) + R(7,11)^{--} \\
R(10,3t+8)^{--} & = & t\cdot R(10,3) + R(10,8)^{--} 
\end{eqnarray}

It turns out that none of the bad pairs (enumerated in the lemma statement above) 
converts to another bad pair on applying a $(7,6)$-hquad shift on $R(7,11)^{--}$ and 
a $(10,3)$-hquad shift on $R(10,8)^{--}$. 
Also, a $7\times6$ rectangle satisfies the criterions of the Chu-Johnsonbaugh Theorem, and so is tileable.
Using these facts and equations (23) and (24),
we conclude that $R(7,6l+11)^{--}$ and $R(10,3t+8)^{--}$ always permit a tromino tiling when the 
missing domino does not occupy the pairs enumerated above. So we need only consider tiling $R(7,6l+8)^{--}$ 
rectangles, where $l\geq0$. We first note that, 
\begin{eqnarray}
R(7,6l+8)^{--} & = & l\cdot R(7,6) + R(7,8)^{--} 
\end{eqnarray}
 
If the missing domino in $R(7,8)^{--}$ does not form a bad pair, then we can tile it, thereby achieving 
a tiling of $R(7,6l+8)^{--}$. If the missing domino forms a bad pair, and is removed after performing 
a $(7,6)$-hquad shift (see Figure 9(a)-(b)), then we do the same.
However, if such a shift fails to make $R(7,8)^{--}$ tileable, 
then the reader can verify that the bad pair must be either $\{(2,1),(2,2)\}$ 
(resp., $\{(6,1),(6,2)\}$, $\{(2,3t+7),(2,3t+8)\}$, $\{(6,3t+7),(6,3t+8)\}$) or $\{(2,3),(2,4)\}$ 
(resp., $\{(2,3t+5),(2,3t+6)\}$, $\{(6,3),(6,4)\}$, $\{(6,3t+5),(6,3t+6)\}$). The bad pair $\{(2,1),(2,2)\}$ 
changes to $\{(2,3),(2,4)\}$ upon such a shift, and the other pairs change correspondingly. In this case, 
we join a removed $7\times6$ rectangle from the left (right) to form $R(7,14)^{--}$, and 
apply the tiling shown in Figure 9(c)-(d) (symmetric cases are also possible). 
We note that such a join is always possible since the bad pair is not one among the pairs 
enumerated in the lemma statement. So we get a tileable $R(7,14)^{--}$ and $(l-1)$ subrectangles of 
dimension $7\times6$, 
which tile $R(7,6l+8)^{--}$ completely.  \hfill\qed 
\end{proof}

We are now ready to prove our major result, 
thereby settling the open problem posed by Ash and Golomb 
in \cite{marshall}, for tiling an $m\times n$ rectangle when a 
domino has been removed from it, where $m,n\geq7$ and $3|(mn-2)$. 
We have the following theorem: 

\begin{theorem}{\bf [Domino-Deficient Rectangle Theorem]}
An $m\times n$ rectangle, where $m,n\geq7$ and $3|(mn-2)$, from which a domino 
has been removed, can always be tiled with trominoes provided the domino does not occupy the 
positions $\{(2,1), (2,2)\}$, $\{(2,n-1), (2,n)\}$, $\{(m-1,1), (m-1,2)\}$, $\{(m-1,n-1), (m-1,n)\}$, 
$\{(1,2),(2,2)\}$, $\{(1,n-1),(2,n-1)\}$, $\{(m-1,2),(m,2)\}$, $\{(m-1,n-1),(m,n-1)\}$, 
$\{(2,3),(2,4)\}$, $\{(2,n-3),(2,n-2)\}$, $\{(m-1,3),(m-1,4)\}$, $\{(m-1,n-3),(m-1,n-2)\}$, 
$\{(3,2),(4,2)\}$, $\{(3,n-1),(4,n-1)\}$, $\{(m-3,2),(m-2,2)\}$, and $\{(m-3,n-1),(m-2,n-1)\}$. 
\end{theorem}
\begin{proof}
Without loss of generality, we assume $m\equiv1$(mod 3), and $n\equiv2$(mod3).
We treat the cases $m=7,10$ individually. 
Note that if $m\geq13$, then $m-6\ge6$, and so the missing domino is not 
present in either the top or bottom 6 rows. So we successively slice a full rectangle of 
height $6$ off the top (bottom) of $R(m,n)^{--}$, 
that is, $R(m,n)^{--}=R(m-6,n)^{--}+R(6,n)$, until the dimension $m\le13$ is reached. 
In the above equation, the last term $R(6,n)$ 
satisfies the criterions of the Chu-Johnsonbaugh Theorem, and so is tileable. 
Following the above procedure, we eventually land up with either $R(7,3t+8)^{--}$ or 
$R(10,3t+8)^{--}$. If the missing domino is not one among the bad pairs for either of these two subrectangles 
(as enumerated in Lemma 3), then this subrectangle can be tiled by trominoes (see Lemma 3). 
Consider the case when the missing domino is one among the bad pairs enumerated in Lemma 3. 
In this case, we join a removed $6\times n$ rectangle from above or below to this subrectangle, to obtain 
$R(13,3t+8)^{--}$ or $R(16,3t+8)^{--}$. 
We divide this subrectangle by one of the following decompositions:

\begin{eqnarray}
R(13,6l+8)^{--}  & = & l\cdot R(13,6) + R(13,8)^{--} \\
R(13,6l+11)^{--} & = & l\cdot R(13,6) + R(13,11)^{--} \\
R(16,3t+8)^{--}  & = & t\cdot R(16,3) + R(16,8)^{--}  
\end{eqnarray}

\begin{figure}[htbp]
\centerline{
{\scalebox{.12}{\includegraphics{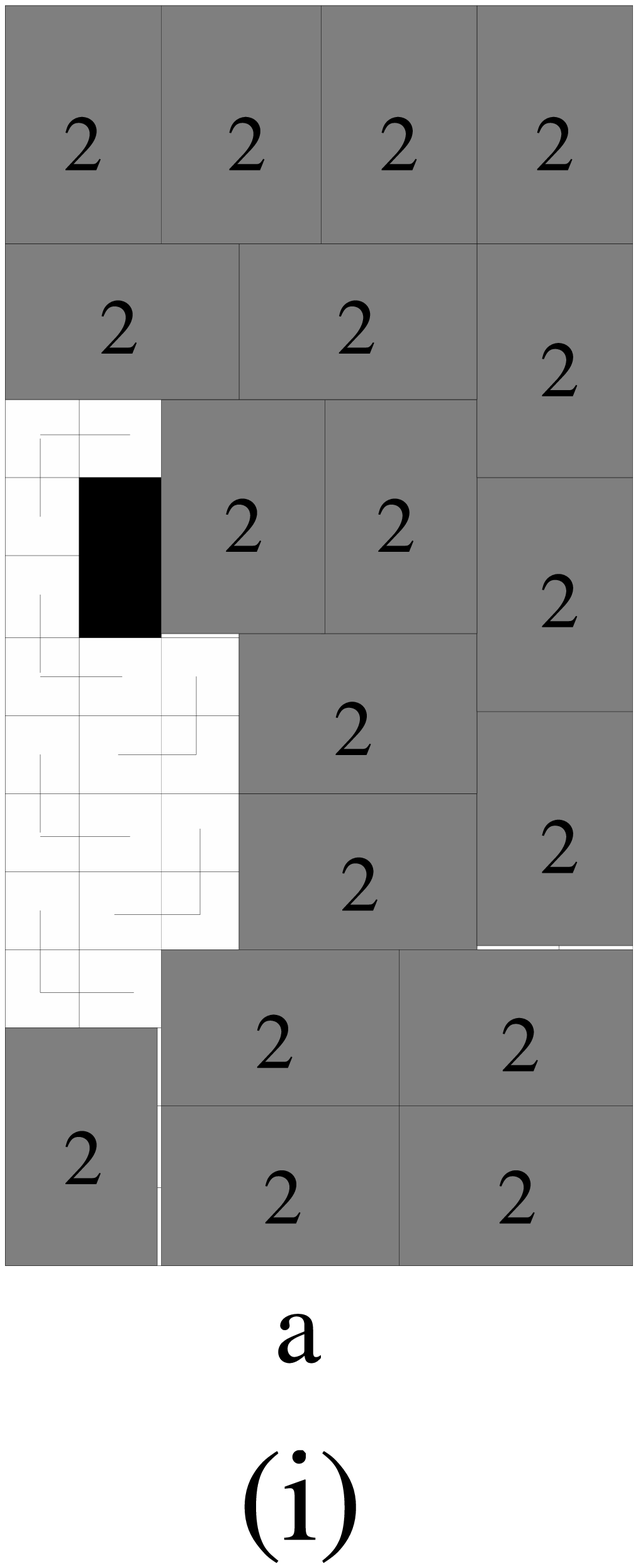}}}
{\scalebox{.12}{\includegraphics{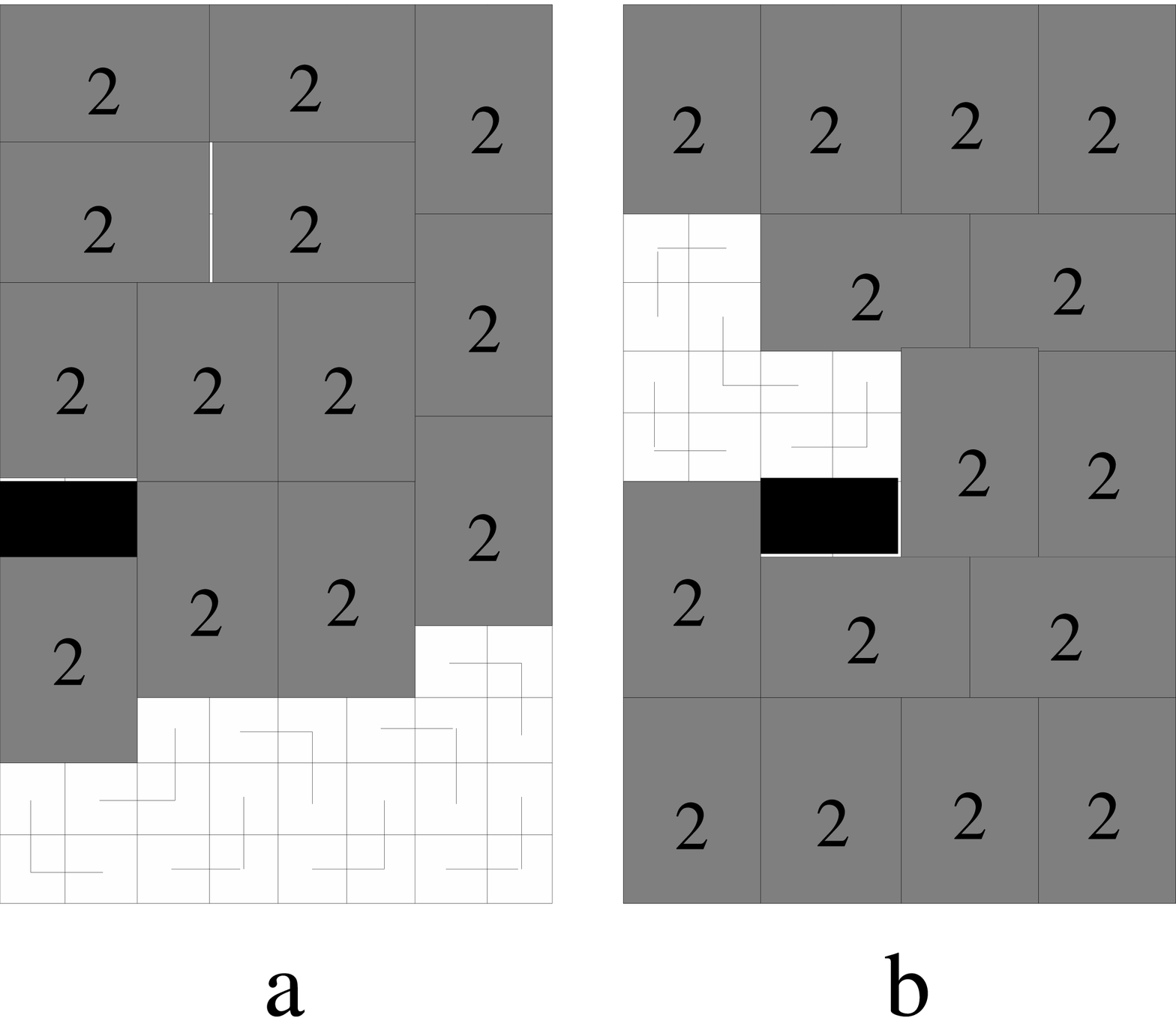}}}
{\scalebox{.12}{\includegraphics{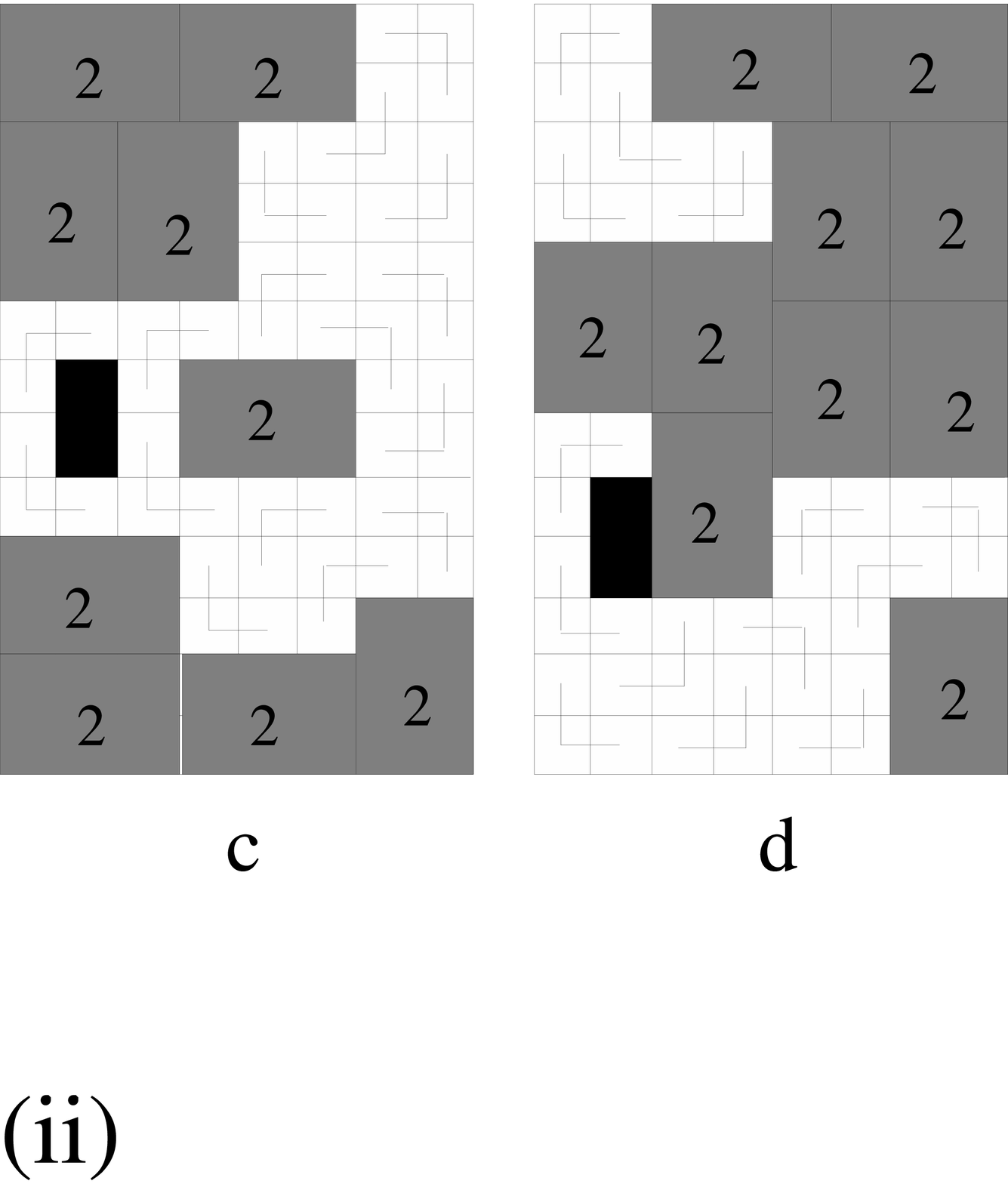}}}
{\scalebox{.12}{\includegraphics{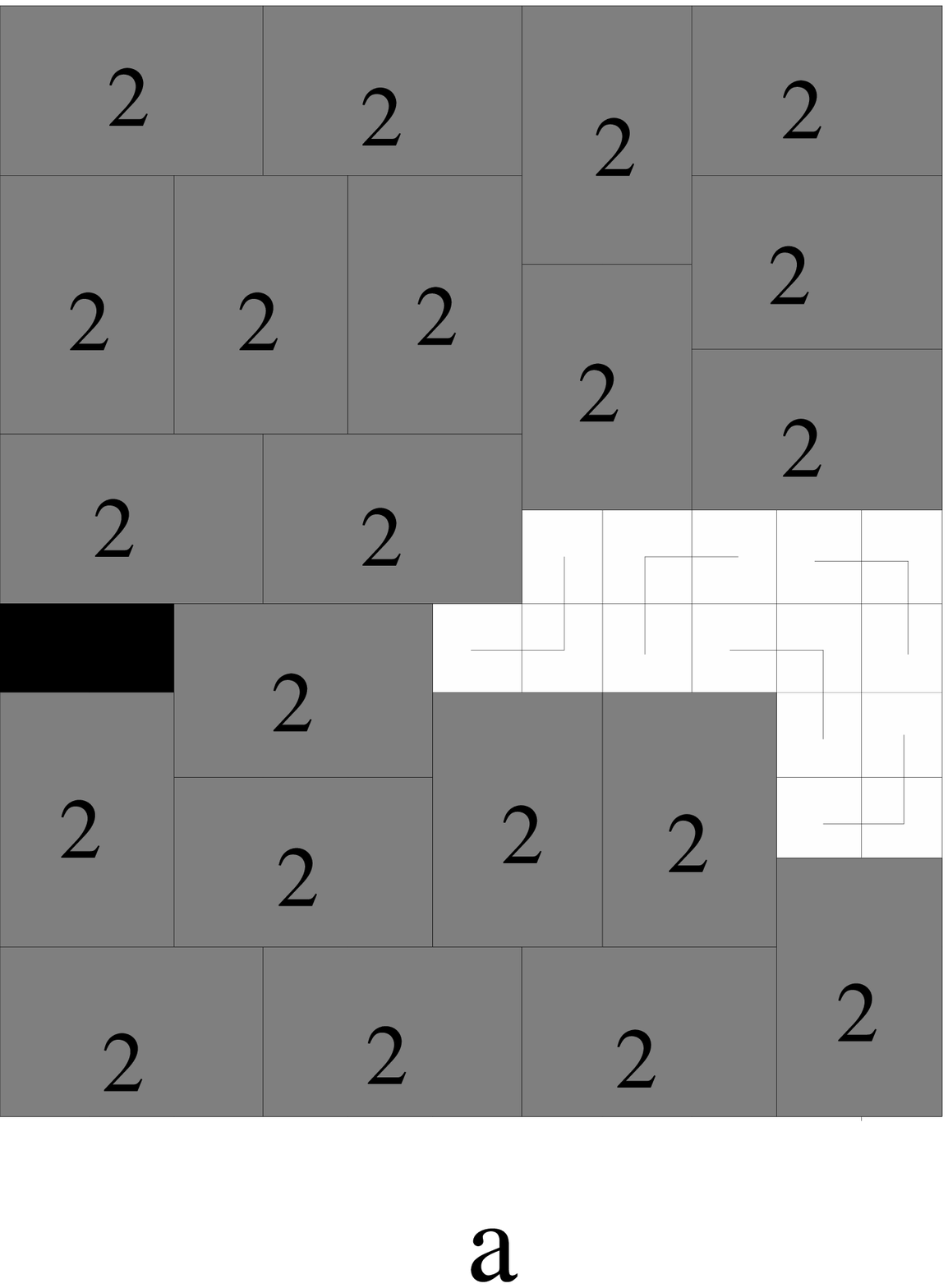}}}
{\scalebox{.12}{\includegraphics{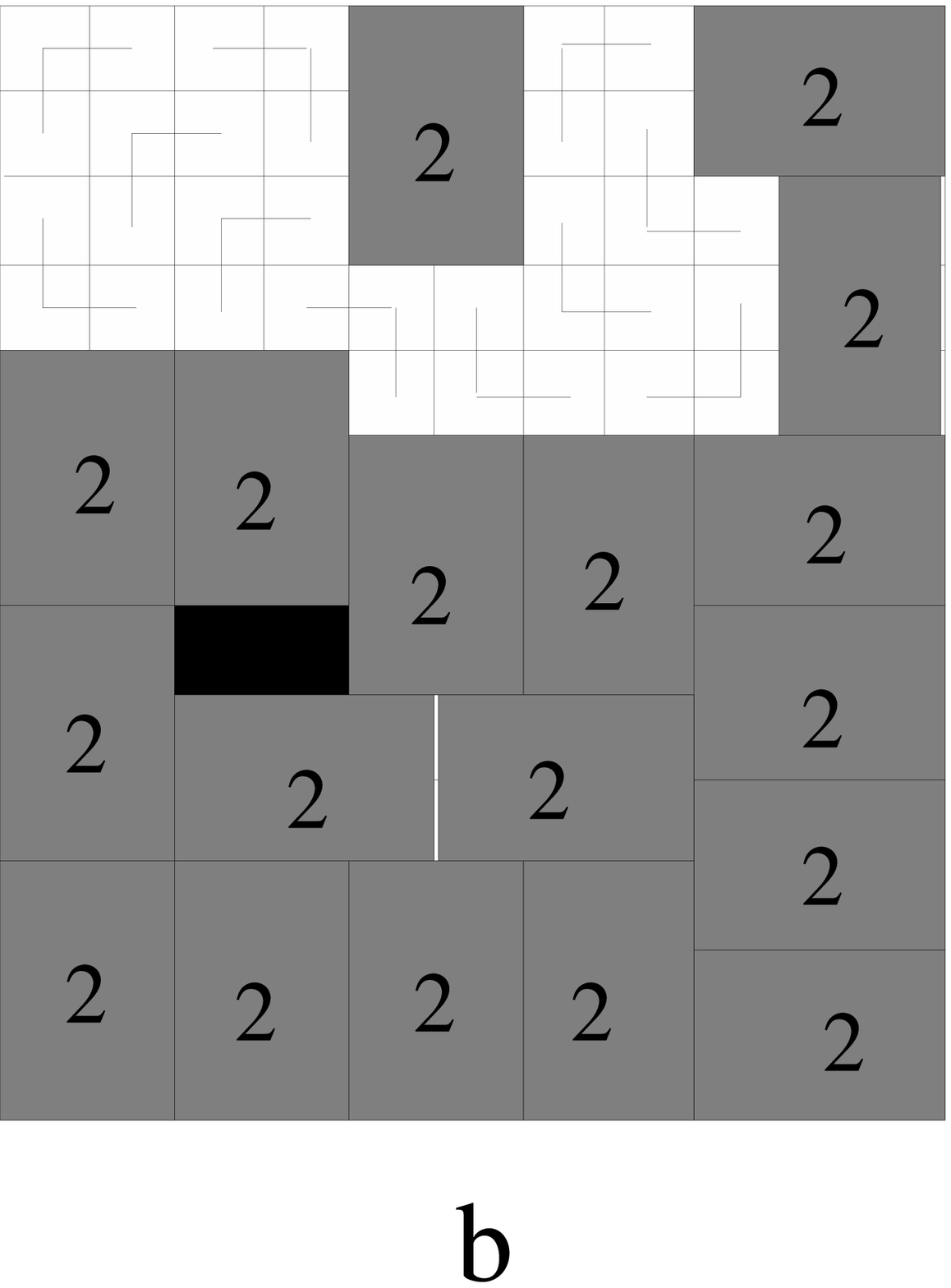}}}
{\scalebox{.12}{\includegraphics{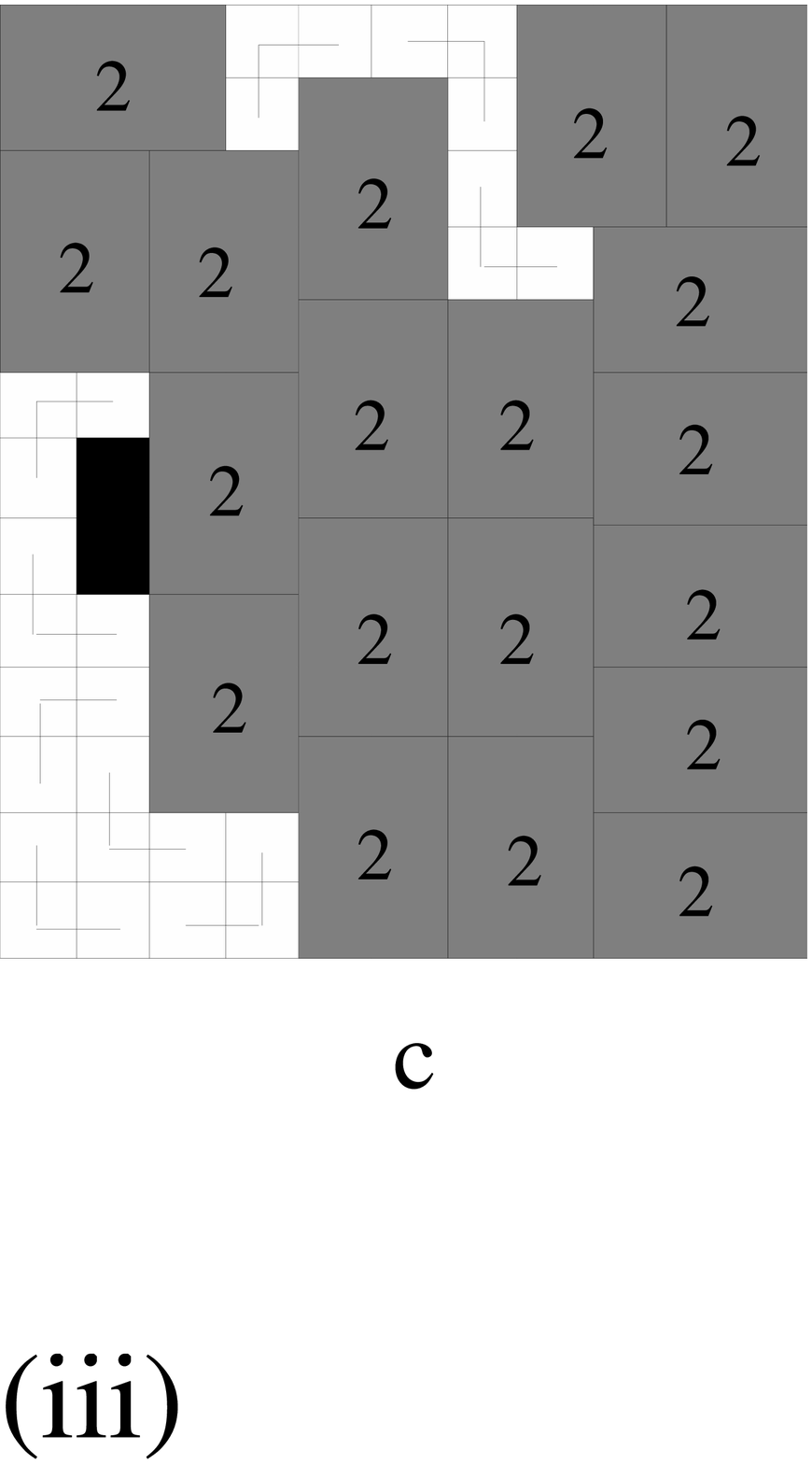}}}
{\scalebox{.12}{\includegraphics{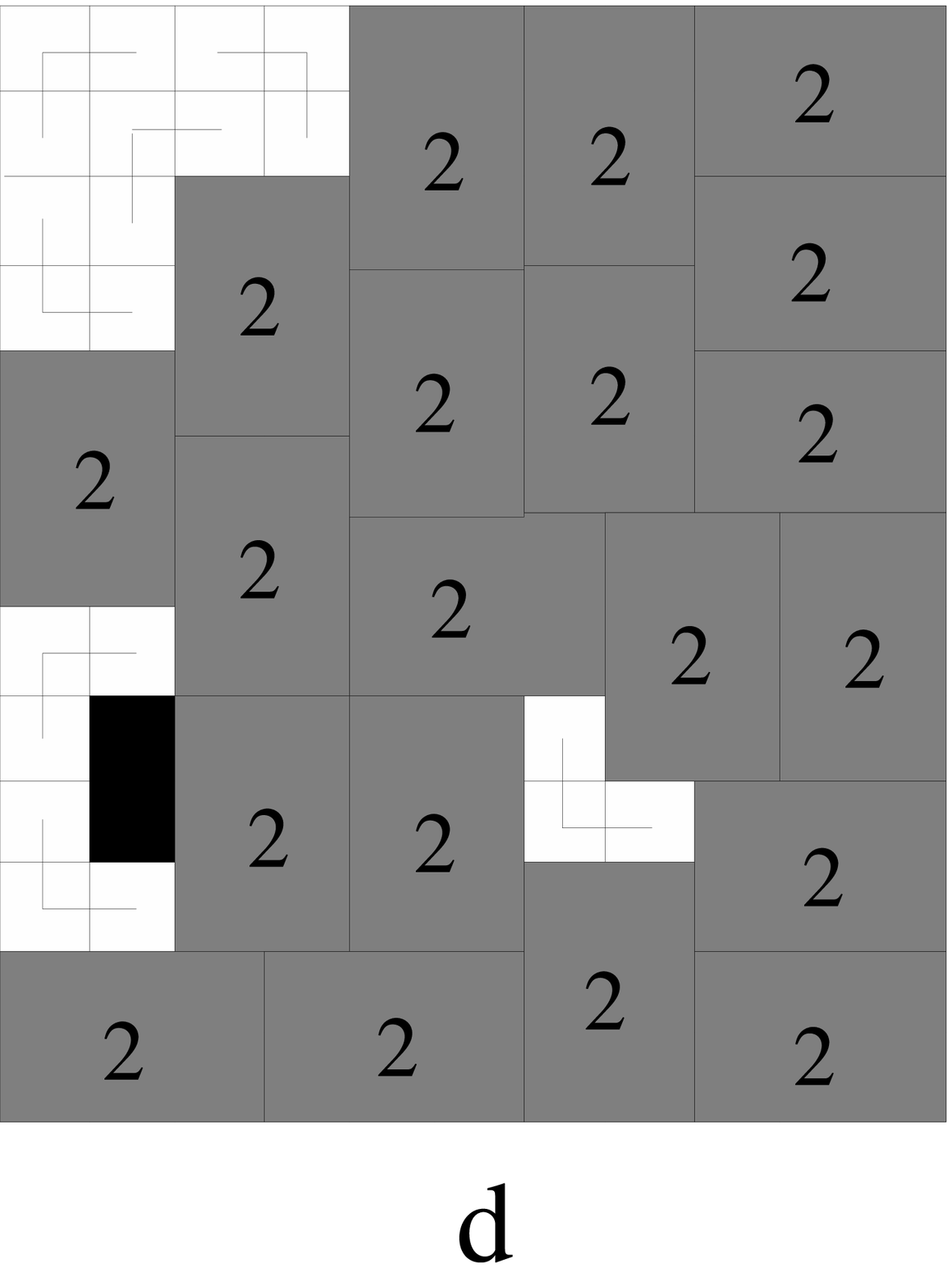}}}
}
\caption{(i) Tilings of $R(16,8)^{--}$. (ii) Tilings of $R(13,8)^{--}$. 
(iii) Tilings of $R(13,11)^{--}$.}
\end{figure}

Both $13\times6$ and $10\times3$ rectangles satisfy 
the conditions of the Chu-Johnsonbaugh Theorem, and so are tileable. 
So we need only consider tilings of $R(13,8)^{--}$, $R(13,11)^{--}$ and 
$R(16,8)^{--}$. Note that the bad cases of domino-removal $\{(2,n-1), (2,n)\}$, 
$\{(m-1,1), (m-1,2)\}$, $\{(m-1,n-1), (m-1,n)\}$, 
are actually the symmetric 
counterparts of the case $\{(2,1), (2,2)\}$, the cases
$\{(1,n-1),(2,n-1)\}$, 
$\{(m-1,2),(m,2)\}$, $\{(m-1,n-1),(m,n-1)\}$, 
are the symmetric 
counterparts of the case $\{(1,2),(2,2)\}$,.... 
and so on, when viewed from the other three 
corners of the given rectangle.
So we need only consider the bad pairs for 
$R(7,8)^{--}$, $R(7,11)^{--}$ and $R(10,8)^{--}$, before joining $R(6,8)$ or $R(6,11)$ 
from the top (bottom), to be one of the cases 
$\{(2,1), (2,2)\}$, $\{(1,2),(2,2)\}$, 
$\{(2,3),(2,4)\}$, and $\{(3,2),(4,2)\}$
(the rest follow from symmetry). 
For $(13,8)^{--}$ and $R(13,11)^{--}$, the corresponding tilings 
are shown in Figure 10(ii)-(iii) for all the four cases of domino removal. 
The situation is slightly different in case of 
$R(16,8)^{--}$. For the bad cases $\{(2,3),(2,4)\}$ and $\{(2,1),(2,2)\}$ in $R(10,8)^{--}$, the 
``badness" is removed if we perform a $(6,8)$-{\it vquad shift}. 
The pairs $\{(1,2),(2,2)\}$ and $\{(3,2),(4,2)\}$ of $R(10,8)^{--}$ actually become 
symmetrical when we join $R(6,8)$ with $R(10,8)^{--}$ from the top (bottom). 
So only one tiling is shown in Figure 10(i) which suffices for both these cases. 
\end{proof}\hfill\qed

The above proof is constructive in nature, i.e., it also identifies the tiling rule 
if there exists one. Based on the above characterization, we suggest a procedure 
for tiling an $m\times n$ domino-deficient rectangle. 
This procedure takes as input the dimensions $m, n$ of the given domino-deficient 
rectangle, and the position of the missing domino, 
all of which can be represented in $\mathcal{O}(\log m + \log n)$ bits. 
It outputs a specific tiling for 
$R(m,n)^{--}$ if there exists one; the 
five steps above may be viewed as a set of rules for generating 
such a tiling. 

\begin{algorithm}
\caption{The Domino-Deficient Tiling Procedure}
\begin{algorithmic}[1]
\STATE Remove the rectangle $R(6\cdot\lfloor i/6\rfloor,n)$ from the top.  \\
\IF {$6|(m-7)$}  
\STATE Remove the rectangle $R(m-6\cdot\lfloor i/6\rfloor-7,n)$ from the bottom\\
\ELSE
\STATE Remove the rectangle $R(m-6\cdot\lfloor i/6\rfloor-10,n)$ from the bottom \\
\ENDIF
\IF {The missing domino does not occupy the bad pairs stated in Lemma 3}
\STATE Identify the corresponding tiling rule 
for $R(7,n)^{--}$ or $R(10,n)^{--}$. \\
\STATE {\bf return}(Tiling Exists !).  \\
\ELSE
\STATE Join $R(6,n)$ from the top or bottom, and identify the tiling rule from Figure 10,  
tiling $R(13,n-8)$ ($R(13,n-11)$ or $R(16,n-8)$) by $R(2,3)$'s. \\
\STATE {\bf return}(Tiling Exists !). \\
\ENDIF
\STATE {\bf return}(No tiling exists !) \\
\end{algorithmic}
\end{algorithm}

\section{Estimating the Number of Domino-Deficient Tilings}

We now proceed to estimate the number of tromino tilings of arbitrary $m\times n$ domino-deficient 
rectangles. A natural question to ask is how does the number of tromino tilings change when the 
number of trominoes are kept the same, but some deficiencies are introduced in the given rectangle? 
We address this question for the case when the deficiency introduced is a domino removal. The first 
comparison between the number of tromino and domino tilings was done by Aanjaneya and Pal in \cite{mpal}. 
They showed that if $N_T(m,n)$ and $N_D(m,n)$ represent the number of tromino and domino tilings of 
an $m\times n$ rectangle, then the following result holds: 

\begin{theorem}
For all rectangles $R(m,n)$, such that $3|mn$ and $m,n>0$, the following inequality holds:
\begin{eqnarray}
N_T(m,n)\leq 2^{\frac{4mn}{3}}\{min[N_D(m,2n),N_D(2m,n)]\}
\end{eqnarray}

where the number of domino tilings of $R(2m,2n)$ is given by the formula,

\begin{eqnarray}
N_D(2m,2n) & = & 4^{mn}\prod_{j=1}^{m}\prod_{k=1}^{n}\{\cos^2\frac{j\pi}{2m+1} + \cos^2\frac{k\pi}{2n+1}\}
\end{eqnarray}
\end{theorem}

We use techniques similar to those in \cite{mpal} for producing an upper bound. For the sake of convenience, 
we review some definitions and notations. 
A {\it monodic tiling} of $R(m,n)$ is a tiling with $\frac{mn}{3}$ dominoes and $\frac{mn}{3}$ monominoes. 
A domino with an arrow (as shown in Figure 11) is called a {\it directed domino}. 
A monodic tiling with directed dominoes is called a {\it directed monodic tiling}. 
A tromino tiling can be converted to a directed monodic tiling by using the one-one mapping 
shown in Figure 11(a)-(d) (by converting every tromino into a combination of a directed domino and a 
monomino). A directed monodic tiling of $R(m,n)$ is {\it valid} if it was obtained from a tromino 
tiling of $R(m,n)$ via the one-one mapping shown in Figure 11. A valid directed monodic tiling can be 
converted back to a tromino tiling by attaching to every directed domino the 
monomino to the right of its arrowhead.  
It can be easily seen that every tromino tiling converts to a {\it unique}
directed monodic tiling by the above procedure. 
As depicted in Figure 11(1), a tromino tiling of $R(m,n)$, after being 
converted to the corresponding directed monodic tiling via the mapping function shown in Figure 11(1)(a)-(d), 
is first made {\it undirected} and {\it coloured}, and then {\it stretched} 
either horizontally or vertically to give a corresponding domino tiling. 
By {\it stretching} we double either the length $n$ or the breadth $m$ of the rectangle $R(m,n)$. 
(Colouring is defined below.)
In this process, every monomino gets converted to a domino, and a domino gets converted into either 
two horizontal dominoes lying side-by-side (in case of a horizontal stretching of $R(m,n)$), or it 
gets converted to two horizontal dominoes one lying on top of the other (in case of a vertical 
stretching of $R(m,n)$). These conversions are shown in Figures 11(1) (i)-(j). So a tromino tiling of 
$R(2,3)$ (as shown in Figure 11(1)(e)) gets converted to a domino tiling of either $R(2,6)$ or 
$R(4,3)$ (as shown in Figures 11(1)(g) and (h)). The reader must note that not every domino tiling 
of $R(m,2n)$ is obtained by stretching a monodic tiling of $R(m,n)$. Note, for example, the domino 
tiling of $R(3,4)$ shown in Figure 11(2). The reader can verify that this domino tiling cannot be 
obtained by stretching (either horizontally or vertically) any monodic tiling of $R(3,2)$. 

\begin{figure}[htbp]
\centerline{
{\scalebox{.25}{\includegraphics{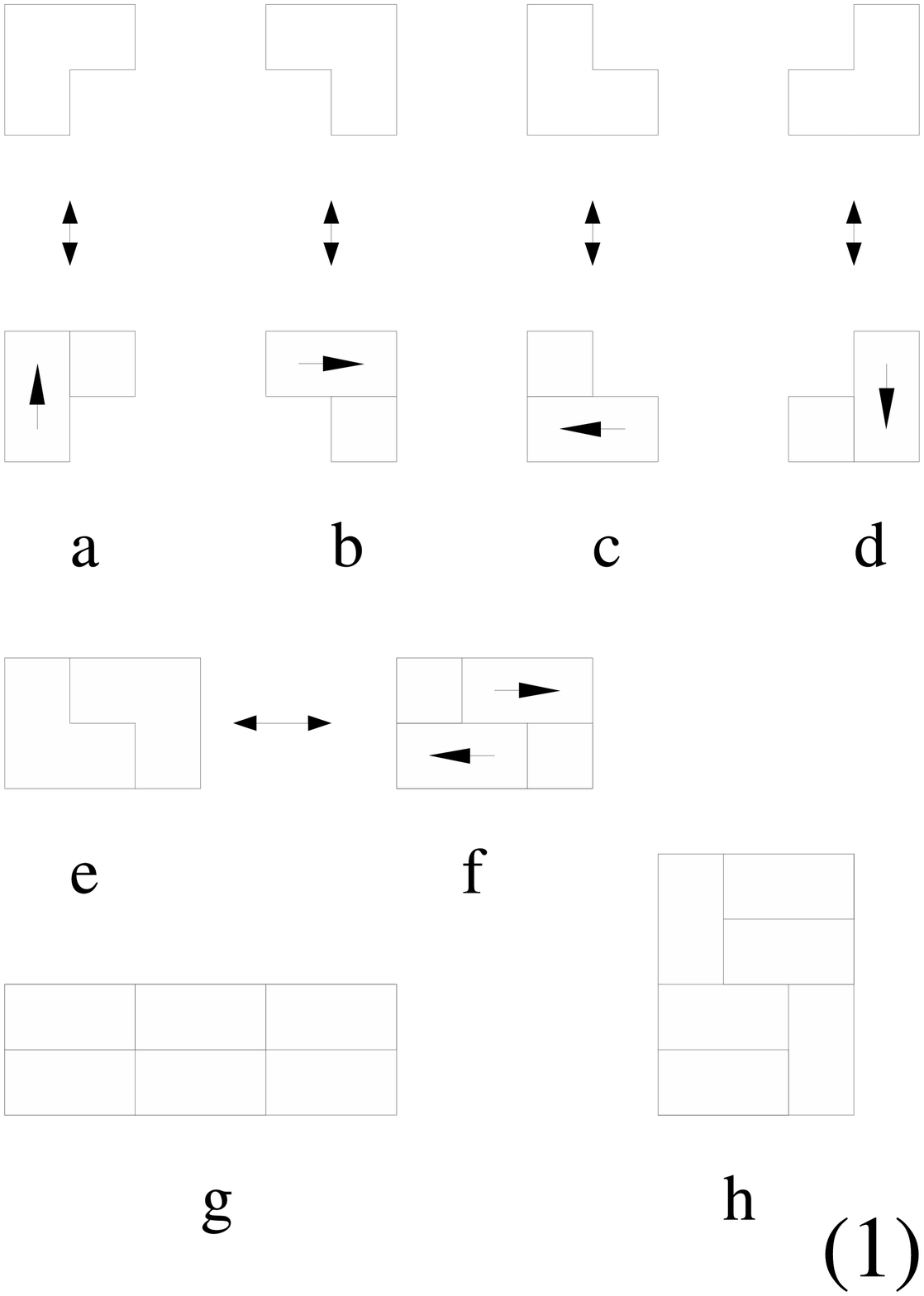}}}
{\scalebox{.23}{\includegraphics{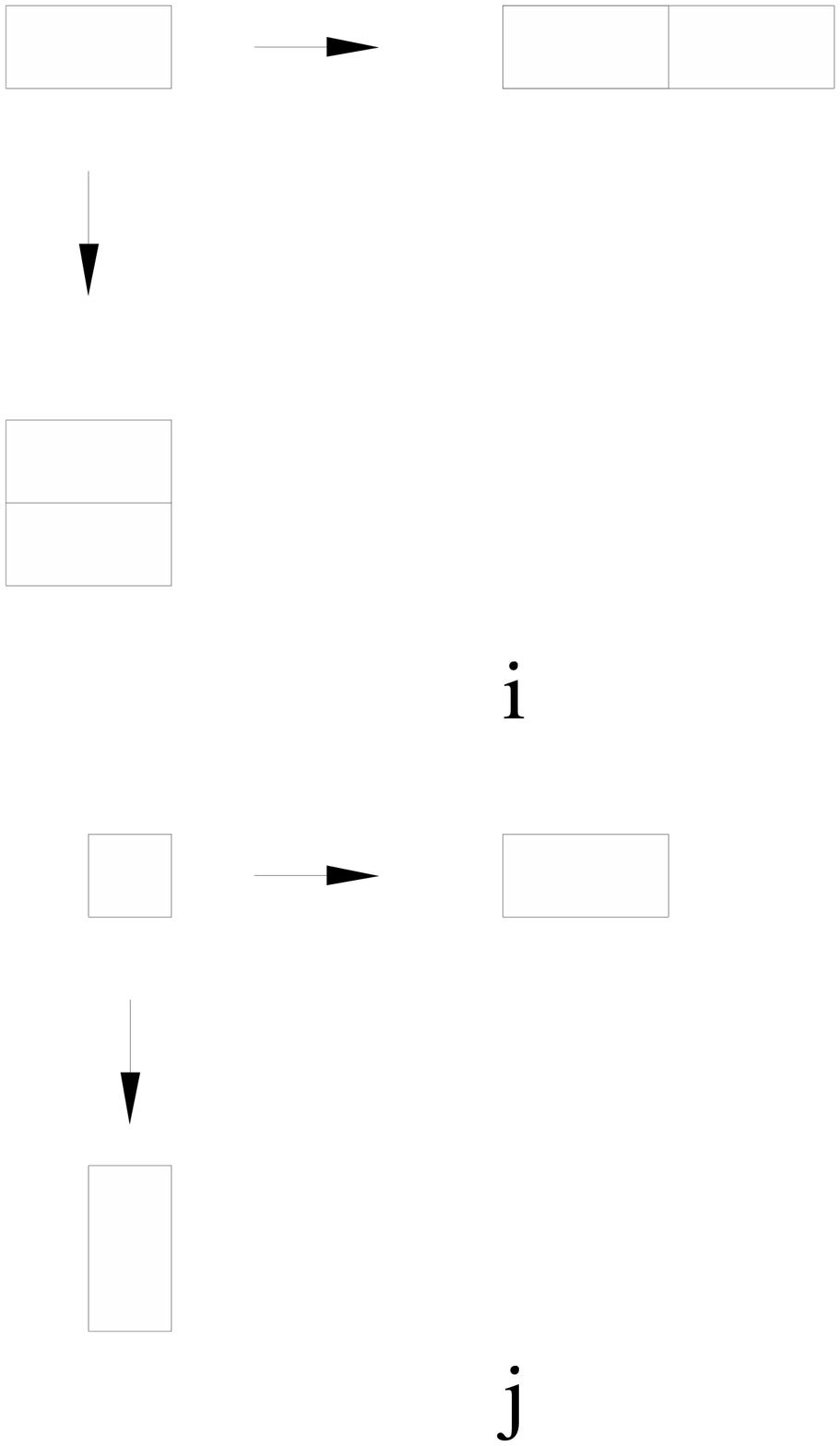}}}
{\scalebox{.15}{\includegraphics{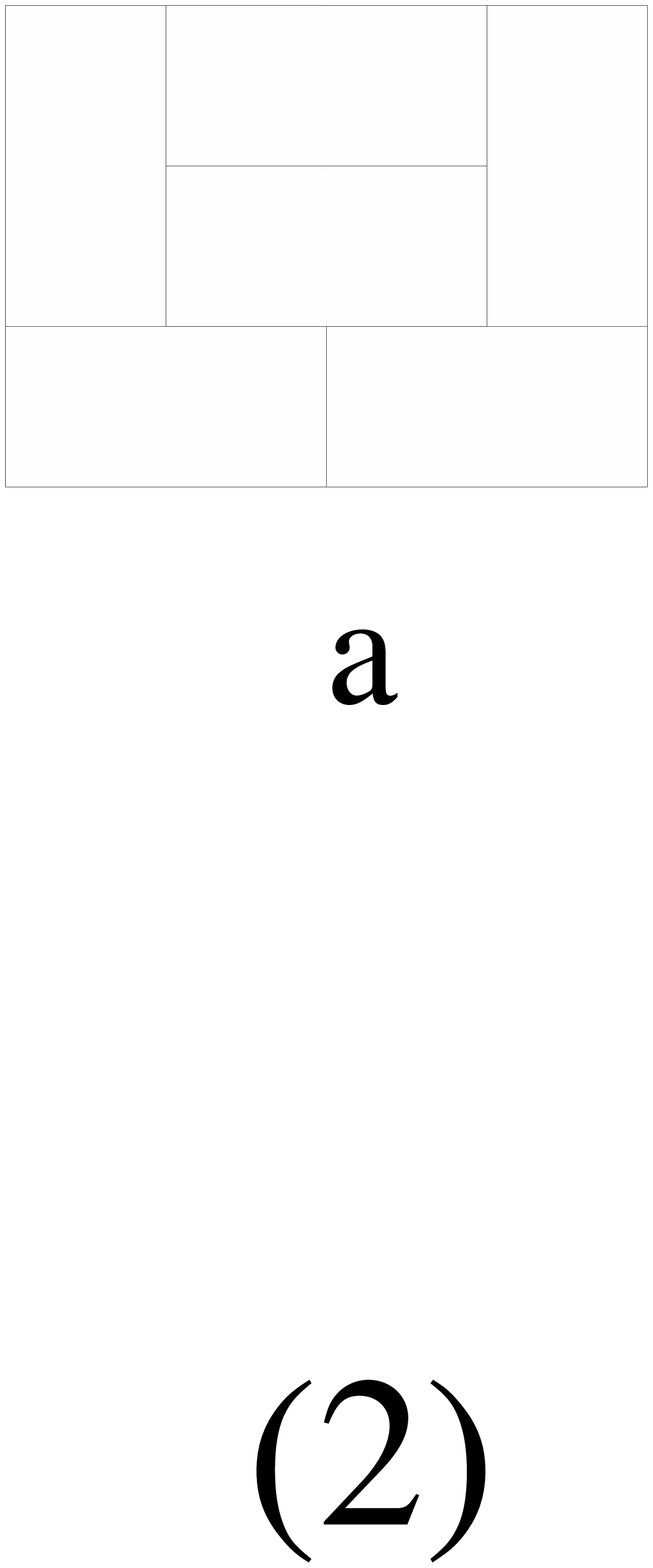}}}
}
\caption{(1) Mapping function from a tromino tiling to a domino tiling. 
(2) A domino tiling which is not obtained via the given mapping function.}
\end{figure}

Now let us define our operations in case of an $m\times n$ domino-deficient rectangle. 
Similar to a monodic tiling, we define a {\it deficient monodic tiling} 
for an $m\times n$ domino-deficient rectangle, as a tiling with $\frac{mn-2}{3}+1$ dominoes and 
$\frac{mn-2}{3}$ monominoes.
While converting a tromino tiling of the given rectangle to a directed monodic tiling, we just 
don't attach any direction with the missing domino. We call the resulting tiling a {\it deficient 
directed monodic tiling}.  The reader can easily see that a deficient directed monodic tiling 
can be converted back to the original domino-deficient tromino tiling by the same procedure described above 
(the missing domino is simply left untouched). 
If we mutiply the total number of deficient monodic tilings of $R(m,n)$ by $2^{\frac{mn-2}{3}}$, we get the 
total number of deficient directed monodic tilings of $R(m,n)$. First, let us focus only on the case when the 
stretching done is horizontal. With slight modification, 
the original streching procedure can also be converted 
to a one-to-one mapping function which converts every 
deficient monodic tiling of $R(m,n)$ to a domino tiling 
of $R(m,2n)$. We call a domino tiling of $R(m,2n)$ {\it valid} if it was obtained by stretching a 
deficient monodic tiling of $R(m,n)$. Any valid domino tiling can be 
converted back to a deficient monodic tiling of 
$R(m,n)$ simply by {\it unstretching} (i.e., compressing lengthwise) the valid domino tiling of $R(m,2n)$.

Suppose we have a deficient monodic tiling of $R(m,n)$. We {\it colour} this tiling in the following manner, 
every monomino is coloured {\it blue} and every domino is coloured {\it red}. We now stretch $R(m,n)$ 
length-wise to produce $R(m,2n)$. The outcome will be a {\it coloured} domino tiling of $R(m,2n)$. Note 
that the dominoe(s) produced by the stretching of a coloured monomino (domino) will be of the same colour 
as that of the monomino (domino). And similarly, we define the unstretching of a coloured domino tiling. 
We have the following lemma.  

\begin{lemma}
Distinct coloured deficient monodic tilings of $R(m,n)$ give rise to distinct coloured domino tilings of $R(m,2n)$.
\end{lemma}
\begin{proof}
We prove the above claim by the method of contradiction. We assume on the contrary that the above claim is false. First suppose that two distinct coloured deficient monodic tilings of $R(m,n)$ give rise to the same coloured domino tiling of $R(m,2n)$ via stretching. Since the two monodic tilings of $R(m,n)$ being considered are distinct, we know that the position of at least one monomino is different in these two tilings. This means that the position of a monomino in the first tiling of $R(m,n)$ is covered by a domino in the second tiling of $R(m,n)$. Let this monomino be $(i,j)$. It can be easily seen that after stretching of $R(m,n)$, the squares $(i,2j)$ and $(i,2j+1)$ correspond to the square $(i,j)$ in $R(m,n)$. These two squares in $R(m,2n)$ will be covered by a blue domino via the first monodic tiling and a red domino via the second. But this is impossible because both the coloured deficient monodic tilings of $R(m,n)$ being considered give rise to the same coloured domino tiling of $R(m,2n)$. Thus, we arrive at a contradiction. 
 
Now suppose that two distinct coloured domino tilings of $R(m,2n)$ give rise to the same coloured deficient 
monodic tiling of $R(m,n)$ after unstretching. Again, since the two coloured domino tilings are distinct, 
the position of at least one square $(i,j)$ (say) in one domino tiling is covered by a red (blue) domino 
and which is covered by a blue (red) domino in the other. The reader can see that after unstretching the 
rectangle $R(m,2n)$, each square $(i,j)$ in $R(m,2n)$ maps to $(i,[j/2])$ in $R(m,n)$ (here [x] represents 
the integer part of x). When we unstretch the two coloured tilings, this means that the square $(i,[j/2])$ 
in $R(m,n)$, corresponding to the square $(i,j)$ in $R(m,2n)$, will be red (blue) via the first tiling 
and blue (red) via the second tiling. Thus, we again arrive at a contradiction. So, we conclude that 
distinct coloured deficient monodic tilings of $R(m,n)$ map to distinct coloured domino 
tilings of $R(m,2n)$.   \hfill \qed
\end{proof}

We have already shown that not all domino tilings of $R(m,2n)$ are valid. Thus, via the two mappings 
just described above, we have defined an injective function from the set of domino-deficient tromino 
tilings of $R(m,n)^{--}$ to a proper subset of coloured domino tilings of $R(m,2n)$. 
Since the direction of stretching is 
unimportant in the above arguments, the reader should convince himself that the above claims also 
hold for vertical stretching of rectangles. 
The number of coloured 
domino tilings of $R(m,2n)$ is simply $2^{mn}$ times the number of domino tilings of $R(m,2n)$ 
(since every domino can be coloured either red or blue). 
Let $N_T(m,n)^{--}$ and $N_D(m,n)$, denote the number of 
tromino tilings of $R(m,n)^{--}$ and the number of domino tilings of $R(m,n)$.
From a given tromino tiling of an $m\times n$ domino-deficient rectangle, we can 
get at most $2^{\frac{mn-2}{3}}.k$ deficient directed monodic tilings, where $k$ is the 
number of deficient monodic tilings of the given rectangle. From each tiling thus obtained, 
we can get a unique coloured domino tiling of $R(m,2n)$, of which there are at most 
$2^{mn}N_D(m,2n)$ of them.  
We summarize these arguments in the following theorem:

\begin{theorem}
For all rectangles $R(m,n)$, such that $3|(mn-2)$ and $m,n>0$, the following inequality holds:
\begin{eqnarray}
N_T(m,n)^{--}\leq 2^{\frac{4mn-2}{3}}\{min[N_D(m,2n),N_D(2m,n)]\}
\end{eqnarray}

where the number of domino tilings of $R(2m,2n)$ is given by the formula,

\begin{eqnarray}
N_D(2m,2n) & = & 4^{mn}\prod_{j=1}^{m}\prod_{k=1}^{n}\{\cos^2\frac{j\pi}{2m+1} + \cos^2\frac{k\pi}{2n+1}\}
\end{eqnarray}
\end{theorem}

The upper bound derived above shows that the number 
of tilings of an $m\times n$ domino-deficient rectangle is at most exponential in both $m$ and $n$. 
So, the number of binary bits required in encoding any tiling 
of the rectangle is upper bounded by a polynomial in $m$ and $n$. 

\section{A step towards general $2$-deficiency in rectangles}

Now consider general $2$-deficiency in rectangles. 
In this section, we denote $2$-deficient $m\times n$ rectangles 
by $R(m,n)^{--}$. 
General $2$-deficiency 
as such is a lot more complicated than domino-deficiency, and till date no characterizations 
exist for any classes of rectangles. In this paper, we 
restrict ourselves to studying general $2$-deficiency in $n\times4$ rectangles. 
For a few examples of the bad pairs for the case when one dimension of the given $m\times n$ rectangle, 
say $m=7$ see \cite{mridul}. 
Apart from the bad pairs shown in Figure 4 of Section 4, Figure 12 shows all bad 
pairs for $R(8,4)^{--}$. 
We encourage the reader to try placing trominoes on $R(8,4)^{--}$ 
with the $2$-deficiencies shown in Figure 12, and verify for himself that the pairs shown are actually bad. 
It turns out that for $R(3t+8,4)^{--}$, where $t\geq0$, 
the corresponding analogues of the pairs shown in Figure 12 are the only bad pairs. We prove this fact in 
the following theorem:

\begin{figure}[htbp]
\centerline{
{\scalebox{.13}{\includegraphics{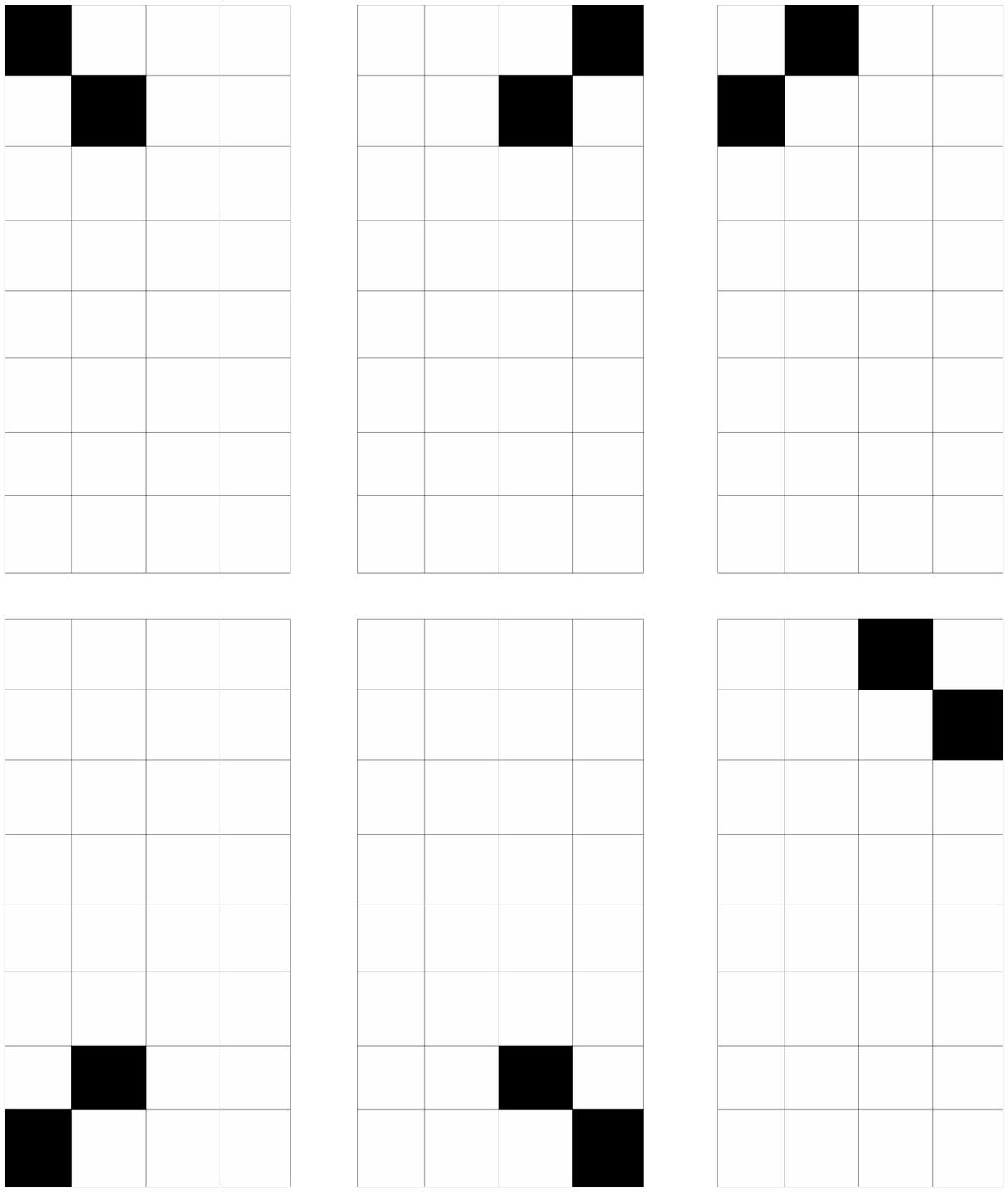}}}
{\scalebox{.13}{\includegraphics{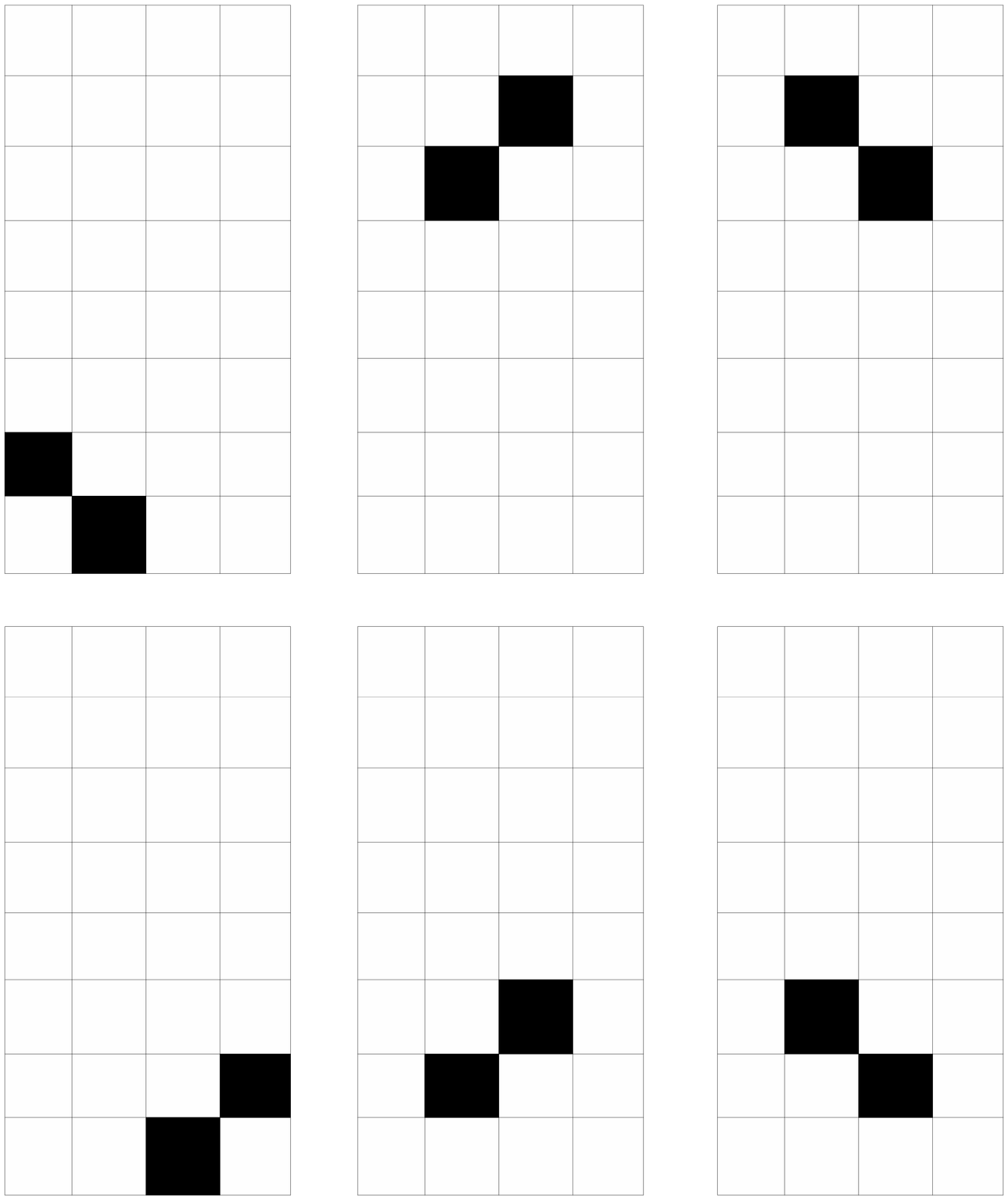}}}
{\scalebox{.13}{\includegraphics{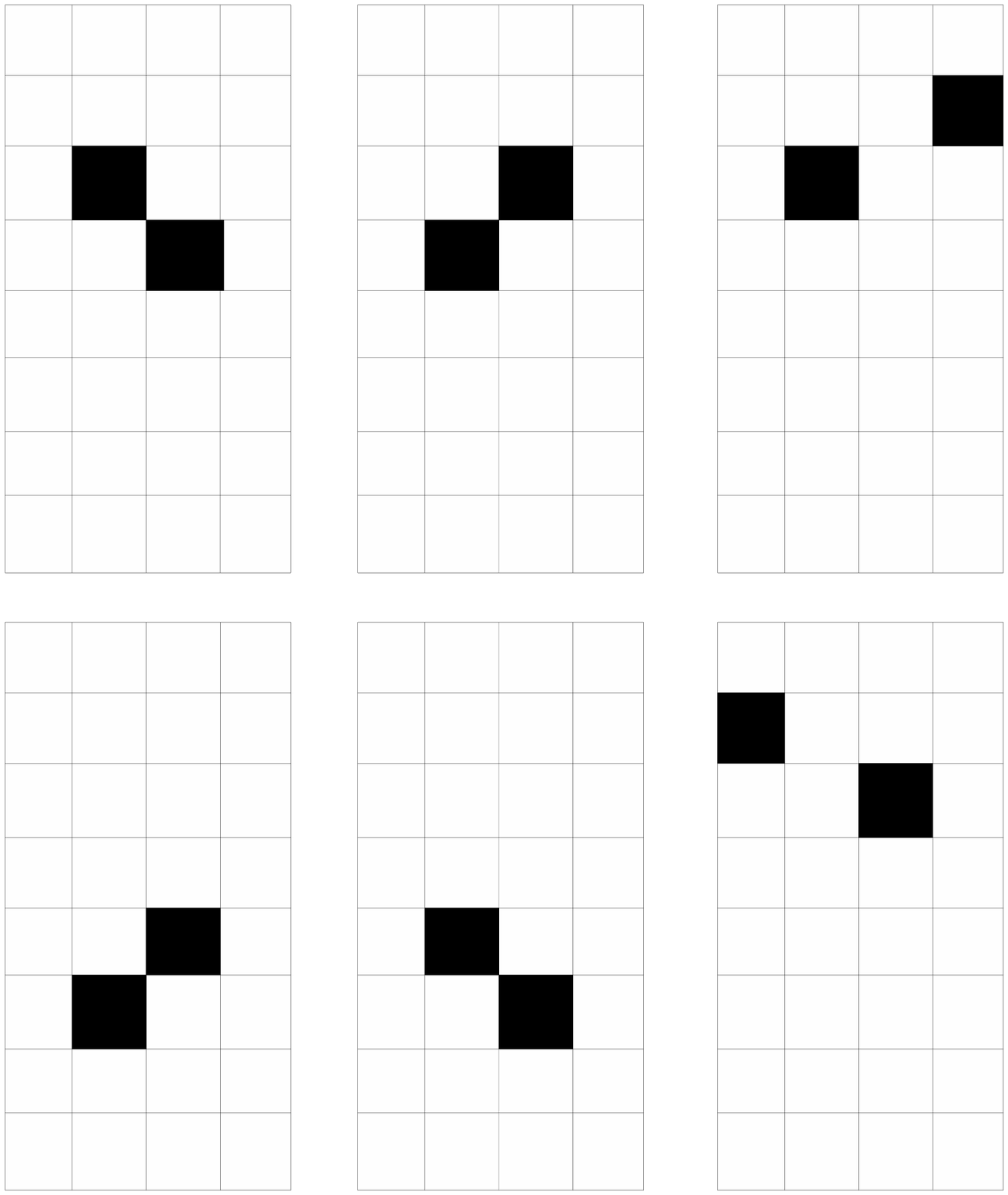}}}
{\scalebox{.13}{\includegraphics{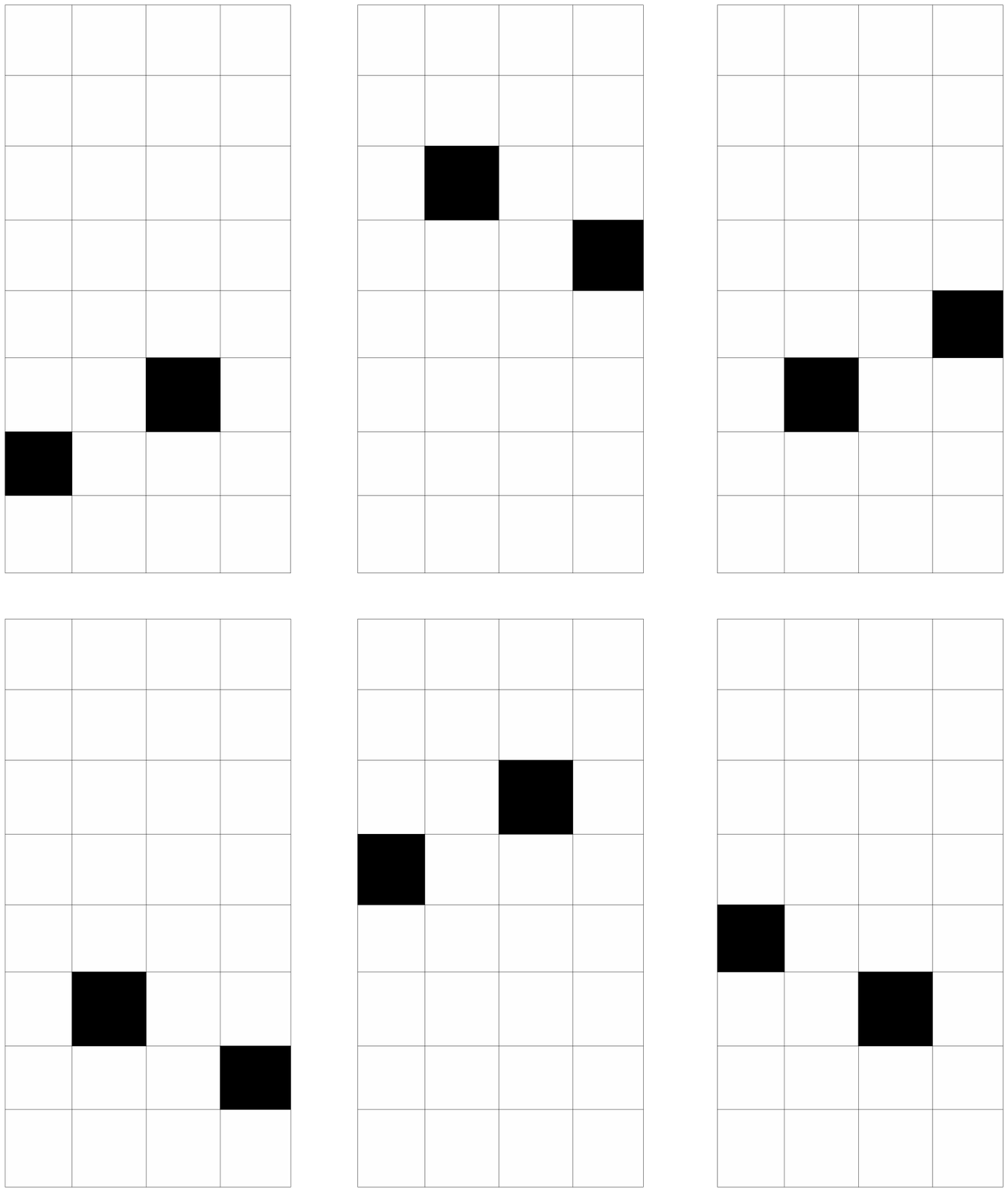}}}
{\scalebox{.13}{\includegraphics{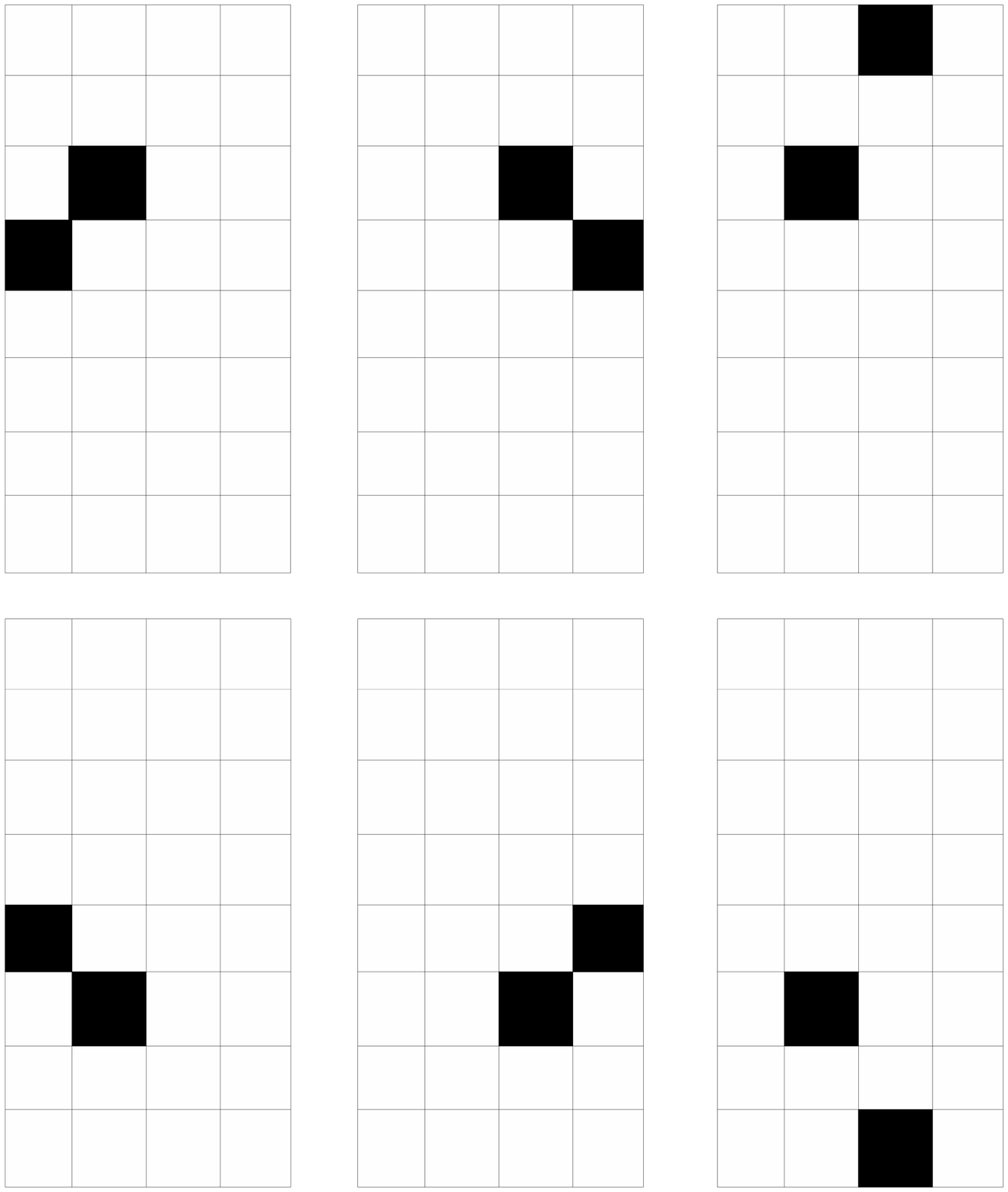}}}
{\scalebox{.13}{\includegraphics{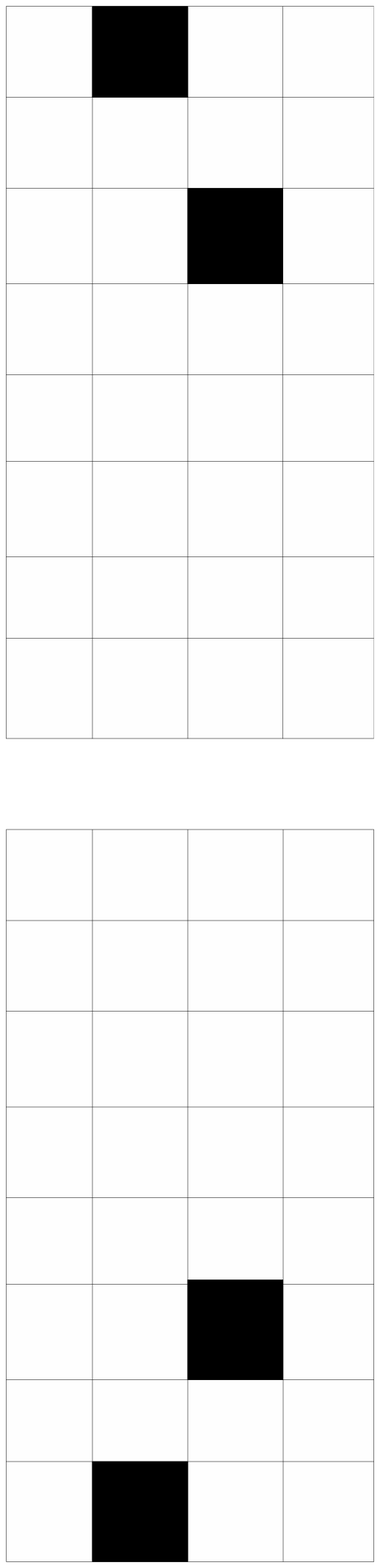}}}
}
\caption{Bad pairs for the general $R(8,4)^{--}$.}
\end{figure}

\begin{theorem}
The only bad pairs (apart from those enumerated in section $4$) for $R(3t+8,4)^{--}$, where $t\geq0$,  
are those corresponding to the pairs in Figure 12. 
\end{theorem}
\begin{proof}
We leave the proof of the badness of the pairs shown in Figure 12 
for the reader. We prove the existence of a tiling in all other cases. 
First consider the case when there exist two distinct $4\times4$ squares 
such that each square contains exactly one missing square, and the remaining area of $R(3t+8,4)$ can be 
divided into $R(3,4)$ rectangles.
Following our additive decomposition notation, we can decompose $R(3t+8,4)^{--}$ as shown in 
equation (33). A $3\times4$ rectangles satisfies the conditions of the 
Chu-Johnsonbaugh Theorem, and so is tileable, while $R(4,4)^{-}$ is 
tileable since it satisfies the conditions of the 
Deficient Rectangle Theorem. 
It follows from above that $R(3t+8,4)^{--}$ 
can be tiled in this case. 

\begin{figure}[htbp]
\centerline{
{\scalebox{.2}{\includegraphics{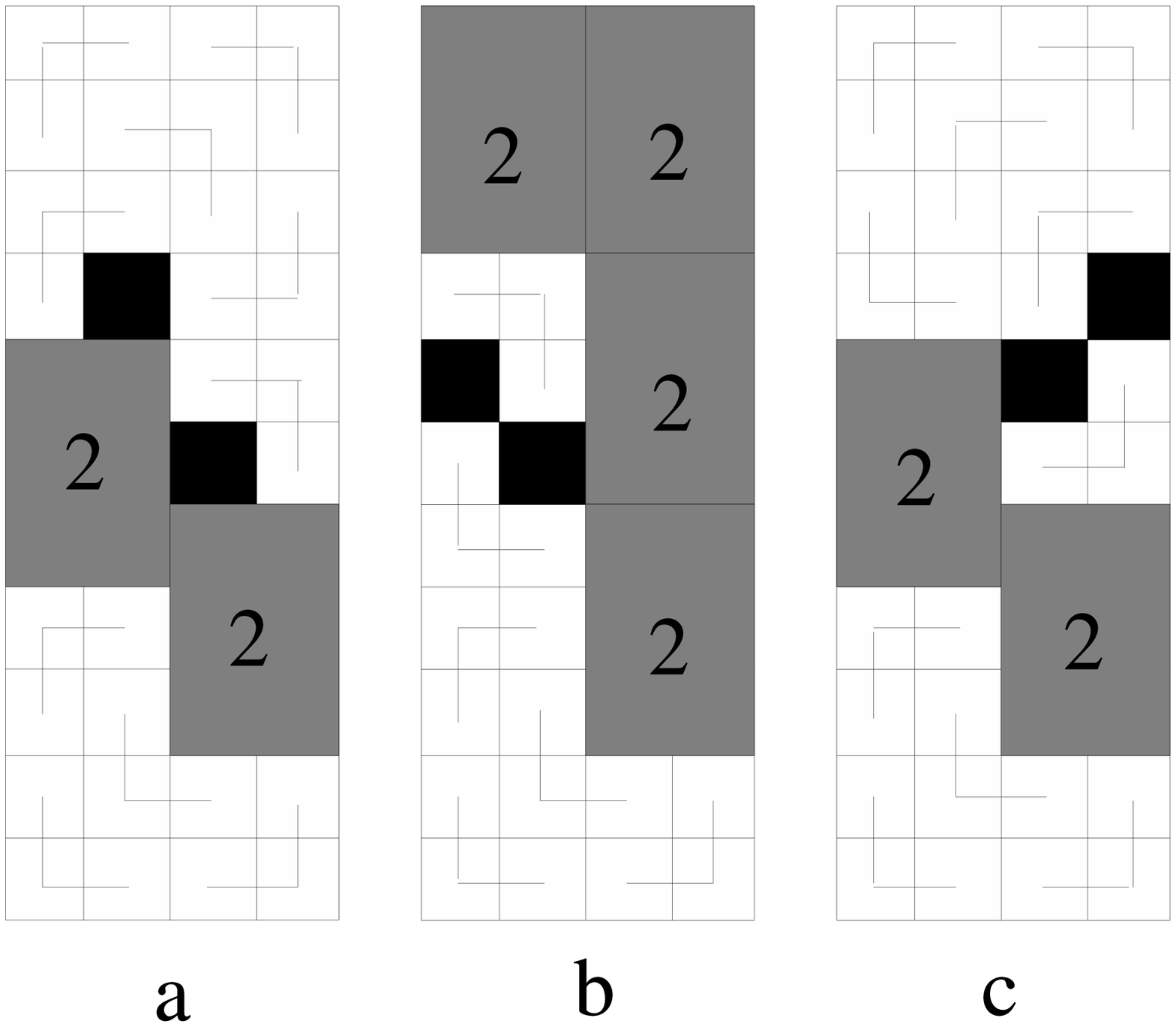}}}
{\scalebox{.2}{\includegraphics{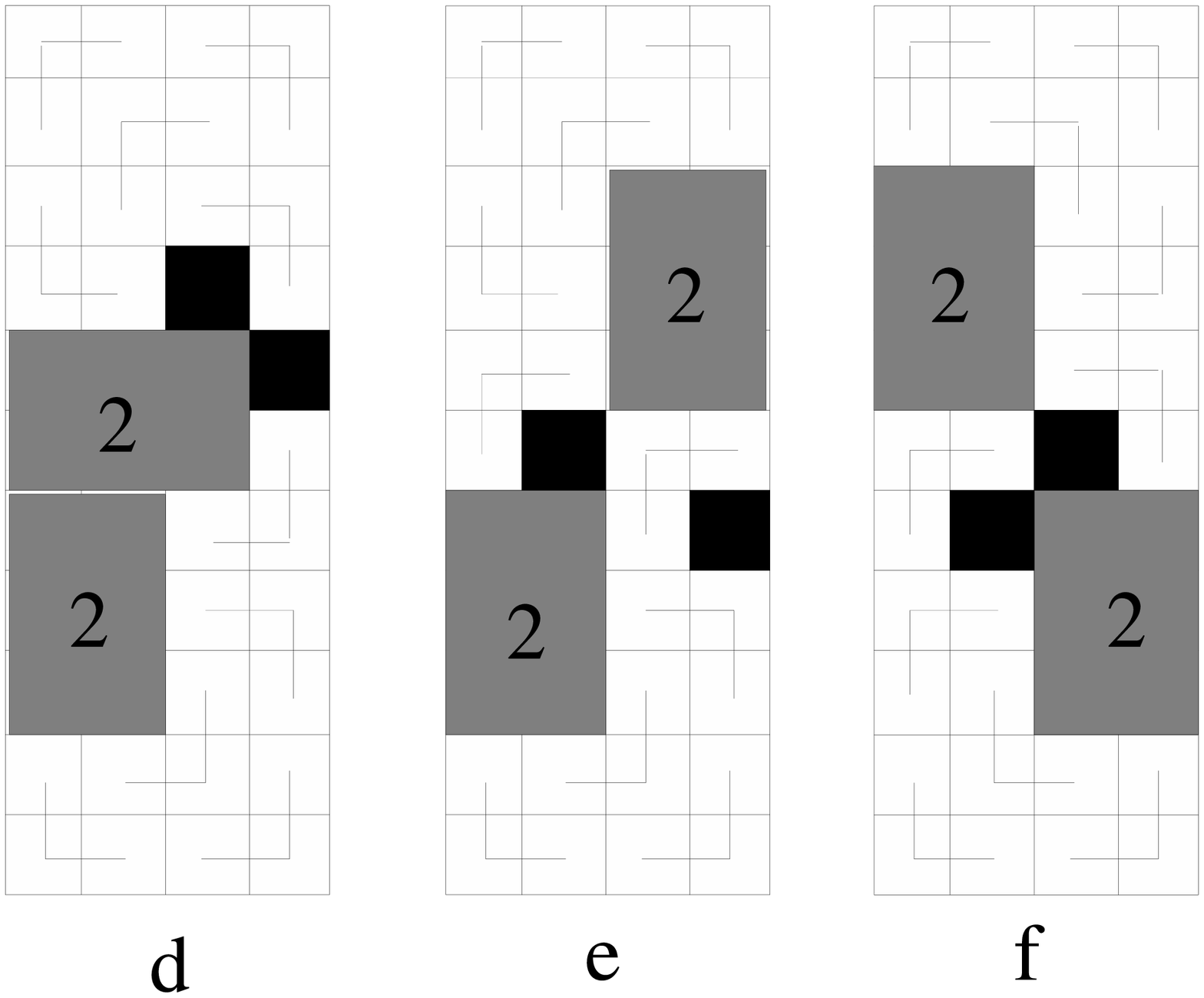}}}
}
\caption{Tiling the general $R(3t+8,4)^{--}$.}
\end{figure}

\begin{eqnarray}
R(3t+8,4)^{--} & = & t\cdot R(3,4) + 2R(4,4)^{-} \\
R(3t+8,4)^{--} & = & t\cdot R(3,4) + R(8,4)^{--}
\end{eqnarray}

We now consider the case when the position of the two missing squares does 
not permit the decomposition in the former case. 
It is easy to see that in this case, we can decompose $R(3t+8,4)^{--}$ as shown in (34). 
If the missing squares in $R(8,4)^{--}$ do not form a bad pair, then we are done. 
Otherwise, the missing squares form one of the bad pairs shown in 
Figure 12. In this case, we join a removed $3\times4$ rectangle from the top (bottom), 
and apply the corresponding tiling rule among the various cases shown in Figure 13. 
So, in this case, we tile $R(11,4)^{--}$ 
as shown, and separately tile $(t-1)$ subrectangles of dimension 
$3\times4$, to get a tiling of $R(3t+8,4)^{--}$.  \hfill\qed
\end{proof}

\section{Final Remarks}

We are currently exploring general 2-deficiency in rectangles 
and wish to characterize bad pairs for arbitrary $m\times n$ rectangles. 
The proof of the last 
theorem is particularly important because it suggests an 
approach for solving general $2$-deficiency 
in large rectangles. When the missing squares are far apart, the given rectangle 
can be broken down into two subrectangles, 
such that each subrectangle contains exactly one missing 
square. Now the Deficient Rectangle Theorem may be used to find a tiling rule, if one exists. 
However, one would have to consider certain cases as ``basis" cases, before using the above argument, 
as we did in the proof of the Domino-Deficient Rectangle Theorem in Section 6. 
We observed that the number 
of bad pairs for these ``basis" cases is extremely large. So, 
analysis using existing techniques becomes extremely complicated. 
Moreover, showing non-existence of a tiling by this method seems to be rather hard. 
We believe that enumeration of bad pairs may be analysed by studying 
possible transformations that map between classes of bad pairs. 
The hquad and vquad shifts are examples of such 
transformations, however, they cannot generate {\it all} 
bad pairs starting from a single one. Furthermore, 
as a sequel to 2-deficiency problems, one may also consider 
the question of $k$-deficiency, for $k\geq3$. 

Another research direction would be to establish a lower bound on the number 
of tromino tilings of $m\times n$ domino-deficient rectangles; it is an 
interesting open problem to determine whether there exists a lower bound comparable 
to our upper bound derived in inequality (31), Section 7. No such 
lower bounds have been established till date. 
Partitioning schemes such as the one used in
the Domino-Deficient Tiling Procedure in Section 6,
lead to lower bounds on the number of tilings of the 
entire rectangle obtained by multiplying lower bounds on the number of tilings for 
each part.  Devising partitioning schemes that lead to good lower bounds 
is an interesting open problem.

\section{Acknowledgements}

I would like to give my sincere thanks to Prof. Sudebkumar Prasant Pal, my mentor, for his 
valuable suggestions, constant encouragement and guidance.  
I would also like to thank the two anonymous referees for their excellent suggestions which greatly 
helped in improving the final version of the paper.

\end{document}